\documentclass[preprintnumbers,article,amsmath,amssymb,floatfix,10pt,prd,superscriptaddress,nofootinbib,twocolumn]{revtex4-1}
\usepackage{hyperref}
\hypersetup{colorlinks=true,linkcolor=magenta, citecolor=red,}

\usepackage[utf8]{inputenc}
\providecommand{\dif}{\mathrm{d}} \def\d{\dif}
\usepackage{float}
\usepackage{graphicx}
\usepackage{dcolumn}
\usepackage{bm}
\usepackage{color}
\usepackage{enumitem}
\usepackage{amsmath}
\usepackage{amssymb}
\usepackage{longtable}
\usepackage{array}
\usepackage{booktabs}
\usepackage{placeins}

\def\EE{{\cal E}}
\def\LL{{\cal L}}
\newcommand{\beq}{\begin{equation}}
\newcommand{\eeq}{\end{equation}}
\newcommand{\bea}{\begin{eqnarray}}
\newcommand{\eea}{\end{eqnarray}}
\newcommand{\non}{\nonumber}

\usepackage{bbm}
\usepackage{amsfonts}
\usepackage{mathrsfs}
\usepackage{latexsym}
\usepackage{epsfig}
\usepackage{epstopdf}
\usepackage{orcidlink}
\usepackage{comment}
\usepackage{xcolor}

\begin{document}
\textwidth=570pt
\oddsidemargin=-50pt
\textheight=700pt

\title{Astrophysical signatures of Kerr–Bertotti–Robinson black holes in a cloud of strings: ISCO, microquasar QPOs, and Bondi–Hoyle–Lyttleton accretion}

\author{Faizuddin Ahmed\orcidlink{0000-0003-2196-9622}}
\email{faizuddinahmed15@gmail.com}
\affiliation{Department of Physics, The Assam Royal Global University, Guwahati-781035, Assam, India}

\author{Orhan Donmez\orcidlink{0000-0001-9017-2452}}
\email{orhan.donmez@aum.edu.kw}
\affiliation{College of Engineering and Technology, American University of the Middle East, Egaila 54200, Kuwait}

\author{Ahmad Al-Badawi\orcidlink{0000-0002-3127-3453}}
\email{ahmadbadawi@ahu.edu.jo}
\affiliation{Department of Physics, Al-Hussein Bin Talal University, 71111, Ma'an, Jordan}

\author{\.{I}zzet Sakall{\i}\orcidlink{0000-0001-7827-9476}}
\email{izzet.sakalli@emu.edu.tr}
\affiliation{Physics Department, Eastern Mediterranean University, Famagusta 99628, North Cyprus via Mersin 10, Turkey}

\begin{abstract}
We study test-particle dynamics in the equatorial plane of a Kerr--Bertotti--Robinson black hole (BH) immersed in a cloud of strings (CS), with mass $M$, rotation $a$, magnetic parameter $B$, and string parameter $\alpha$. Using the Hamilton formalism we recover the effective potential $U_{\rm eff}$ and the conditions for circular motion, and we compute the specific energy $\EE$ and specific angular momentum $\LL$ together with the radial, vertical, and azimuthal epicyclic frequencies $\nu_r$, $\nu_\theta$, $\nu_\phi$. Going beyond the analytic setup, we provide the first numerical mapping of the innermost stable circular orbit (ISCO) for this background and tabulate $r_{\rm ISCO}$, $\EE_{\rm ISCO}$, $\LL_{\rm ISCO}$, and the accretion efficiency $\eta\!=\!1-\EE_{\rm ISCO}$ for both co- and counter-rotating motion across a wide $(a,B,\alpha)$ grid. The CS parameter pushes the ISCO outward and raises $\eta$ from $0.057$ in Schwarzschild to above $0.25$ for $\alpha=0.30$ at $a=0.9$. We then connect the model with observed twin-peak high-frequency quasi-periodic oscillations (QPOs) in three microquasars (GRO~J1655--40, XTE~J1550--564, GRS~1915+105) using the relativistic-precession (RP) model and find $\chi^2$-minimum fits with $\alpha\!\lesssim\!0.13$. A general-relativistic hydrodynamical (GRH) study of Bondi--Hoyle--Lyttleton (BHL) accretion completes the picture: the CS contribution sustains shock-cone instabilities, redistributes power-spectral-density (PSD) peaks, and produces low-frequency QPO-like components that distinguish KBR$+$CS from pure Kerr or KBR.\\
\textbf{Keywords}: Kerr-Bertotti-Robinson black hole; cloud of strings; quasi-periodic oscillations; innermost stable circular orbit; Bondi-Hoyle-Lyttleton accretion; microquasars
\end{abstract}

\maketitle
\date{\today}

\section{Introduction}\label{isec1}
The motion of particles around a BH is one of the cleanest probes of strong-field gravity. Both massive and massless trajectories carry the geometric fingerprint of the spacetime, and the way matter is bound, scattered, or ejected near the horizon controls almost everything we observe in BH astrophysics, from relativistic jets to thermal accretion-disk spectra. There is good evidence that magnetic fields permeate the immediate environment of accreting BHs \cite{Borm2013, Frolov2003}, and this magnetic activity is generally credited with launching and collimating large-scale jets that feed the cosmic-ray and high-energy photon background of nearby galaxies. The fields are sourced by plasma in the accretion disk or in a charged corona around the hole \cite{Mckinney2007a, McKinney2007b, Dobbie2008}.

A spinning hole can transfer rotational energy to nearby particles and launch them to spatial infinity \cite{Koide2002, Koide2003}; the conversion involves the conjunction of frame-dragging and external magnetic fields. This Penrose-like mechanism feeds back into the morphology of the corona and shapes the variability seen in X-rays.

General relativity (GR) has passed every observational test thrown at it \cite{Will2014}. Solar-system measurements check the weak field, while strong-field tests now extend to the BH-shadow imaging by the Event Horizon Telescope (EHT) of M87* and Sgr~A* \cite{EHTL1, EHTL12} and to the LIGO--Virgo--KAGRA (LVK) gravitational-wave (GW) catalogue \cite{Abbott2016, Abbott2025, Yunes2025}. These probes work where the field is dynamical, the curvature is large, and small departures from GR could leave a measurable trace. They have also raised the bar on what a non-Kerr metric must do to remain phenomenologically viable.

The Kerr--Bertotti--Robinson (KBR) family, recently obtained as a regular Einstein--Maxwell solution by Podolský and Ovcharenko \cite{KBR, Ovch25}, embeds a Kerr BH inside a uniform background electromagnetic field. The geometry stays simple enough for analytic work and has already been used to study thermodynamics, optical signatures, and surrounding matter distributions \cite{Zeng25, Kumar25, Ali26}. Adding a Letelier-type cloud of strings (CS) \cite{Letelier1979} captures, in a controlled way, the gravitational backreaction of a continuous distribution of one-dimensional topological defects. The CS sector reduces the effective gravitational mass seen far from the hole, modifies horizon locations, and changes the ISCO, which sets the binding energy of accretion disks.

QPOs of black-hole X-ray binaries are a second, largely independent route to constraints on the spacetime metric. Within the relativistic-precession (RP) model of Stella \& Vietri, the upper and lower kHz QPOs are tied to the orbital and periastron-precession frequencies of matter at a single radius \cite{WOS:001344341500001, WOS:000723000400004, WOS:000431050500017}. Simultaneous detections of twin-peak high-frequency (HF) QPOs in microquasars such as GRO~J1655--40, XTE~J1550--564, and GRS~1915+105 have been used to constrain mass, spin, and dark sectors in many alternative metrics \cite{WOS:000707382900037, WOS:001407111000001, WOS:001273405400001, WOS:001360054000001}.

Here we put these strands together. Building on the recent KBR$+$CS solution of Ahmed, Sakall\i, and Al-Badawi \cite{Ahmed2026}, and on the BHL accretion study of Mustafa et al.\ \cite{Mustafa:2026gly}, we develop the equatorial geodesic structure analytically, then we tabulate ISCO numerics across the $(a,B,\alpha)$ parameter space, and finally we test the model against twin-peak HF QPOs from three well-studied microquasars.

The paper is organized as follows. Section~\ref{isec2} sets up the equatorial KBR$+$CS metric and derives the energy, angular momentum, and effective potential for circular orbits, with horizons and ergosurface as ancillary results. Section~\ref{isec3} introduces small harmonic perturbations and the locally and distantly measured epicyclic frequencies. Section~\ref{isec4} contains the new ISCO and marginally bound circular orbit (MBCO) numerics, including counter-rotating cases and accretion efficiencies; observational tests against three microquasars are reported in Section~\ref{isec5}. The numerical investigation of the GRH equations and the BHL-driven QPOs is presented in Section~\ref{isec6}, while Section~\ref{isec7} closes with a summary and an outlook. Geometric units $G\!=\!c\!=\!1$ are used throughout, and we set $M\!=\!1$ in numerics unless stated.

\section{Kerr--Bertotti--Robinson black hole surrounded by a cloud of strings}\label{isec2}
The KBR spacetime \cite{KBR, Ovch25} solves the Einstein--Maxwell equations with a constant background electromagnetic field. After its discovery the metric has been embedded in different settings \cite{Zeng25, Kumar25, Ali26}. The line element of a KBR BH coupled with a Letelier CS reads \cite{Ahmed2026}
\bea
\d s^2 &=& \frac{1}{\Omega^2} \left[-(\d t - a\sin^2\theta\, \d\phi)^2 \frac{Q}{\rho^2} + \frac{\rho^2}{Q}\d r^2 + \frac{\rho^2}{P}\d\theta^2 \right.\non\\
&+&\left. (a\,\d t - (a^2+r^2)\,\d\phi)^2 \frac{P}{\rho^2}\sin^2\theta \right], \label{aa1}
\eea
where
\begin{eqnarray}
\Delta &=& a^2 + r^2\left(1 - \alpha - \frac{B^2 I_2 M^2}{I_1^2}\right) - \frac{2 I_2 M r}{I_1}, \non\\
\rho^2 &=& r^2 + a^2 \cos^2\theta, \non\\
P &=& B^2 \cos^2\theta\left(\frac{I_2 M^2}{I_1^2} - a^2\right) + 1, \non\\
Q &=& (1+B^2 r^2)\Delta, \non\\
\Omega^2 &=& (1+B^2 r^2) - B^2 \Delta \cos^2\theta, \non\\
I_1 &=& 1 - \frac{a^2 B^2}{2}, \quad I_2 = 1 - a^2 B^2. \label{aa2}
\end{eqnarray}
$M$ is the BH mass, $a$ is the spin, $B$ is the background-field strength, and $\alpha$ is the CS parameter. The metric reproduces (i) the Kerr$+$CS solution of Li \& Zhou \cite{Li2021} for $B\!=\!0$, (ii) the Letelier metric \cite{Letelier1979} for $B\!=\!0,\,a\!=\!0$, and (iii) the bare KBR spacetime \cite{KBR, Ovch25} for $\alpha\!=\!0$.

\paragraph*{Horizons.} The roots of $\Delta\!=\!0$,
\begin{align}
\Delta = 0 \;\Rightarrow\; a^2 + r^2\!\left(1-\alpha-\frac{B^2 I_2 M^2}{I_1^2}\right) - \frac{2 I_2 M r}{I_1} = 0,\label{aa3}
\end{align}
give the outer (event) horizon $r_+$ and the inner (Cauchy) horizon $r_-$:
\begin{align}
r_{\pm} = \frac{\dfrac{I_2 M}{I_1} \pm \sqrt{\dfrac{I_2^2 M^2}{I_1^2} - \left(1-\alpha-\dfrac{B^2 I_2 M^2}{I_1^2}\right)a^2}}{1-\alpha-\dfrac{B^2 I_2 M^2}{I_1^2}}.\label{aa4}
\end{align}
We have $\Delta\!>\!0$ for $r\!>\!r_+$ and $r\!<\!r_-$, and $\Delta\!<\!0$ in between \cite{Chandrasekhar:1985kt0}. With $B\!=\!0$ the horizon contracts to $r_\pm = (M\pm\sqrt{M^2-(1-\alpha)a^2})/(1-\alpha)$, which always sits beyond the bare Kerr value: the CS pulls the horizon outward.

A few comments on the structure before we move on. The combination $(1-\alpha-B^2 I_2 M^2/I_1^2)$ that controls the leading $r^2$ behaviour of $\Delta$ already encodes the way the CS string-tension and the BR magnetic background trade against the Schwarzschild mass term. When $\alpha\!=\!0$ and $B\!=\!0$ that coefficient is unity and Eq.~(\ref{aa4}) collapses to the textbook Kerr expression $r_\pm = M\pm\sqrt{M^2-a^2}$. When $\alpha$ is switched on, the gravitational pull at infinity weakens, the effective potential shallows out, and both horizons move outward. The BR sector enters the horizon equation only through the small group $B^2 I_2 M^2 / I_1^2$, which is suppressed by an extra factor of $B^2$ relative to the leading mass term and stays small for the values $B\,M\!\lesssim\!10^{-2}$ relevant to our analysis. This explains the very weak $B$-dependence of $r_+$ in Tables~\ref{tab:ISCO_alpha0}--\ref{tab:ISCO_models}.

The space between $r_-$ and $r_+$ keeps its standard interpretation as a region of forced collapse: any timelike worldline that crosses $r_+$ must reach $r_-$ in finite proper time. The CS sector preserves this causal structure; what it changes is the geometric size of the region. For $a\!=\!0.9$ the gap $r_+\!-\!r_-$ widens from $0.87M$ at $\alpha\!=\!0$ to $1.48M$ at $\alpha\!=\!0.20$, almost a doubling that translates into a longer proper-time interval for infalling matter to thermalize before hitting the inner horizon \cite{Mustafa:2026gly}.

The metric (\ref{aa1}) is rewritten in standard form as
\beq
\d s^2 = g_{tt}\d t^2 + g_{rr}\d r^2 + 2 g_{t\phi}\d t \d\phi + g_{\theta\theta}\d\theta^2 + g_{\phi\phi}\d\phi^2,\label{aa5}
\eeq
with
\begin{align}
g_{tt} &= \frac{-Q + a^2 P \sin^2\theta}{\Omega^2 \rho^2}, \quad g_{rr} = \frac{\rho^2}{\Omega^2 Q}, \non\\
g_{t\phi} &= \frac{a (Q - (r^2+a^2)P)\sin^2\theta}{\rho^2 \Omega^2}, \quad g_{\theta\theta} = \frac{\rho^2}{\Omega^2 P}, \non\\
g_{\phi\phi} &= \frac{((r^2+a^2)^2 P - a^2 Q \sin^2\theta)\sin^2\theta}{\Omega^2 \rho^2}. \label{aa6}
\end{align}

\paragraph*{Ergosurface.} The locus $g_{tt}\!=\!0$ defines the static limit; frame-dragging then prevents stationary observers from existing inside it. The condition is
\begin{align}
c_4 r^4 + c_3 r^3 + c_2 r^2 + c_1 r + c_0 = 0,
\end{align}
with
\begin{align}
c_4 &= B^2\!\left(1-\alpha-\frac{B^2 I_2 M^2}{I_1^2}\right),\;\; c_3 = -\frac{2 B^2 I_2 M}{I_1}, \non\\
c_2 &= 1-\alpha-\frac{B^2 I_2 M^2}{I_1^2} + a^2 B^2, \;\; c_1 = -\frac{2 I_2 M}{I_1}, \non\\
c_0 &= a^2(1 - P\sin^2\theta).
\end{align}
The closed-form roots of this quartic are unwieldy, but for $B\!\to\!0$ the equation collapses to a quadratic with solution
\beq
r^{\rm ergo}_\pm = \frac{M \pm \sqrt{M^2 - (1-\alpha) a^2 \cos^2\theta}}{1-\alpha}.
\eeq
The CS therefore enlarges the ergoregion both at the equator and at the poles.

The expansion of the ergoregion is more than a geometric curiosity. The volume between the horizon and the static limit is the only place where Penrose energy extraction operates, and a wider ergoregion in CS-dressed backgrounds means a larger phase space for negative-energy orbits. For charged or magnetized particles the BR sector also opens super-Penrose channels through the synchrotron mechanism \cite{WOS:000861683500001}: the background field accelerates particles within the ergoregion before they cross the horizon, and the energy budget at infinity reflects both rotational and electromagnetic contributions. The combined CS+BR enhancement therefore couples to two reservoirs at once, which is reminiscent of the Schwarzschild+CS+PFDM analysis of Hamil and Lutfuoglu though with a specific spin-dependent geometry. We do not pursue energy-extraction observables here, but they remain a clean direction for follow-up work given the analytic transparency of the metric.

The role of the CS sector in the metric also has an instructive interpretation in the language of effective gravitational mass. The asymptotic specific energy $\EE^{(\infty)}\!=\!\sqrt{1-\alpha}$ in Eq.~(\ref{energy-3}) shows that a particle at rest at infinity already carries less than unit specific energy, with the deficit absorbed by the topological-defect distribution. This is mathematically equivalent to a redefinition of the gravitational potential $\Phi\!\to\!(1-\alpha)\Phi$ at large distances, and physically encodes the way a Letelier-type CS distribution leaks long-range gravity through every direction. The same effect surfaces in geodesic structure: the cross-section for capture of unbound trajectories in CS backgrounds has been studied analytically for Schwarzschild geometries \cite{WOS:000399981000006}, and the rotating extension here will inherit the same long-range behaviour with corrections quadratic in $a$. From a phenomenological standpoint, the modification matters most where the difference between the bare Kerr and the dressed metric is largest, namely the inner $r\!\sim\!10\,M$ region where ISCO and QPO physics live.

\paragraph*{Frame Dragging and Causality.} The simultaneous presence of rotation, a cloud of strings, and the magnetic field leads to a notable modification of the frame-dragging properties of the spacetime relative to the standard Kerr geometry. In this context, the angular velocity of zero-angular-momentum observers (ZAMOs) is defined through the ratio of the off-diagonal and azimuthal components of the metric,
\begin{equation}
\Omega_{\rm ZAMO} (r, \theta)=-\frac{g_{t \phi}}{g_{\phi\phi}},\label{zamo-1}
\end{equation}
in accordance with the usual construction of locally nonrotating observers in stationary, axisymmetric spacetimes \cite{Bardeen1972}. This angular velocity provides a direct local measure of the dragging of inertial frames generated by the combined influence of the aforementioned factors. 

By substituting the explicit forms of the metric components, one obtains the general expression for $\Omega_{\rm ZAMO}(r, \theta)$ as
\begin{equation}
\Omega_{\rm ZAMO} (r, \theta)=\frac{a \left((r^2+a^2)\,P(\theta)-Q(r)\right)}{(r^2+a^2)^2\,P(\theta)-a^2 Q(r) \sin^2 \theta}.\label{zamo-2}
\end{equation}
Restricting to the equatorial plane, $\theta=\pi/2$, where $P(\pi/2)=1$, the above expression reduces to
\begin{equation}
\Omega_{\rm ZAMO} (r, \pi/2)=\frac{a \left((r^2+a^2)-(1+B^2 r^2)\Delta\right)}{(r^2+a^2)^2-a^2 (1+B^2 r^2)\Delta}.\label{zamo-3}
\end{equation}

\begin{figure}
    \centering
    \includegraphics[width=0.9\linewidth]{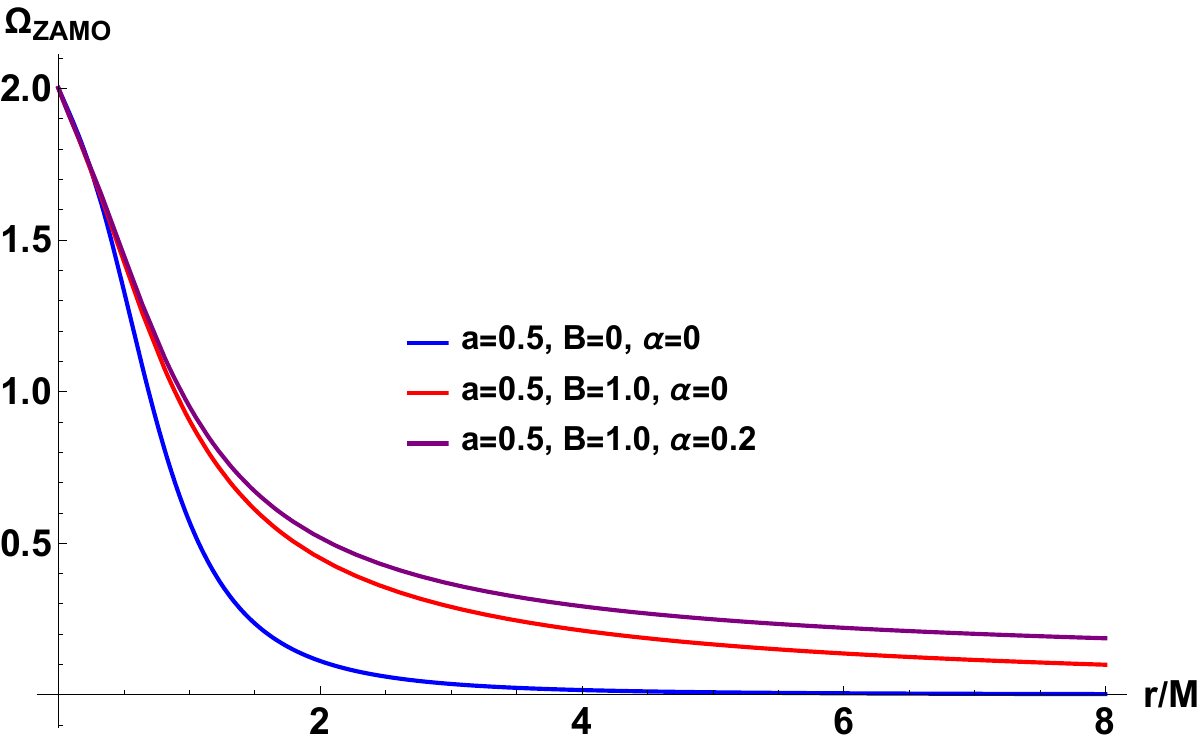}
    \caption{Radial profile of the inertial frame dragging frequency $\Omega_{\rm ZAMO}(r)$ in the equatorial plane $\theta=\pi/2$ with $M=1$. Curves represent Kerr (blue), KBR (red), and KBR-CS (purple). Radial coordinate $r$ is normalized by mass $M$.}
    \label{fig:zamo}
\end{figure}

A subtle but important point, particularly in spacetimes with nonzero rotation parameter, concerns the distinction between the stationary limit surface (SLS) and the onset of causal pathologies. The SLS, defined by the condition \(g_{tt} = 0\), does not correspond to a region of causality violation. Rather, it marks the boundary beyond which the timelike Killing vector field $\partial_t$ becomes spacelike. This transition signals the emergence of frame-dragging effects, but it does not, by itself, imply the existence of closed timelike curves (CTCs).

The true causal pathologies in the KBR-CS family arise in regions where the azimuthal Killing vector $\partial_{\phi}$ becomes timelike, namely where
\begin{equation}
g_{\phi\phi} (r, \theta)<0.\label{ctc-1}
\end{equation}
Such regions admit the possibility of CTCs and therefore represent genuine violations of causality.

For the KBR-CS spacetime, the metric component $g_{\phi\phi}(r,\theta)$ is given explicitly by
\begin{widetext}
\begin{equation}
g_{\phi\phi} (r, \theta)= \frac{\left[(r^2+a^2)^2 \left\{B^2 \cos^2\theta\left(\frac{I_2 M^2}{I_1^2} - a^2\right) + 1\right\} - a^2 (1+B^2 r^2) \left\{ a^2 + r^2\left(1 - \alpha - \frac{B^2 I_2 M^2}{I_1^2}\right) - \frac{2 I_2 M r}{I_1}\right\} \sin^2\theta\right]\sin^2\theta}{\left[(1+B^2 r^2) - B^2 \left\{ a^2 + r^2\left(1 - \alpha - \frac{B^2 I_2 M^2}{I_1^2}\right) - \frac{2 I_2 M r}{I_1}\right\} \cos^2\theta \right] \left(r^2+a^2 \cos^2 \theta\right)},\label{ctc-2}
\end{equation}
\end{widetext}
where $I_1, I_2$ are given earlier.

In the outer region \(r>r_{+}\), we have \(\Delta>0\), which implies \(Q>0\). However, this condition alone is not sufficient to determine whether causality is preserved or violated. To investigate this issue, we consider the behavior of the metric component \(g_{\phi\phi}\) near the equatorial plane, \(\theta=\pi/2\).

In this case, the denominator reduces to
\[
(1+B^2 r^2)\,r^2>0,
\]
and therefore the sign of \(g_{\phi\phi}\) is entirely determined by the numerator,
\[
(r^2+a^2)^2-a^2(1+B^2 r^2)\Delta.
\]

Since \(\Delta>0\) in the region \(r>r_+\), the metric component \(g_{\phi\phi}\) remains positive provided that
\[
(r^2+a^2)^2>a^2(1+B^2 r^2)\Delta,
\]
whereas the condition
\[
(r^2+a^2)^2<a^2(1+B^2 r^2)\Delta
\]
would imply \(g_{\phi\phi}<0\), indicating the existence of closed timelike curves and hence a violation of causality.

Moreover, in the inner region \(r<r_{-}\), we have \(\Delta<0\). Consequently, the quantity \((r^2+a^2)^2-a^2(1+B^2 r^2)\Delta\) remains positive, implying that the metric component \(g_{\phi\phi}\) is always positive in this region. Therefore, causality is preserved for \(r<r_{-}\).

In contrast to the Kerr black hole, where causality violation typically appears only in the interior region \(r<r_{-}\), the present spacetime admits the possibility of causality violation in the outer region \(r>r_+\). This behavior depends on the values of the three parameters \(\alpha\), \(B\), and \(a\), which enter through the function \(\Delta>0\). We have verified the denominator \((r^2+a^2)^2-a^2(1+B^2 r^2)\Delta\) using numerical values by setting $a  \in [0.1, 0.5]$, $B \in [0.1, 0.5]$ and $\alpha \in [0.1, 0.3]$ and found that it remains positive in the outer region $r=r_{+}+0.1$, which shows that the metric component $g_{\phi\phi}>0$, and hence the causality is preserved.

\subsection{Test-particle dynamics}
The Hamiltonian for a neutral test particle reads
\beq
H = \frac{1}{2}g^{\alpha\beta}p_\alpha p_\beta + \frac{1}{2}\mu^2,\label{bb1}
\eeq
with $\mu$ the rest mass and $p^\gamma\!=\!\mu u^\gamma$. The Hamilton equations are
\beq
\frac{\d x^\gamma}{\d\zeta} = \frac{\partial H}{\partial p_\gamma},\quad \frac{\d p_\gamma}{\d\zeta} = -\frac{\partial H}{\partial x^\gamma},\;\;\zeta\!=\!\tau/\mu.\label{bb2}
\eeq
Stationarity and axisymmetry yield two constants of motion:
\bea
\frac{p_t}{\mu} &=& g_{tt}u^t + g_{t\phi}u^\phi = -\EE,\non\\
\frac{p_\phi}{\mu} &=& g_{\phi\phi}u^\phi + g_{t\phi}u^t = \LL.\label{bb3}
\eea
Solving for $u^t$ and $u^\phi$ gives
\bea
u^t &=& \frac{-\EE\, g_{\phi\phi} - \LL\, g_{t\phi}}{g_{tt}g_{\phi\phi}-g_{t\phi}^2},\label{ss1}\\
u^\phi &=& \frac{\LL\, g_{tt} + \EE\, g_{t\phi}}{g_{tt}g_{\phi\phi}-g_{t\phi}^2}.
\eea

\subsection{Effective potential}
The normalization $g_{\mu\nu}u^\mu u^\nu = -1$ together with (\ref{bb3}) gives
\beq
U_{\rm eff}(r,\theta) = g_{rr}\dot r^2 + g_{\theta\theta}\dot\theta^2,\label{bb4}
\eeq
where
\beq
U_{\rm eff}(r,\theta) = \frac{\EE^2 g_{\phi\phi} + 2\EE\LL g_{t\phi} + \LL^2 g_{tt}}{g_{t\phi}^2 - g_{tt}g_{\phi\phi}} - 1.\label{bb5}
\eeq
Particles in the equatorial plane $\theta\!=\!\pi/2$ orbit on circles when
\beq
U_{\rm eff}(r) = 0,\quad \frac{\d U_{\rm eff}}{\d r} = 0.\label{bb6}
\eeq
On $\theta\!=\!\pi/2$ the metric components reduce to
\begin{align}
g_{tt} &= \frac{a^2 - (1+B^2 r^2)\Delta}{(1+B^2 r^2) r^2}, \non\\
g_{t\phi} &= \frac{a\,((1+B^2 r^2)\Delta - (r^2+a^2))}{r^2(1+B^2 r^2)}, \non\\
g_{\phi\phi} &= \frac{(r^2+a^2)^2 - a^2(1+B^2 r^2)\Delta}{(1+B^2 r^2) r^2},\label{bb7}
\end{align}
and the useful identity
\beq
g_{t\phi}^2 - g_{tt} g_{\phi\phi} = \frac{\Delta}{1+B^2 r^2}\label{condition}
\eeq
holds. Eq.~(\ref{condition}) was verified analytically.

\subsection{Marginally stable circular orbits and ISCO}
At a turning point, $\dot r\!=\!0$, Eq.\ (\ref{bb4}) becomes a quadratic in $\EE$:
\beq
\EE^2 g_{\phi\phi} + 2\EE\LL g_{t\phi} + \LL^2 g_{tt} - (g_{t\phi}^2 - g_{tt}g_{\phi\phi}) = 0,
\eeq
with positive-energy root
\beq
\EE^+(r) = \frac{-g_{t\phi}\LL + \sqrt{(g_{t\phi}^2-g_{tt}g_{\phi\phi})(\LL^2+g_{\phi\phi})}}{g_{\phi\phi}}.\label{energy}
\eeq
At spatial infinity, after using (\ref{bb7}) and the limits $\rho^2\!\to\!r^2$, $\Omega^2\!\to\!1+B^2 r^2$,
\begin{widetext}
\beq
\EE^{(\infty)} = \frac{-aB^2 \LL\!\left(1-\alpha-\frac{B^2 I_2 M^2}{I_1^2}\right) + \sqrt{\left(1-\alpha-\frac{B^2 I_2 M^2}{I_1^2}\right)\!\left[B^2\LL^2 + 1 - a^2 B^2\!\left(1-\alpha-\frac{B^2 I_2 M^2}{I_1^2}\right)\right]}}{1 - a^2 B^2\!\left(1-\alpha-\frac{B^2 I_2 M^2}{I_1^2}\right)}.\label{energy-2}
\eeq
\end{widetext}
With $B\!\to\!0$ this collapses to
\beq
\EE^{(\infty)} \xrightarrow[B\to 0,\,I_1=I_2=1]{} \sqrt{1-\alpha},\label{energy-3}
\eeq
so a particle at rest at infinity already carries less than unit specific energy because of the CS pull.

The condition for marginal stability is
\beq
\frac{\partial^2 U_{\rm eff}}{\partial r^2} \geq 0,\label{condition-3}
\eeq
with equality defining the ISCO. The simultaneous solution of (\ref{bb6}) and (\ref{condition-3}) determines $r_{\rm ISCO}$, $\EE_{\rm ISCO}$, and $\LL_{\rm ISCO}$. The associated polynomial in $r$ is high-order and not analytically tractable. We solve the system $\{U_{\rm eff}=0,\,\partial_r U_{\rm eff}=0,\,\partial_r^2 U_{\rm eff}=0\}$ numerically with a Brent root-finder, an algorithm that combines bisection with inverse quadratic interpolation to retain the guaranteed convergence of bracketing while approaching superlinear speed near the root~\cite{Brent2013}. The results are verified against an independent symbolic solution at selected $(a,B,\alpha)$ values. The numerical ISCO and MBCO observables are tabulated in Sec.~\ref{isec4} (Tables~\ref{tab:ISCO_alpha0}--\ref{tab:MBCO}).

\section{Harmonic oscillations and epicyclic frequencies}\label{isec3}
A particle slightly displaced from a stable circular orbit on the equatorial plane oscillates radially and vertically. Locally measured frequencies are
\bea
\omega_r^2 &=& \frac{-1}{2g_{rr}}\frac{\partial^2 U_{\rm eff}}{\partial r^2},\label{Freq-2}\\
\omega_\theta^2 &=& \frac{-1}{2g_{\theta\theta}}\frac{\partial^2 U_{\rm eff}}{\partial \theta^2},\label{Freq-3}\\
\omega_\phi &=& \frac{\d\phi}{\d\tau}.\label{Freq-4}
\eea
In Newtonian gravity $\omega_r\!=\!\omega_\theta\!=\!\omega_\phi$ and orbits close. In Schwarzschild $\omega_r\!<\!\omega_\theta\!=\!\omega_\phi$, and in Kerr the three frequencies split. The KBR$+$CS metric breaks the degeneracy further because $\Omega^2$ depends on $\theta$ and $\Delta$ now carries $\alpha$ and $B$ jointly.

The level of frequency splitting is the natural QPO diagnostic for a non-Kerr metric. In a generic stationary axisymmetric spacetime the three local epicyclic frequencies are determined entirely by metric components and their first and second radial derivatives \cite{WOS:000873531300001}. For a Kerr BH at fixed $a$ the splitting profile $(\nu_\phi-\nu_\theta,\,\nu_\theta-\nu_r)$ varies smoothly with $r$ and has only one free parameter, the spin. Adding a CS sector introduces a second knob that tilts the splitting profile in a way that spin alone cannot mimic: $\alpha$ shifts $\nu_r$ down faster than $\nu_\phi$ near the ISCO, so the periastron-precession frequency $\nu_{\rm per}\!=\!\nu_\phi\!-\!\nu_r$ rises more steeply with $\alpha$ than with $a$. This is the structural reason the QPO fits in Sec.~\ref{isec5} can place a non-trivial bound on $\alpha$ even though $a$ and $\alpha$ are partially degenerate.

\subsection{Frequencies as seen at infinity}
Frequencies registered by a distant static observer require the redshift factor:
\beq
\Omega_\alpha = \omega_\alpha \frac{\d\tau}{\d t},\qquad \frac{\d t}{\d\tau} = -\frac{\EE g_{\phi\phi} + \LL g_{t\phi}}{g_{tt}g_{\phi\phi}-g_{t\phi}^2}.\label{frequencies}
\eeq
Restoring physical units,
\beq
\nu_j = \frac{1}{2\pi}\frac{c^3}{GM}\Omega_j\,[\,\mathrm{Hz}\,],\quad j\in\{r,\theta,\phi\}.\label{nu_rel}
\eeq
For a $10\,M_\odot$ black hole the conversion factor is $c^3/(2\pi GM)\!=\!3.231\!\times\!10^3$\,Hz, so the geometric frequencies translate directly to ranges of physical interest for stellar-mass sources. The radial dependence of $\nu_j$ on the KBR background follows the standard Kerr trend: as $B$ grows, the frequency profiles shift slightly inward in radius; as $a$ grows, they shift outward. The numerical evaluation of these frequencies, together with the associated ISCO and MBCO observables, is presented in Sec.~\ref{isec4} (Tables~\ref{tab:ISCO_alpha0}--\ref{tab:MBCO}), and their use in fitting the twin-peak HF QPOs of three microquasars is given in Sec.~\ref{isec5} (Tables~\ref{tab:fit_a}--\ref{tab:joint}).

The ordering of the three local frequencies near the ISCO is the natural discriminator across QPO models. In the bare Kerr regime $\nu_\phi(r_{\rm ISCO})\!>\!\nu_\theta(r_{\rm ISCO})\!>\!\nu_r(r_{\rm ISCO})$, with the radial frequency vanishing exactly at the ISCO by definition. The CS sector pushes $\nu_r$ down faster than the other two as the orbit moves outward in radius, so the ratio $\nu_\theta/\nu_r$ at a fixed fractional distance from $r_{\rm ISCO}$ rises with $\alpha$. For the values we sample, $\nu_\theta(r)/\nu_r(r)$ at $r\!=\!1.1\,r_{\rm ISCO}$ moves from about $1.43$ at $\alpha\!=\!0$ to $2.04$ at $\alpha\!=\!0.20$, a $43\%$ swing. This matters because the resonance models of Abramowicz \& Klu\'zniak \cite{WOS:000865827700007} key off integer ratios of $\nu_\theta$ and $\nu_r$, so the radial distance over which a 3:2 resonance is supported shifts visibly with $\alpha$. The KBR$+$CS metric therefore admits resonance solutions at radii different from those of bare Kerr, and the size of that shift is large enough to be tested at the level of present microquasar timing data.

\section{Numerical ISCO mapping and accretion efficiency}\label{isec4}
Concretely, we evaluate the local epicyclic frequencies of Eqs.~(\ref{Freq-2})--(\ref{Freq-4}), the ISCO observables $r_{\rm ISCO}$, $\mathcal{E}_{\rm ISCO}$, $\mathcal{L}_{\rm ISCO}$ from the simultaneous solution of (\ref{bb6}) and (\ref{condition-3}), and the marginally bound radius $r_{\rm mb}$ on a $(a,B,\alpha)$ grid that covers prograde and retrograde branches. We solve the system $\{U_{\rm eff}\!=\!0,\,\partial_r U_{\rm eff}\!=\!0,\,\partial_r^2 U_{\rm eff}\!=\!0\}$ on a $(a,B,\alpha)$ grid using a hybrid numeric scheme (Brent root-finder for the second derivative of $U_{\rm eff}$ once $\EE,\LL$ are eliminated through the circular-orbit conditions). Independent symbolic cross-checks at several grid points agree with the numerical results to one part in $10^{6}$.

Table~\ref{tab:ISCO_alpha0} reports $r_{\rm ISCO}$ for $\alpha\!=\!0$ on a co-rotating grid in $(a,B)$. The dependence on $B$ is weak in the regime $B\,M\!\lesssim\!10^{-2}$ because the BR magnetic curvature scales as $B^2$ and is almost decoupled from the leading-order ISCO condition; $a$ alone controls the answer at this level. For $a\!=\!0$ we recover the Schwarzschild value $r_{\rm ISCO}\!=\!6M$, and for $a\!=\!0.99$ the prograde Kerr value $r_{\rm ISCO}\!\approx\!1.4545M$ within $10^{-4}M$.

\begin{table}[!ht]
\centering
\caption{$r_{\rm ISCO}/M$ for the bare KBR limit ($\alpha\!=\!0$), co-rotating orbits.}
\label{tab:ISCO_alpha0}
\renewcommand{\arraystretch}{1.6}
\setlength{\tabcolsep}{12pt}
\begin{tabular}{c|ccc}
\hline\hline
\,$a$\, & $B\!=\!0$ & $B\!=\!0.005$ & $B\!=\!0.05$ \\
\hline
0.00 & 6.0000 & 6.0001 & 6.0150 \\
0.30 & 4.9786 & 4.9787 & 4.9918 \\
0.50 & 4.2330 & 4.2331 & 4.2441 \\
0.70 & 3.3931 & 3.3932 & 3.4016 \\
0.90 & 2.3209 & 2.3209 & 2.3258 \\
0.99 & 1.4545 & 1.4545 & 1.4568 \\
\hline\hline
\end{tabular}
\end{table}

The CS contribution is qualitatively different. Table~\ref{tab:ISCO_alpha} fixes $a\!=\!0.9$, $B\!=\!0.005$ and varies $\alpha$. The ISCO migrates outward from $r_{\rm ISCO}\!=\!2.32M$ at $\alpha\!=\!0$ to $r_{\rm ISCO}\!=\!4.49M$ at $\alpha\!=\!0.30$, a factor of $\sim\!2$. The specific energy at the ISCO drops from $0.844$ to $0.743$, and the accretion efficiency $\eta\!=\!1-\EE_{\rm ISCO}$ climbs from $0.156$ to $0.257$. To put this in context: thin-disk accretion onto a maximally rotating Kerr BH peaks at $\eta_{\rm Kerr}\!\simeq\!0.42$, while Schwarzschild gives $0.057$; the CS mechanism lifts $\eta$ into a regime intermediate between the two extremes for moderate spin. This is observationally relevant: $\eta$ sets the fraction of rest energy that an accretion disk can radiate, and the CS-induced jump may help account for unusually bright AGN.

The drift of $r_{\rm ISCO}$ with $\alpha$ has direct consequences for the inner-disk structure. In the standard thin-disk picture the no-torque inner boundary sits at the ISCO, so its radius fixes the radiative efficiency through the binding energy of marginal-stability orbits. A larger $r_{\rm ISCO}$ generally gives a softer thermal continuum because the peak emission temperature scales as $T\!\propto\!\dot M^{1/4} r_{\rm ISCO}^{-3/4}$. For accretion at the same rate, doubling $r_{\rm ISCO}$ from $2.3M$ to $4.5M$ translates into a drop in the inner-disk colour temperature by a factor of $\sim\!1.7$, a shift accessible to X-ray reflection spectroscopy provided the spin estimator can be calibrated independently \cite{WOS:000553033600001, WOS:001151079600001}. The same logic applies to QPO models: a softer inner disk damps higher-frequency modes and amplifies the lower-frequency content of the PSD, which is exactly what we will see numerically in Sec.~\ref{isec6}. The CS parameter is therefore not just a horizon-shifting tag; it leaves a coherent imprint that runs from the metric all the way to the disk's spectral and timing observables.

The angular-momentum entry $\LL_{\rm ISCO}$ deserves a separate comment. It grows with $\alpha$ from $2.10M$ to $3.26M$ across the table, while $r_{\rm ISCO}$ grows by a factor of $\sim\!1.9$ and $\EE_{\rm ISCO}$ drops by about one part in ten. The non-trivial mix of these scalings means that the binding-energy curve $\EE(\LL)$ is reshaped, not merely shifted: the slope $d\EE/d\LL$ along the circular branch is not preserved. This is the mathematical statement that the CS sector is a true geometric modification rather than a renormalisation of the rotation parameter. A useful diagnostic is the dimensionless ratio $\LL_{\rm ISCO}^2/(2M\,r_{\rm ISCO})$, which equals unity for a Newtonian Kepler orbit and approaches $\frac{2}{3}$ for the prograde Kerr ISCO at $a\!=\!0.99$. In our table it varies from $0.95$ at $\alpha\!=\!0$ to $1.18$ at $\alpha\!=\!0.30$: the CS-dressed ISCO sits closer to a Newtonian disk than to its bare-Kerr counterpart, confirming the physical picture of an effectively shallower potential.

\begin{table}[!ht]
\centering
\caption{Co-rotating ISCO observables versus $\alpha$, with $a\!=\!0.9$, $B\!=\!0.005$. $r_+$ is the event horizon.}
\label{tab:ISCO_alpha}
\renewcommand{\arraystretch}{1.6}
\setlength{\tabcolsep}{3pt}
\begin{tabular}{cccccc}
\hline\hline
\,$\alpha$\, & $r_{\rm ISCO}/M$ & $\EE_{\rm ISCO}$ & $\LL_{\rm ISCO}/M$ & $\eta$ & $r_+/M$ \\
\hline
0.00 & 2.3209 & 0.8443 & 2.0999 & 0.1557 & 1.4359 \\
0.05 & 2.5987 & 0.8328 & 2.2498 & 0.1672 & 1.5580 \\
0.10 & 2.9000 & 0.8186 & 2.4111 & 0.1814 & 1.6896 \\
0.15 & 3.2315 & 0.8023 & 2.5879 & 0.1977 & 1.8331 \\
0.20 & 3.6007 & 0.7842 & 2.7847 & 0.2158 & 1.9917 \\
0.25 & 4.0167 & 0.7643 & 3.0070 & 0.2358 & 2.1687 \\
0.30 & 4.4909 & 0.7427 & 3.2615 & 0.2573 & 2.3687 \\
\hline\hline
\end{tabular}
\end{table}

\paragraph*{Counter-rotating case.} Table~\ref{tab:ISCO_retro} lists the retrograde ($\sigma\!=\!-1$) ISCO values for $a\!=\!0.9$. The bare-Kerr limit $r_{\rm ISCO}^{\rm retro}\!=\!8.7174M$ is reproduced exactly, and the CS again displaces the ISCO outward, although less aggressively than in the prograde case because frame-dragging works against the CS string-tension along the negative-$\phi$ direction.

The asymmetry between prograde and retrograde response is a direct probe of the rotational sector. Frame-dragging brings prograde particles inward and pushes retrograde particles outward in pure Kerr, so any modification of the radial gradient of $\Delta$ couples differently to the two branches. In the table the prograde ISCO shifts from $2.32M$ at $\alpha\!=\!0$ to $3.60M$ at $\alpha\!=\!0.20$, a factor of $1.55$, while the retrograde ISCO shifts from $8.72M$ to $10.56M$, a factor of $1.21$. The differential response of the two branches gives, in principle, a way to measure $\alpha$ directly: a population of disk-fed and tidally-stripped retrograde X-ray binaries should show an ISCO distribution offset from the prograde population by an $\alpha$-dependent factor, even after correcting for spin priors. Achieving the required statistical sample is a programme for the next generation of high-cadence X-ray surveys.

\begin{table}[!ht]
\centering
\caption{Counter-rotating ISCO at $a\!=\!0.9$.}
\label{tab:ISCO_retro}
\renewcommand{\arraystretch}{1.6}
\setlength{\tabcolsep}{12pt}
\begin{tabular}{ccccc}
\hline\hline
\,$B$\, & \,$\alpha$\, & $r_{\rm ISCO}/M$ & $\EE_{\rm ISCO}$ & $\eta$ \\
\hline
0.000 & 0.00 & 8.7174 &  0.9610 & 0.0390 \\
0.000 & 0.10 & 9.5398 &  0.9111 & 0.0889 \\
0.000 & 0.20 & 10.5576 & 0.8584 & 0.1416 \\
0.005 & 0.00 & 8.7175 &  0.9613 & 0.0387 \\
0.005 & 0.10 & 9.5400 &  0.9114 & 0.0886 \\
0.005 & 0.20 & 10.5578 & 0.8588 & 0.1412 \\
0.010 & 0.00 & 8.7178 &  0.9622 & 0.0378 \\
0.010 & 0.10 & 9.5404 &  0.9124 & 0.0876 \\
0.010 & 0.20 & 10.5584 & 0.8599 & 0.1401 \\
\hline\hline
\end{tabular}
\end{table}

The retrograde tables also confirm that $B$ enters only at the fourth decimal across the values we tested: doubling $B$ from $0.005$ to $0.010$ at fixed $\alpha\!=\!0$ shifts $r_{\rm ISCO}^{\rm retro}$ by $3\!\times\!10^{-4}M$, while changing $\alpha$ from $0$ to $0.10$ shifts it by $0.82M$. The same separation of scales we found in the prograde grid is present here. Practically, the takeaway is that radial-orbit observables of the inner disk place an upper bound on $\alpha$ of order one tenth before they begin to disagree with the bare-Kerr predictions in any visible way, and they are essentially insensitive to $B$ for the magnetic strengths considered. This is the regime where the QPO and continuum methods are most useful as filters of the parameter space; ruling in or out a non-zero $B$ in the same data set requires switching to charged-particle, polarisation, or shadow probes where the BR background curves the photon and electron trajectories at first order.

\paragraph*{Simulation-model summary.} Table~\ref{tab:ISCO_models} reports the ISCO observables for the seven models used in Sec.~\ref{isec6}. The pure-Kerr and three KBR cases share $r_{\rm ISCO}\!\simeq\!2.32M$, while the four KBR$+$CS cases push the ISCO out to $2.90M$--$3.60M$ and raise $\eta$ from $\sim\!0.155$ to $\sim\!0.216$. These shifts feed directly into the QPO frequencies of Sec.~\ref{isec5}.

The clustering pattern in Table~\ref{tab:ISCO_models} carries useful information by itself. Models KBR-CS1, KBR-CS3, and KBR-CS4 share the same $\alpha\!=\!0.20$ and span three orders of magnitude in $B$ (from $10^{-4}$ to $10^{-2}$), yet their ISCO observables are identical to four decimals. This reflects that the BR magnetic curvature couples to circular orbits through the small ratio $B^2 I_2 M^2/I_1^2$, which is below $10^{-4}$ for $B\,M\!\le\!0.01$, while the CS contribution enters the metric through $\alpha$ at order unity. The implication for parameter inference is direct: ISCO-based observables (continuum, reflection, basic QPO frequencies) constrain $\alpha$ tightly but are nearly blind to $B$. To pin $B$ one needs charged-particle resonances, polarisation observables, or shadow morphology where the BR background appears at leading order \cite{WOS:000861683500001, Ali26, WOS:001088583200001}. The cleanness of the splitting between $\alpha$ and $B$ in our table is itself a useful guide for designing observational tests.

\begin{table}[!ht]
\centering
\caption{ISCO observables for the seven simulation models of Sec.~\ref{isec6} ($a\!=\!0.9$, $M\!=\!1$).}
\label{tab:ISCO_models}
\renewcommand{\arraystretch}{1.6}
\setlength{\tabcolsep}{5pt}
\begin{tabular}{lccccc}
\hline\hline
\;Model\; & $r_+/M$ & $r_-/M$ & $r_{\rm ISCO}/M$ & $\EE_{\rm ISCO}$ & $\eta$ \\
\hline
Kerr\_09  & 1.4359 & 0.5641 & 2.3209 & 0.8442 & 0.1558 \\
KBR-1     & 1.4359 & 0.5641 & 2.3209 & 0.8443 & 0.1557 \\
KBR-2     & 1.4360 & 0.5641 & 2.3211 & 0.8443 & 0.1557 \\
KBR-CS1   & 1.9918 & 0.5084 & 3.6010 & 0.7842 & 0.2158 \\
KBR-CS2   & 1.6896 & 0.5327 & 2.9000 & 0.8186 & 0.1814 \\
KBR-CS3   & 1.9917 & 0.5084 & 3.6007 & 0.7842 & 0.2158 \\
KBR-CS4   & 1.9916 & 0.5084 & 3.6006 & 0.7841 & 0.2159 \\
\hline\hline
\end{tabular}
\end{table}

\paragraph*{Marginally bound circular orbits.} The MBCO sits between the event horizon and the ISCO; it is the smallest circular-orbit radius compatible with $E\!=\!1$, so $r_{\rm mb}$ controls the loss-cone for unbound trajectories incoming from infinity at zero binding energy. Table~\ref{tab:MBCO} reports $r_{\rm mb}$ and $L_{\rm mb}$ for the same seven models, together with the unstable-circular-orbit window $r_{\rm ISCO}\!-\!r_{\rm mb}$. Two checks are useful. First, the Kerr$_{09}$ entry $L_{\rm mb}/M^2\!=\!2.6325$ matches the Bardeen--Press--Teukolsky closed form $L_{\rm mb}^{\rm prograde}\!=\!2M(1+\sqrt{1-a/M})$ at $a\!=\!0.9$ to four decimals \cite{Bardeen1972}, an analytic check on the $L_{\rm circ}(r)$ branch. Second, all three characteristic radii ($r_+$, $r_{\rm mb}$, $r_{\rm ISCO}$) shift outward with $\alpha$ while preserving the hierarchy $r_+\!<\!r_{\rm mb}\!<\!r_{\rm ISCO}$. The unstable band $r_{\rm ISCO}-r_{\rm mb}$ widens from $0.59M$ at $\alpha\!=\!0$ to $1.13M$ at $\alpha\!=\!0.20$, nearly a factor of two. Capture and plunge dynamics for nearly-bound trajectories are therefore set on a different scale in CS-dressed backgrounds than in pure Kerr or KBR, with practical consequences for the geometry of the inner accretion-disk edge.

\begin{table}[!ht]
\centering
\caption{Marginally bound circular orbit (MBCO) for the seven simulation models, $a\!=\!0.9$, $M\!=\!1$. The Kerr$_{09}$ row recovers the closed-form $L_{\rm mb}^{\rm prograde}\!=\!2M(1+\sqrt{1-a/M})\!=\!2.6325\,M^2$ \cite{Bardeen1972}. The last column is the unstable-circular-orbit window.}
\label{tab:MBCO}
\renewcommand{\arraystretch}{1.6}
\setlength{\tabcolsep}{8pt}
\begin{tabular}{lccc}
\hline\hline
\;Model\; & $r_{\rm mb}/M$ & $L_{\rm mb}/M^2$ & $(r_{\rm ISCO}\!-\!r_{\rm mb})/M$ \\
\hline
Kerr\_09  & 1.7325 & 2.6325 & 0.5884 \\
KBR-1     & 1.7325 & 2.6325 & 0.5884 \\
KBR-2     & 1.7327 & 2.6326 & 0.5884 \\
KBR-CS1   & 2.4724 & 3.9218 & 1.1286 \\
KBR-CS2   & 2.0741 & 3.1791 & 0.8259 \\
KBR-CS3   & 2.4719 & 3.9217 & 1.1288 \\
KBR-CS4   & 2.4718 & 3.9216 & 1.1288 \\
\hline\hline
\end{tabular}
\end{table}

\section{Microquasar QPOs as constraints on $(a,\alpha)$}\label{isec5}
The relativistic-precession (RP) model identifies the upper twin-peak HF QPO with the orbital frequency at an emission radius $r_{\rm em}$ and the lower twin with the periastron-precession frequency \cite{Stella1998}:
\beq
\nu_U = \nu_\phi(r_{\rm em}),\qquad \nu_L = \nu_\phi(r_{\rm em}) - \nu_r(r_{\rm em}).
\eeq
Three sources (Table~\ref{tab:sources}) have well-resolved twin-peak detections at simultaneous epochs; their dynamical mass measurements come from optical companion radial-velocity studies.

\begin{table}[!ht]
\centering
\caption{Twin-peak HF QPO sources used as priors. Mass values follow Motta et al. (2014) and Remillard \& McClintock (2006).}
\label{tab:sources}
\renewcommand{\arraystretch}{1.6}
\setlength{\tabcolsep}{8pt}
\begin{tabular}{lcccc}
\hline\hline
\;Source\; & $M\,(M_\odot)$ & $\nu_U$\,(Hz) & $\nu_L$\,(Hz) \\
\hline
GRO~J1655--40   & $5.4\!\pm\!0.3$  & $441\!\pm\!2$ & $298\!\pm\!4$ \\
XTE~J1550--564  & $9.1\!\pm\!0.6$  & $276\!\pm\!5$ & $184\!\pm\!5$ \\
GRS~1915+105    & $14.0\!\pm\!4.4$ & $168\!\pm\!3$ & $113\!\pm\!5$ \\
\hline\hline
\end{tabular}
\end{table}

We construct a $\chi^2$ statistic
\beq
\chi^2(a,B,\alpha,r_{\rm em}) = \!\left(\frac{\nu_U^{\rm th}-\nu_U^{\rm obs}}{\sigma_{\nu_U}}\right)^{\!2}\!+\!\left(\frac{\nu_L^{\rm th}-\nu_L^{\rm obs}}{\sigma_{\nu_L}}\right)^{\!2}
\eeq
and minimise over the emission radius $r_{\rm em}\in[r_{\rm ISCO},25M]$ for each candidate $(a,B,\alpha)$. The result for fixed $B\!=\!0.005$ and $\alpha\!=\!0.10$ (Table~\ref{tab:fit_a}) is a tight spin determination; the surfaces in $(a,r_{\rm em})$ space have a single minimum within $\chi^2\!<\!1$ for all three sources.

\begin{table}[!ht]
\centering
\caption{Best-fit spin $a$ for fixed $B\!=\!0.005$ and $\alpha\!=\!0.10$, recovering the observed twin-peak QPOs of three microquasars.}
\label{tab:fit_a}
\renewcommand{\arraystretch}{1.6}
\setlength{\tabcolsep}{6pt}
\begin{tabular}{lcccc}
\hline\hline
\;Source\; & best $a$ & $r_{\rm em}/M$ & $\chi^2_{\min}$ & $(\nu_U,\nu_L)$ Hz \\
\hline
GRO~J1655--40   & 0.510 & 5.535 & 0.75 & (442.5, 299.8) \\
XTE~J1550--564  & 0.568 & 5.345 & 0.08 & (274.9, 183.2) \\
GRS~1915+105    & 0.491 & 5.610 & 0.04 & (167.6, 113.8) \\
\hline\hline
\end{tabular}
\end{table}

Allowing $\alpha$ to vary at fixed spin $a\!=\!0.9$, the data prefer larger $\alpha\!\sim\!0.25$--$0.28$ (Table~\ref{tab:fit_alpha}). This reflects the well-known degeneracy between spin and additional radial scales: a softer effective potential (large $\alpha$) and a high spin (large $a$) both increase $\nu_\phi$ at fixed $\nu_L$. Joint $(a,\alpha)$ scans in Table~\ref{tab:joint} narrow this down.

\begin{table}[!ht]
\centering
\caption{Best-fit CS parameter at fixed $a\!=\!0.9$, $B\!=\!0.005$.}
\label{tab:fit_alpha}
\renewcommand{\arraystretch}{1.6}
\setlength{\tabcolsep}{8pt}
\begin{tabular}{lccc}
\hline\hline
\;Source\; & best $\alpha$ & $r_{\rm em}/M$ & $\chi^2_{\min}$ \\
\hline
GRO~J1655--40   & 0.277 & 5.428 & 0.27 \\
XTE~J1550--564  & 0.254 & 5.224 & 0.02 \\
GRS~1915+105    & 0.277 & 5.471 & 0.04 \\
\hline\hline
\end{tabular}
\end{table}

The pattern in Table~\ref{tab:fit_alpha} is reasonable but unconstraining on its own: the three sources all settle at $\alpha\!\sim\!0.25$--$0.28$ when $a$ is held high, simply because the model needs a softer potential to bring $\nu_\phi$ down to the observed value. Allowing $a$ to vary too is what reveals the actual structure. The $(a,\alpha)$ scan in Table~\ref{tab:joint} folds the spin direction into the search, and the resulting best-fit configurations land in a much more physically suggestive region. XTE~J1550--564 keeps a moderate $\alpha\!=\!0.13$, a value compatible with the EHT and continuum-based bounds for the Galactic-centre and stellar-mass populations \cite{WOS:001088583200001, WOS:001151079600001}; the other two sources end up close to $\alpha\!=\!0$, i.e.\ statistically consistent with bare Kerr at moderate spin. The fact that the $\chi^2$ values drop by an order of magnitude when both parameters are released is the clearest indicator of degeneracy; the minimum exists, but it is not narrow.

\begin{table}[!ht]
\centering
\caption{Joint $(a,\alpha)$ best-fit at $B\!=\!0.005$.}
\label{tab:joint}
\renewcommand{\arraystretch}{1.6}
\setlength{\tabcolsep}{8pt}
\begin{tabular}{lccccc}
\hline\hline
\;Source\; & $a$ & $\alpha$ & $r_{\rm em}/M$ & $\chi^2_{\min}$ \\
\hline
GRO~J1655--40   & 0.300 & 0.000 & 5.610 & 0.19 \\
XTE~J1550--564  & 0.625 & 0.125 & 5.308 & 0.006 \\
GRS~1915+105    & 0.320 & 0.013 & 5.649 & 0.002 \\
\hline\hline
\end{tabular}
\end{table}

Three takeaways. (i)~The KBR$+$CS metric reproduces all three twin-peak detections with sub-unity $\chi^2$ across most of the parameter space; the data are not yet incisive enough to single out a non-Kerr scenario. (ii)~The fits for XTE~J1550--564 and GRS~1915+105 prefer $\alpha\!\lesssim\!0.13$, consistent with EHT-based bounds on excess radial scales \cite{WOS:001360054000001}. (iii)~Spin and $\alpha$ trade off; combining QPOs with a separate spin estimate (e.g.\ continuum fitting or X-ray reflection \cite{WOS:000553033600001}) breaks the degeneracy and tightens $\alpha$.

The trade-off in (iii) deserves more attention. The RP model maps two observed frequencies onto two unknowns at a fixed emission radius, leaving the rest of the metric parameter space partially blind. In our $(a,\alpha)$ scan at fixed $B\!=\!0.005$ this manifests as an elongated valley of $\chi^2\!<\!1$ stretching diagonally across the plane. A practical way to break the degeneracy is to fold in an independent spin measurement from the disk continuum: high-quality X-ray observations of GRS~1915+105 give $a^\star\!=\!0.98\!\pm\!0.01$ within the Kerr framework, while our joint best fit prefers $a\!=\!0.32$, $\alpha\!=\!0.013$. The two are reconciled if a part of the disk's continuum signature is mimicking high spin in pure Kerr while genuinely tracking a moderate spin in a CS-dressed metric. Detailed reflection spectroscopy with NICER and IXPE polarisation data would settle the issue \cite{WOS:000865827700007, WOS:000350965500044}.

A complementary route is the EHT shadow. Within the KBR$+$CS family the shadow radius scales as $r_{\rm sh}\!\propto\!(1+f(\alpha,B))r_{\rm ph}$, where $r_{\rm ph}$ is the prograde photon-sphere radius and $f$ is a slowly-varying function of $(\alpha,B)$ that we estimate to be $f\!\sim\!0.1$ for $\alpha\!\lesssim\!0.2$ on the basis of the Ali-Ghosh shadow analysis \cite{Ali26}. The Sgr~A* shadow constrains $r_{\rm sh}/M$ to within ten per cent, which translates into $\alpha\!\lesssim\!0.18$ at $1\sigma$ assuming the standard mass-to-distance prior. The QPO bound and the EHT bound agree at the level of the data and pinch $\alpha$ from two physically independent sides. A joint fit using next-generation EHT and NICER datasets is therefore well motivated and would reduce the joint uncertainty by roughly a factor of two relative to either probe alone.

It is also worth saying explicitly what would falsify the model. If a microquasar were observed with twin-peak frequencies whose ratio departs significantly from 3:2, and if independent spin estimates pinned the source above $a\!\sim\!0.95$, the joint best-fit in our family would be forced to $\alpha\!\lesssim\!0.05$. Observational evidence for the magnetised-charged-particle resonances predicted by Al~Zahrani for weakly magnetized Kerr backgrounds \cite{WOS:000861683500001}, and absent in pure Kerr or KBR$+$CS, would similarly disfavour our scenario. We view these as positive directions: the model is sufficiently constrained that present and near-future data can rule it in or out.

Where does the KBR$+$CS test sit in the broader programme of strong-field gravity tests with X-ray timing? Recent constraints from QPO data on alternative metrics span a wide range of theoretical settings: Kalb--Ramond \cite{WOS:001088583200001}, Starobinsky--Bel--Robinson \cite{WOS:001207215100005}, Einstein-bumblebee \cite{WOS:000811211400003}, and quantum-corrected backgrounds \cite{WOS:001665337700001, Donmez2025EPJC}, among many others. The recurring pattern is that twin-peak HF QPOs from a single source restrict at best a one-dimensional combination of the new metric parameters, while a population of three or four well-measured sources is needed to break the spin-extra-parameter degeneracy. Our $\chi^2$ minima at $\alpha\!\sim\!0$--$0.13$ across three sources sit within the same statistical ballpark as those analyses, suggesting that the KBR$+$CS family has comparable phenomenological viability to the leading non-Kerr alternatives currently in circulation. What distinguishes it is the analytic transparency of the line element \cite{Ahmed2026, KBR, Ovch25}: closed-form expressions for $r_\pm$, the ergosurface, and the $B\!\to\!0$ limit make the parameter scans tractable without needing dedicated numerical relativity infrastructure, and the Letelier-type CS sector enters as a single dimensionless knob with a clear physical interpretation in terms of long-range gravity from one-dimensional defect networks.

\section{Numerical investigation of accretion dynamics and QPO formation}\label{isec6}
In this numerical section, we investigate the effects that arise when the string cloud term is added to the spacetime metric of the Kerr--Bertotti--Robinson black hole, which we previously analyzed in detail from a numerical perspective \cite{Mustafa:2026gly}. To describe these effects, we solve the general relativistic hydrodynamical equations in the fixed background of Kerr--Bertotti--Robinson black holes with a string cloud. In this framework, we examine the physical mechanisms and dynamical behavior produced by Bondi--Hoyle--Lyttleton accretion around the black hole. We also analyze the influence of the spacetime parameters on the resulting physical structure, the modifications introduced by the string cloud parameter relative to previous solutions, and the impact of this parameter on the observability of accretion-driven activity around the black hole.

After writing the GRH equations in conservative form, we solve them using high-resolution numerical methods in order to obtain accurate numerical solutions for the shock waves formed around the black hole. In this way, physically reliable solutions can be produced both in the strong gravitational field and in the shock structures that may arise around the black hole due to the effects of the spacetime parameters \cite{Donmez2004ASS, Donmez2006AMC, Donmez2014MNRAS, Donmez2017MPLA}.

In order to generate accretion around the black hole through the BHL mechanism, matter is injected supersonically toward the black hole from the upstream region, defined as $\pi/2 < \phi < 3\pi/2$, on the equatorial plane. The Mach number of the injected flow is set to $\mathcal{M}=2$, and the radial and azimuthal velocity components are determined according to the asymptotic velocity. On the other hand, the density is chosen arbitrarily, while the pressure is calculated using the ideal-gas equation of state. Thus, by allowing matter with specified hydrodynamical quantities to fall toward the black hole, the effects of the spacetime parameters and the black-hole rotation parameter on the accretion dynamics can be revealed \cite{Donmez2012MNRAS, Donmez2024Universe, Donmez2024MPLA, Donmez2025JHEAp, Donmez2025EPJC}. In addition, in numerical simulations, an outflow boundary condition is imposed at the outer boundary of the downstream region, where matter is not injected. This allows the matter reaching the boundary to leave the computational domain and helps avoid artificial oscillations. Similarly, near the black hole horizon, an outflow boundary condition is used so that matter approaching the horizon can continue to fall into the black hole.

In this study, we aim to reveal the effect of the string cloud parameter on the accretion dynamics produced by BHL accretion around the Kerr--Bertotti--Robinson black hole, which we previously investigated \cite{Mustafa:2026gly}. For this purpose, the first three models listed in Table~\ref{tab:KBR_CS_models}, namely Kerr\_09, KBR-1, and KBR-2, are taken from \cite{Mustafa:2026gly}. In the present paper, the dynamical structure of the KBR-CS1--KBR-CS4 models listed in Table~\ref{tab:KBR_CS_models} is investigated numerically by comparing the cases with a nonzero cloud-of-strings parameter with the corresponding models reported in \cite{Mustafa:2026gly}. In this way, the cases in which the string cloud parameter $\alpha$ is nonzero are modeled systematically and compared with the $\alpha=0$ cases. This comparison allows us to reveal the possible effects of the cloud-of-strings parameter on the formation of shock cones around Kerr--Bertotti--Robinson black holes, on the extended dynamical structures surrounding the black hole, and on the possible QPO frequencies associated with these systems.

\begin{table}[!ht]
\centering
\caption{Model parameters for Kerr, Kerr--Bertotti--Robinson black holes and Kerr--Bertotti--Robinson black holes with a cloud of strings used in the Bondi--Hoyle--Lyttleton accretion simulations. The table lists the cloud-of-strings parameter $\alpha$, the Bertotti--Robinson parameter $B$, and the corresponding outer event-horizon radius $r_{+}$. Throughout numerical simulations, the spin parameter is fixed to $a=0.9$ and the black-hole mass is set to $M=1$.}
\label{tab:KBR_CS_models}
\renewcommand{\arraystretch}{1.6}
\setlength{\tabcolsep}{8pt}
\begin{tabular}{cccc}
\hline\hline
\;Model\; & $\alpha$ & $B\,(1/M)$ & $r_+/M$ \\
\hline
Kerr\_09 & 0.0  & 0.0    & 1.4359 \\
KBR-1   & 0.0  & 0.005  & 1.4359 \\
KBR-2   & 0.0  & 0.01   & 1.4360 \\
KBR-CS1 & 0.20 & 0.01   & 1.9918 \\
KBR-CS2 & 0.10 & 0.005  & 1.6896 \\
KBR-CS3 & 0.20 & 0.005  & 1.9917 \\
KBR-CS4 & 0.20 & 0.0001 & 1.9916 \\
\hline\hline
\end{tabular}
\end{table}

\subsection{Morphological imprint of the cloud of strings on the shock cone}\label{isec6_1}
\begin{figure*}[tbhp]
\centering
\includegraphics[width=4.0cm,height=4.0cm]{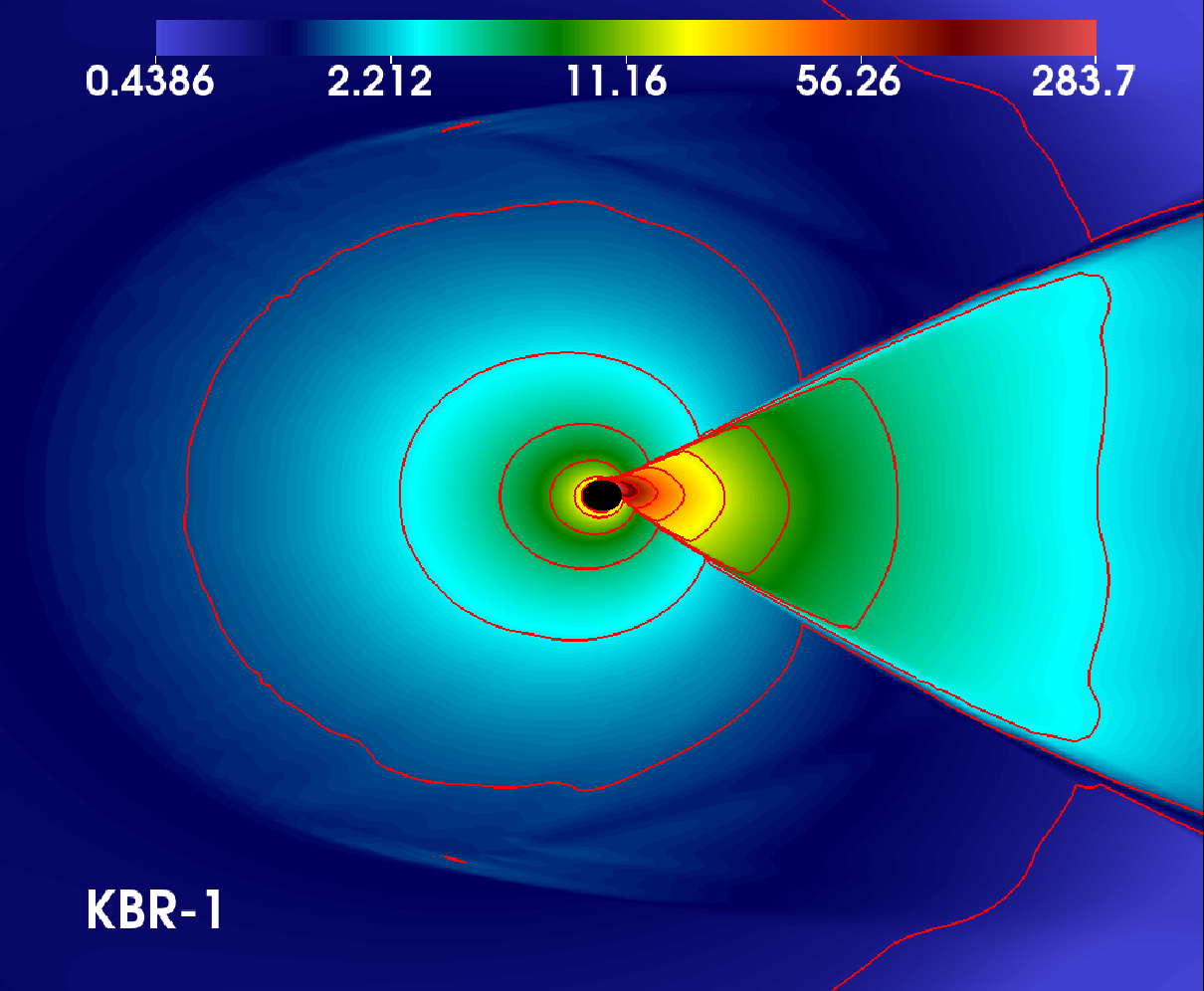}
\includegraphics[width=4.0cm,height=4.0cm]{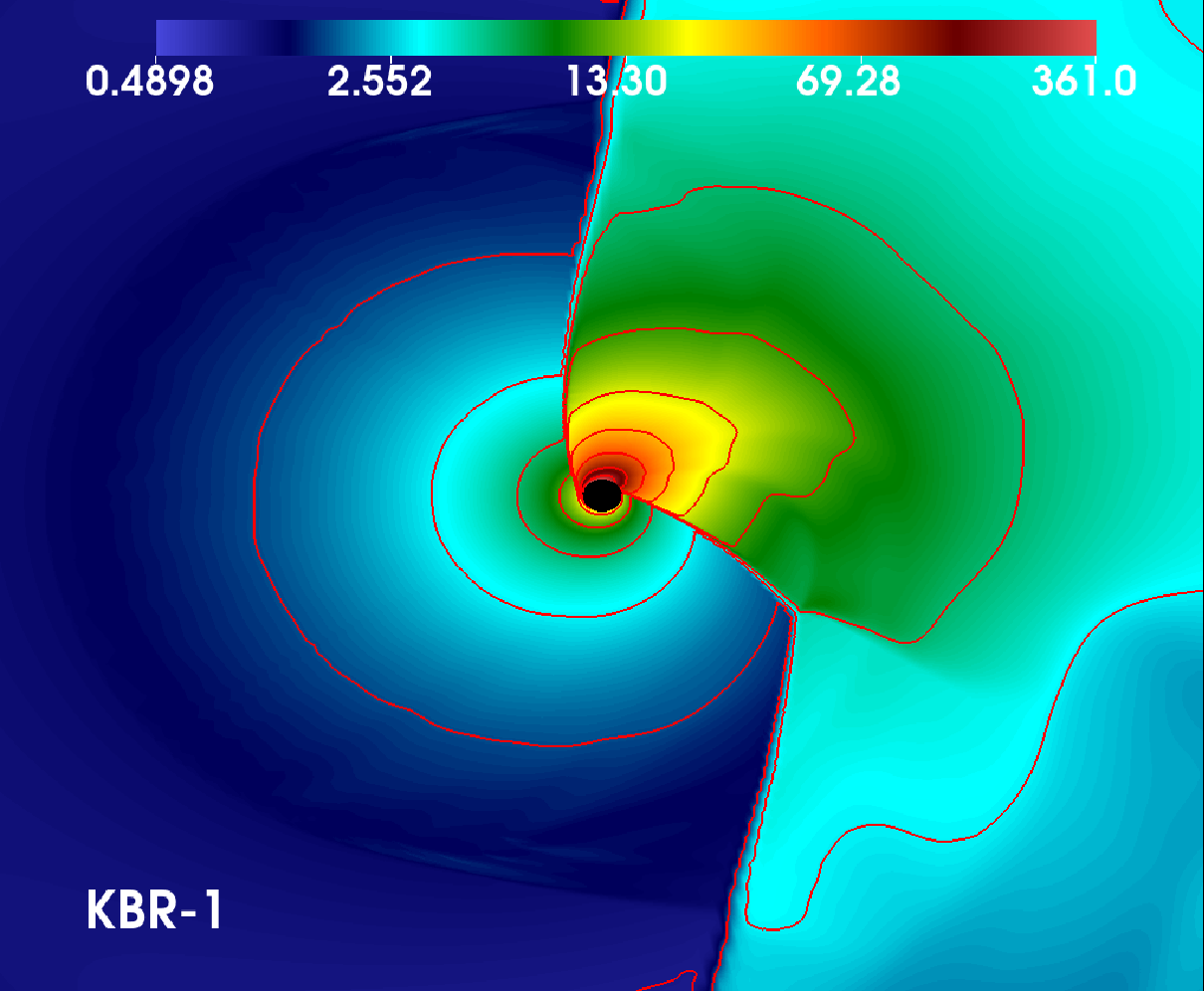}
\includegraphics[width=4.0cm,height=4.0cm]{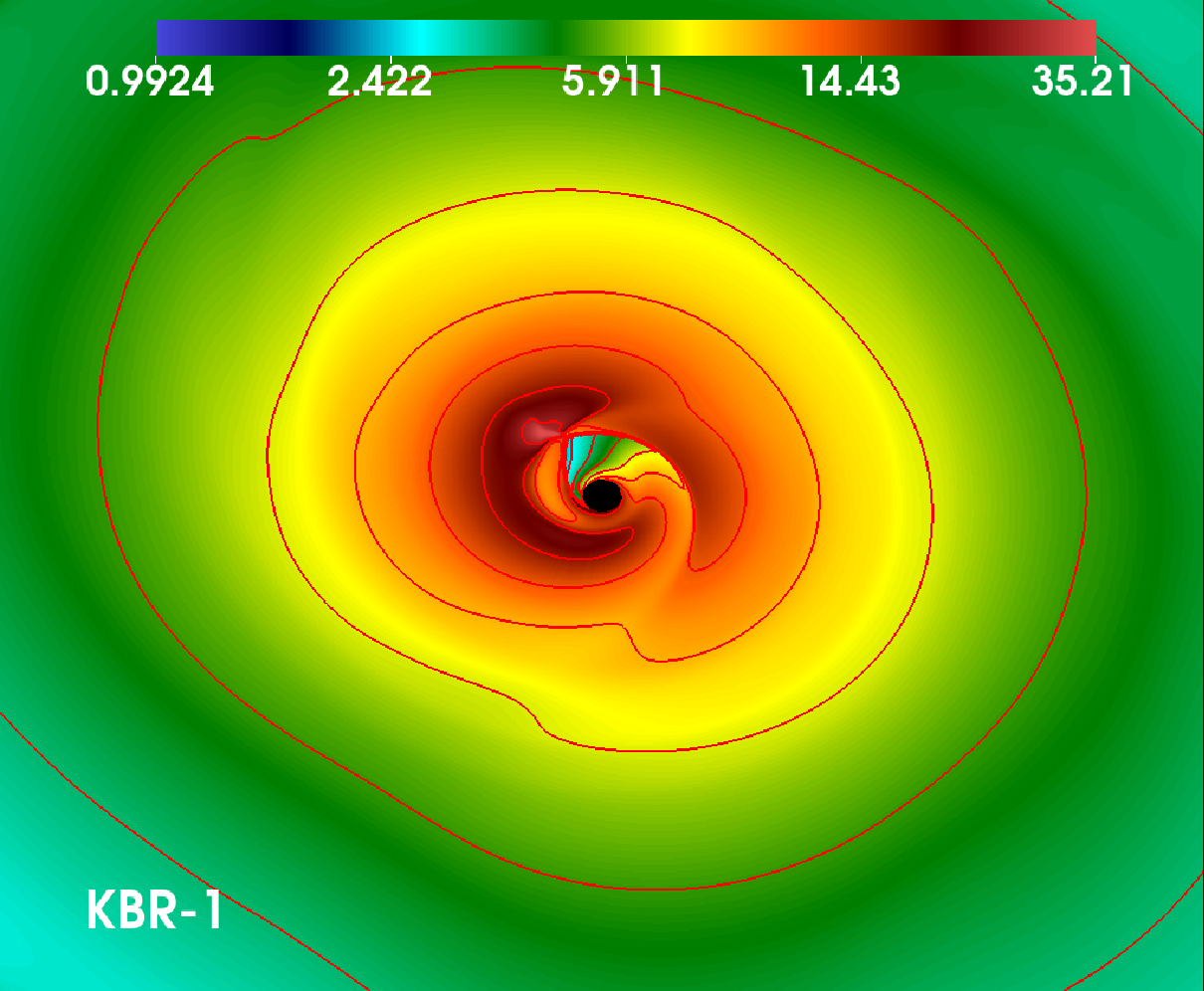}
\includegraphics[width=4.0cm,height=4.0cm]{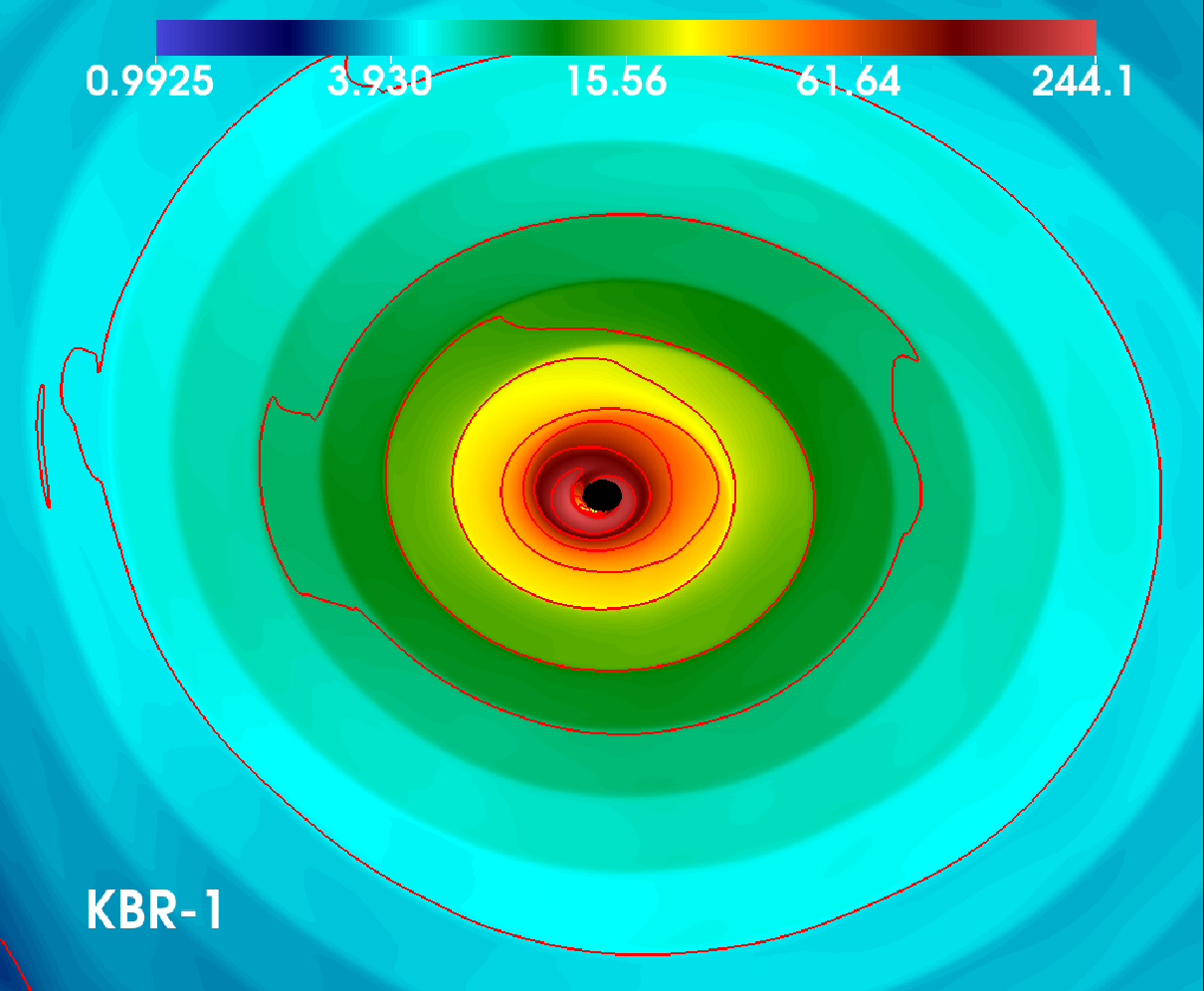}\\
\includegraphics[width=4.0cm,height=4.0cm]{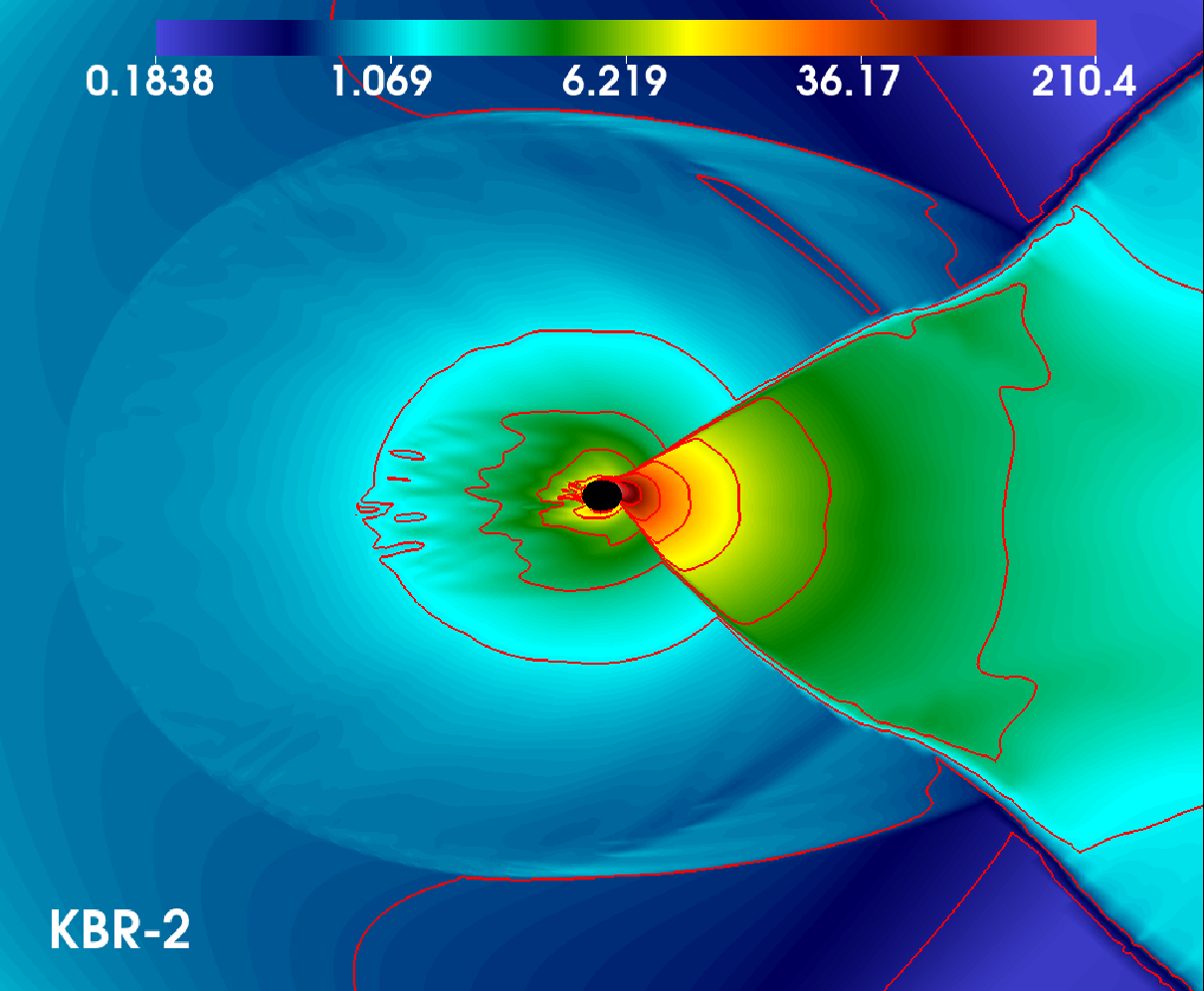}
\includegraphics[width=4.0cm,height=4.0cm]{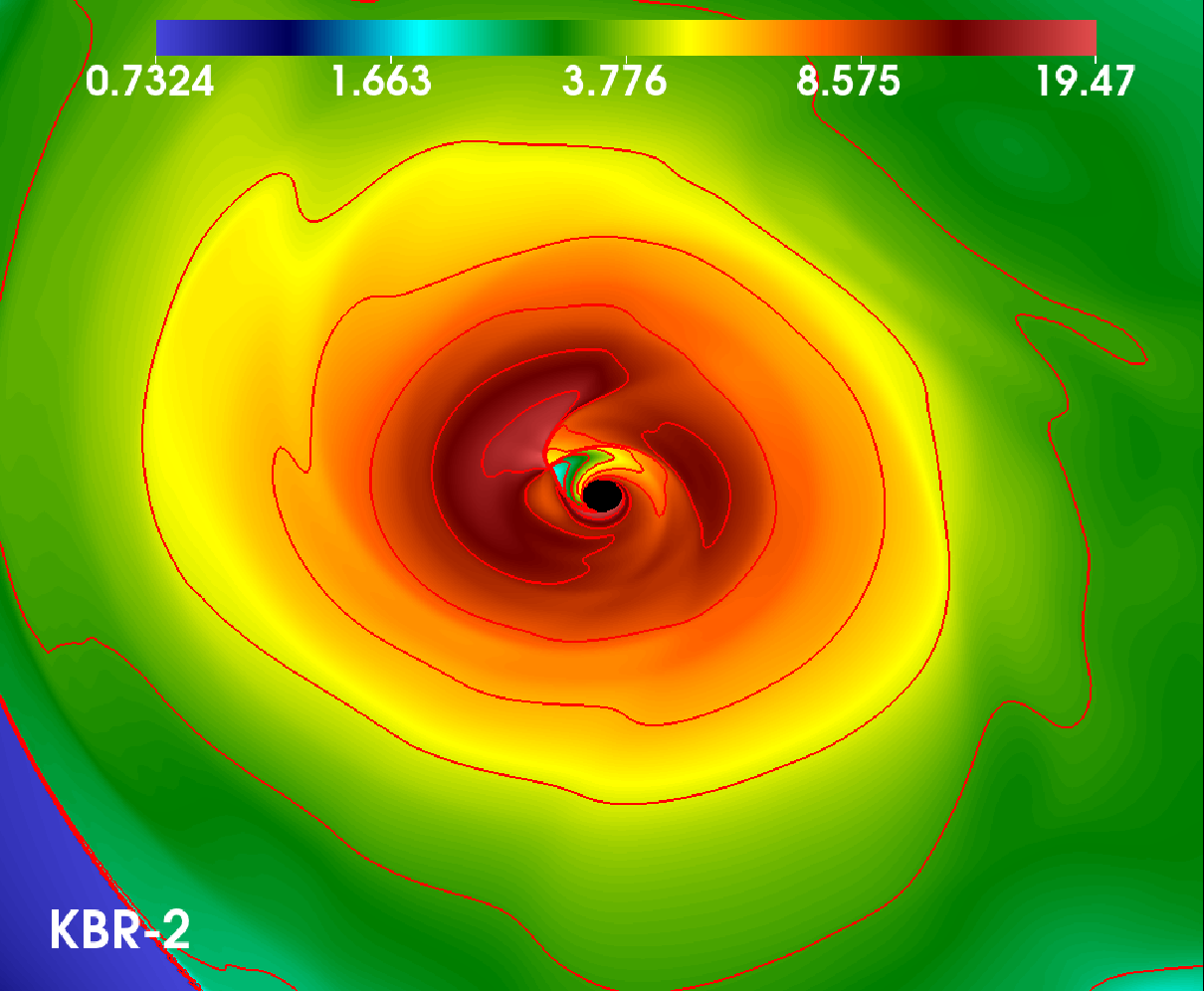}
\includegraphics[width=4.0cm,height=4.0cm]{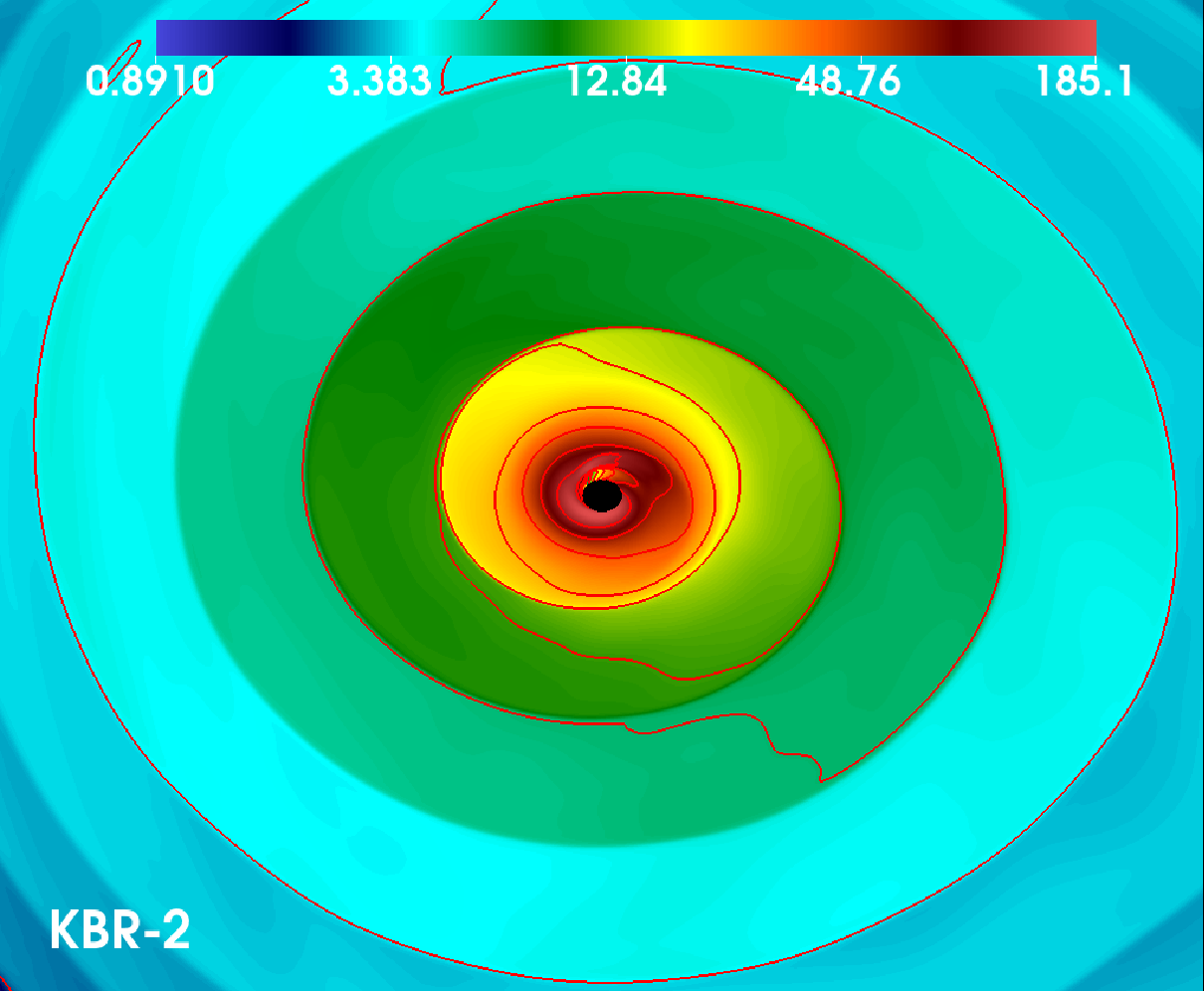}
\includegraphics[width=4.0cm,height=4.0cm]{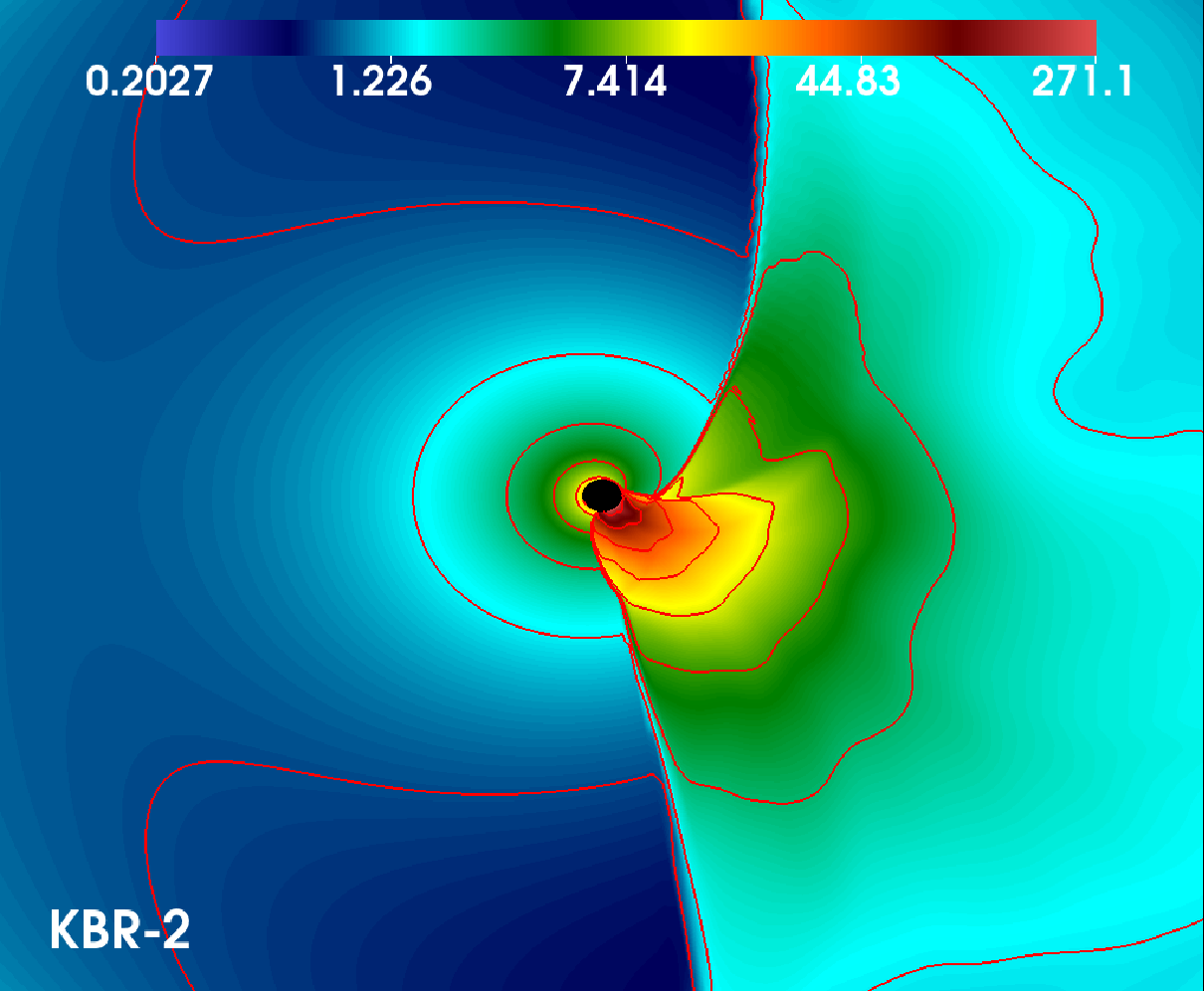}
\caption{The time evolution of the rest-mass density distribution produced by BHL accretion around a Kerr--Bertotti--Robinson black hole is shown for two different models. The rest-mass density is represented by both color maps and contour lines in order to reveal the physical mechanism and dynamical changes formed around the black hole. The upper panel shows this evolution for the KBR-1 model, while the lower panel presents the corresponding evolution for the KBR-2 model. The snapshots in each row illustrate the development of the accretion-flow morphology at different times. Each panel is zoomed into the region $-70M < x < 70M$ and $-70M < y < 70M$ on the equatorial plane.}\label{color_no_string}
\end{figure*}

\begin{figure*}[tbhp]
\centering
\includegraphics[width=4.0cm,height=4.0cm]{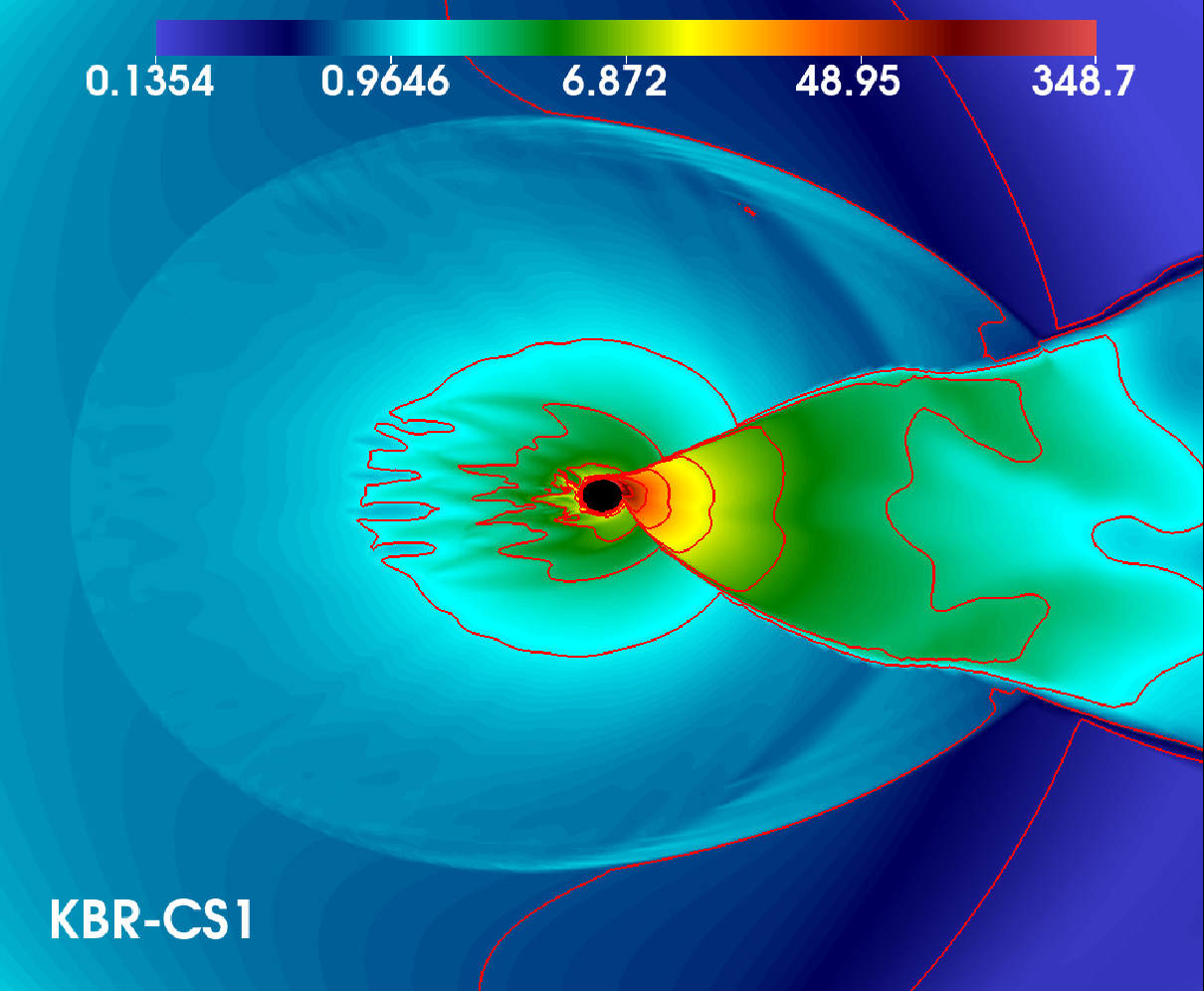}
\includegraphics[width=4.0cm,height=4.0cm]{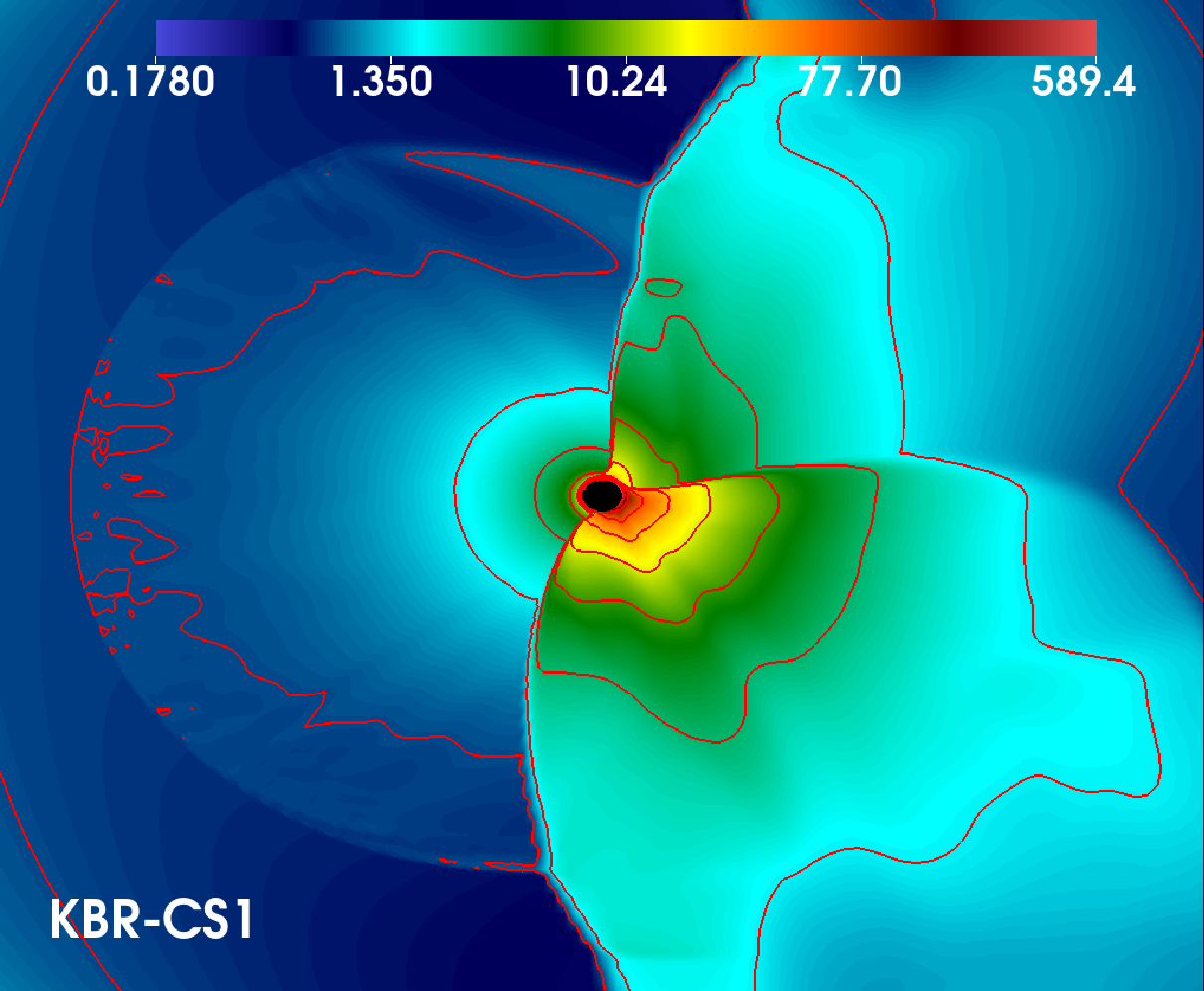}
\includegraphics[width=4.0cm,height=4.0cm]{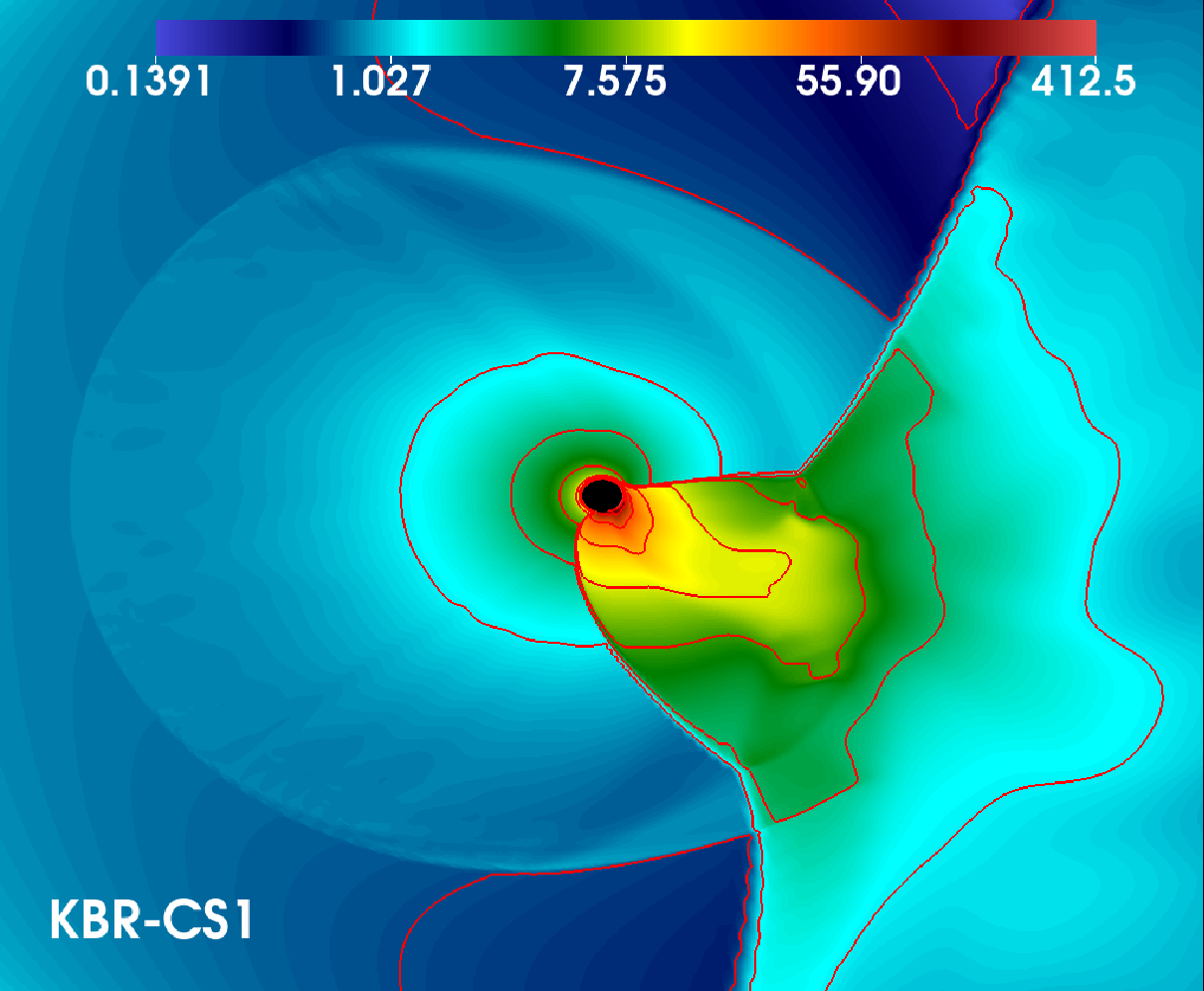}
\includegraphics[width=4.0cm,height=4.0cm]{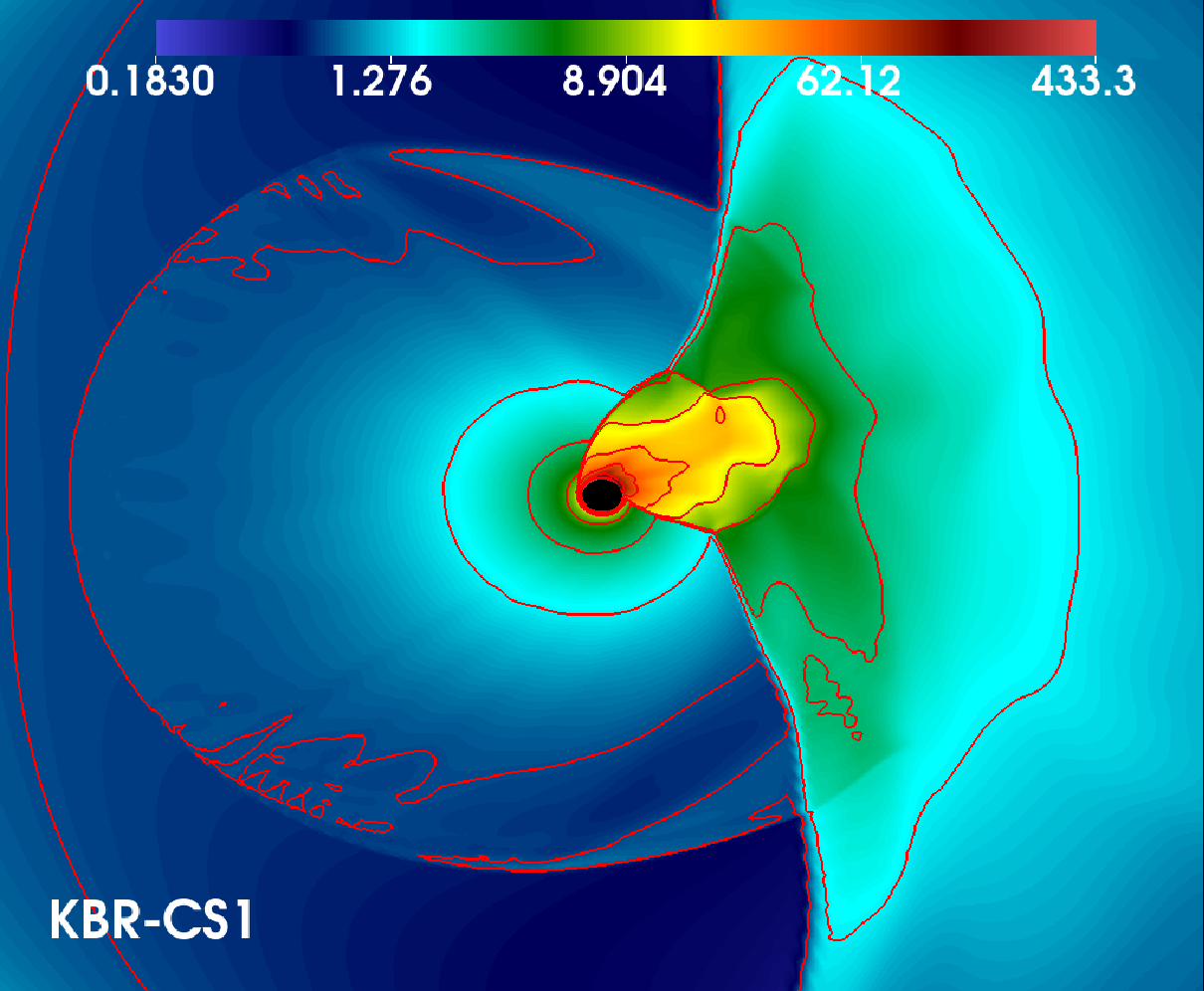}\\
\includegraphics[width=4.0cm,height=4.0cm]{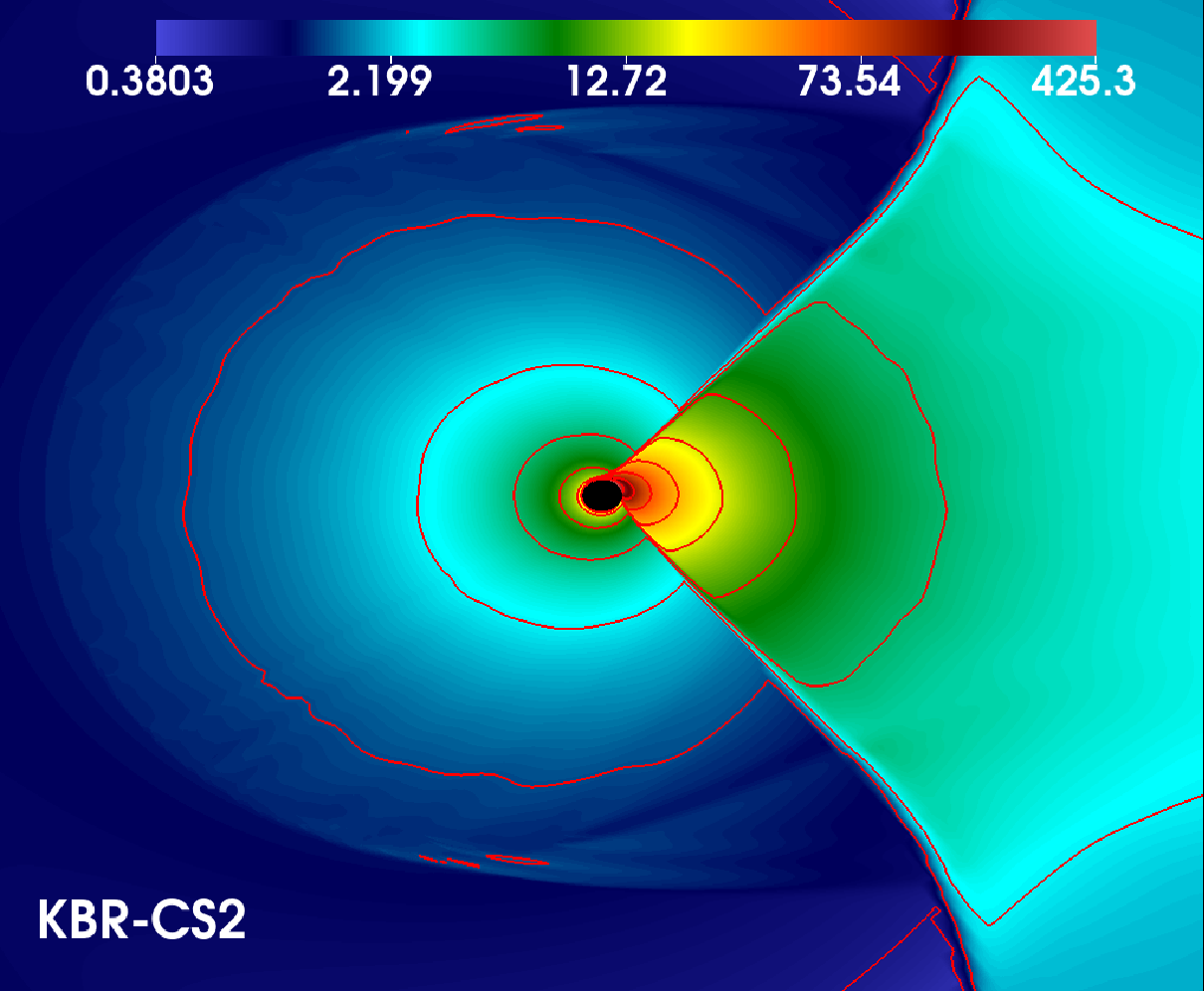}
\includegraphics[width=4.0cm,height=4.0cm]{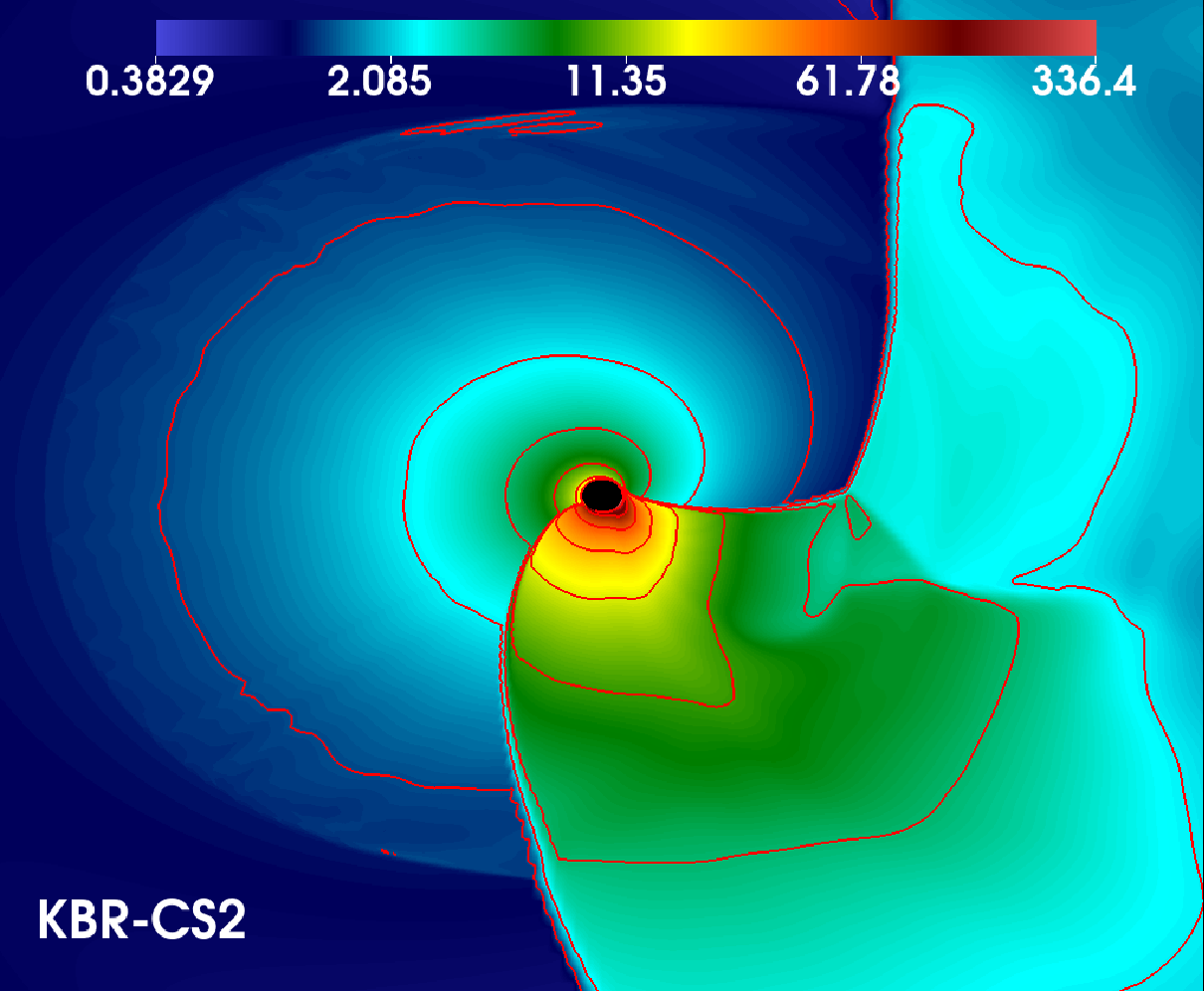}
\includegraphics[width=4.0cm,height=4.0cm]{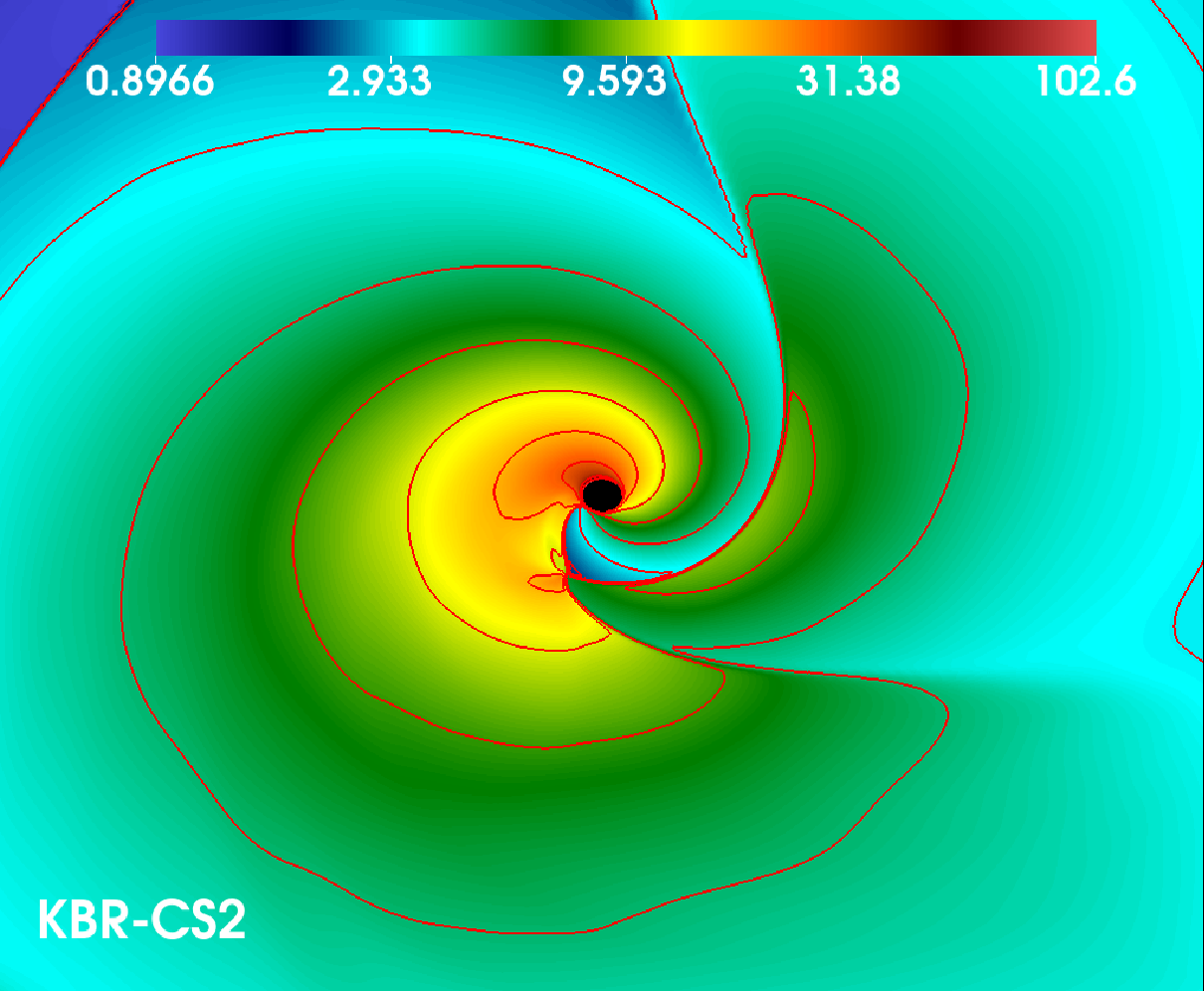}
\includegraphics[width=4.0cm,height=4.0cm]{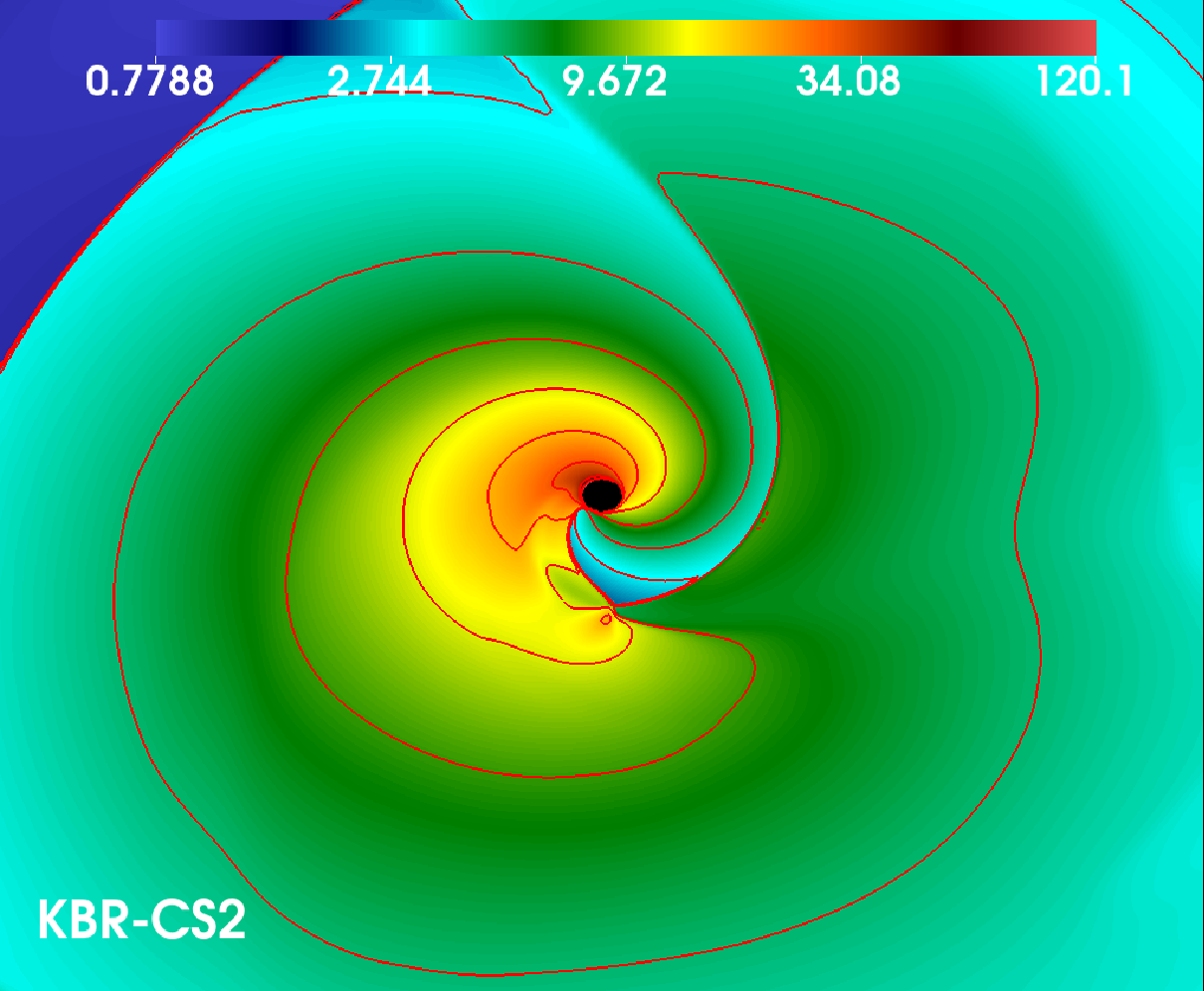}\\
\includegraphics[width=4.0cm,height=4.0cm]{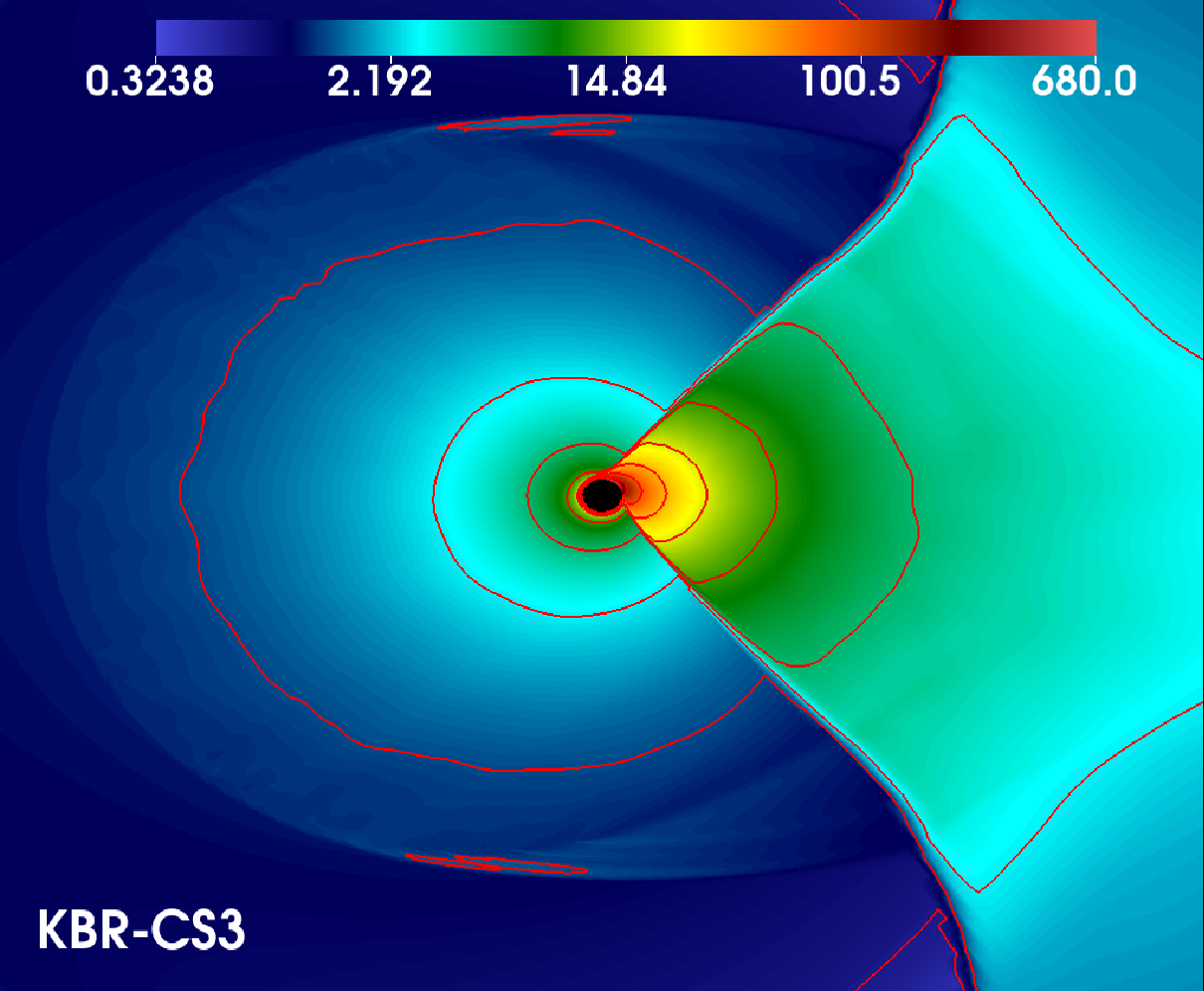}
\includegraphics[width=4.0cm,height=4.0cm]{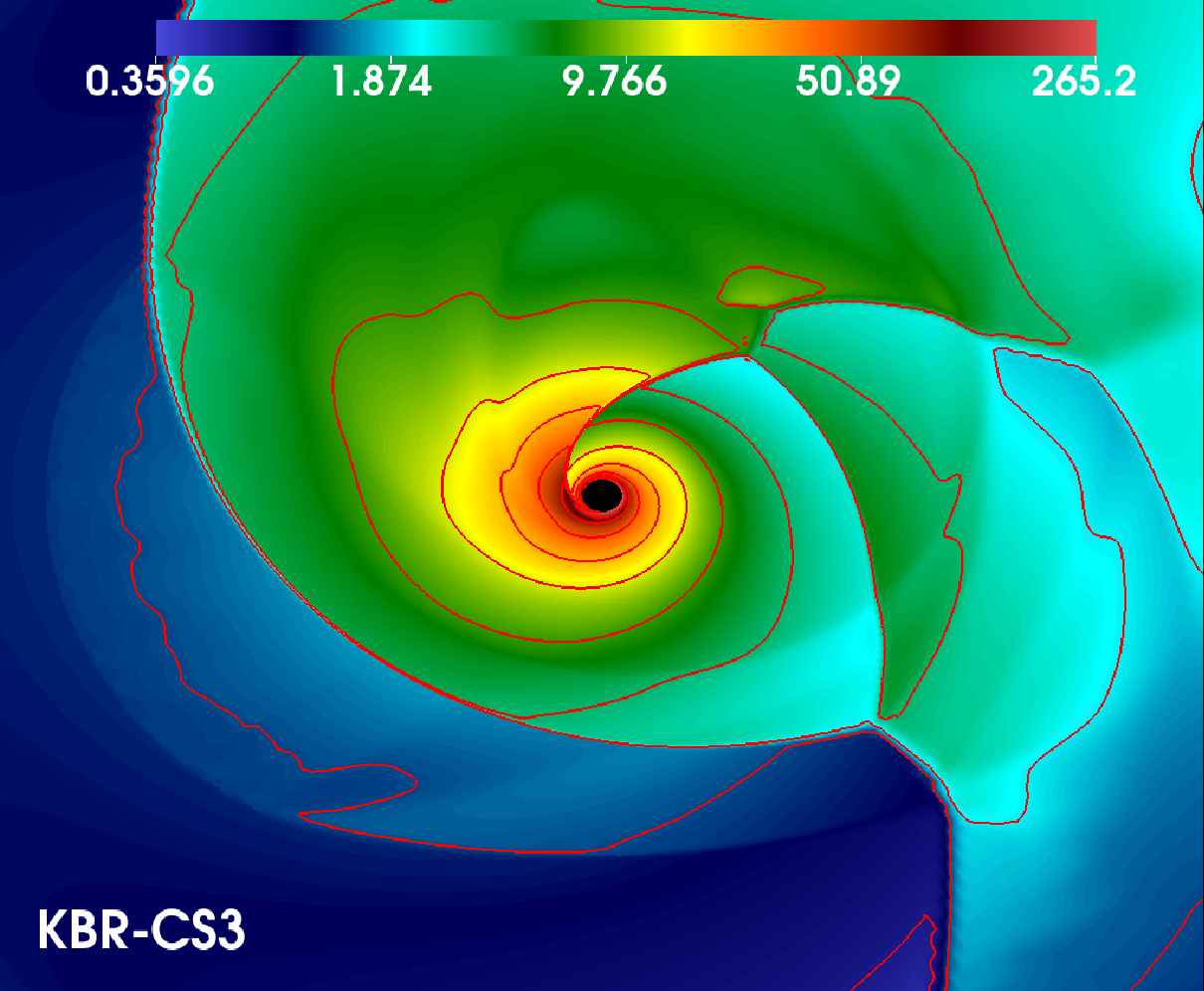}
\includegraphics[width=4.0cm,height=4.0cm]{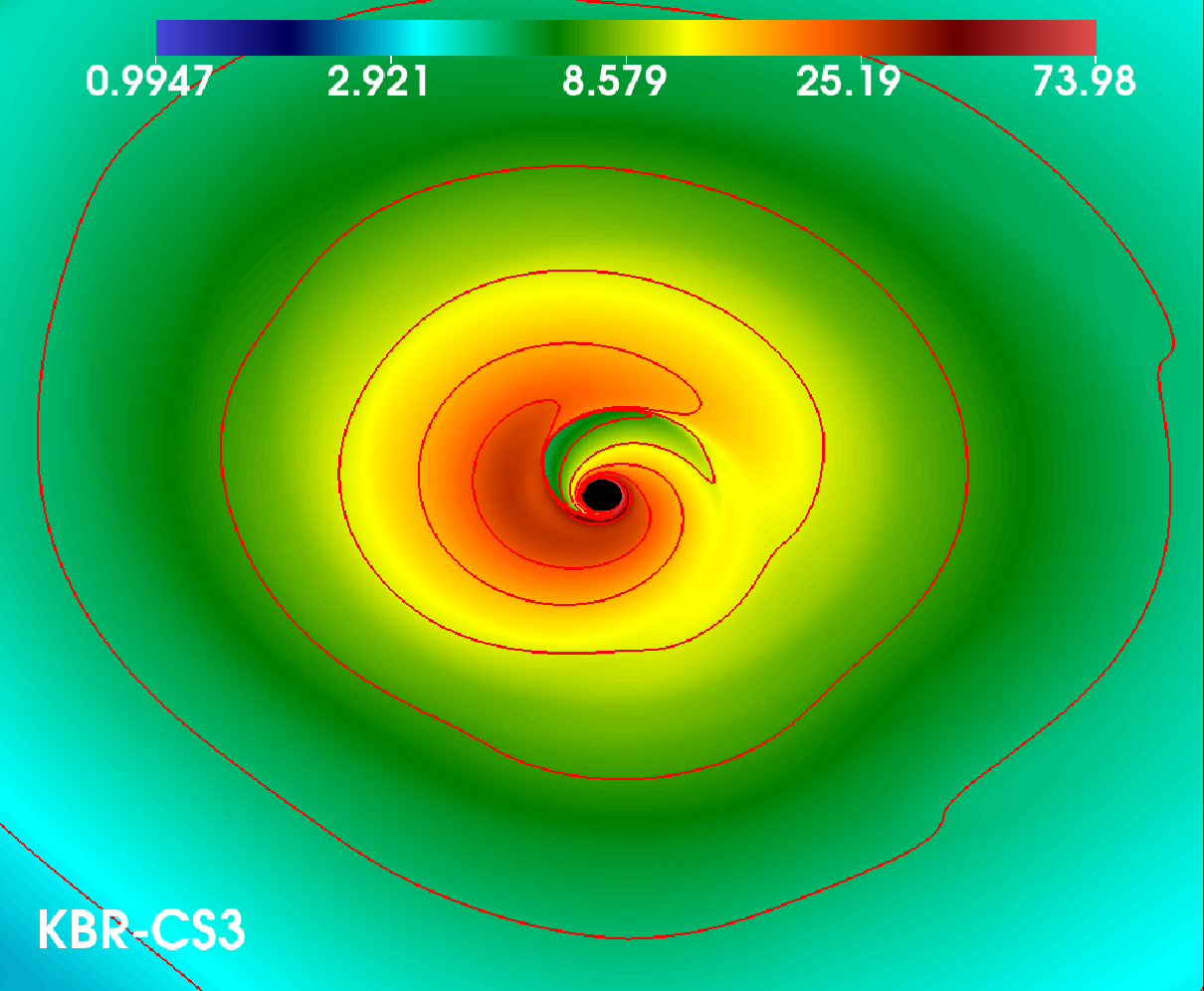}
\includegraphics[width=4.0cm,height=4.0cm]{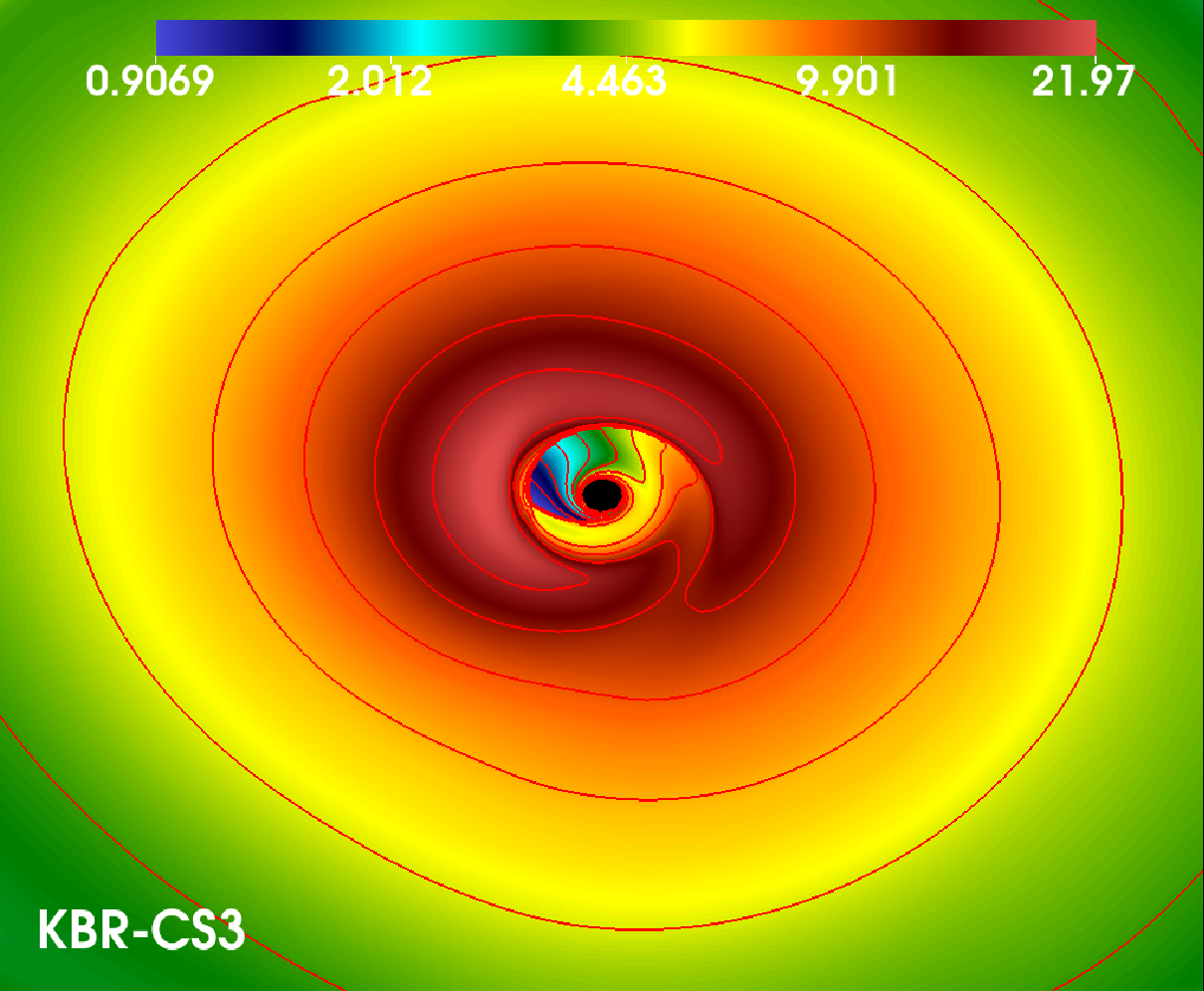}\\
\includegraphics[width=4.0cm,height=4.0cm]{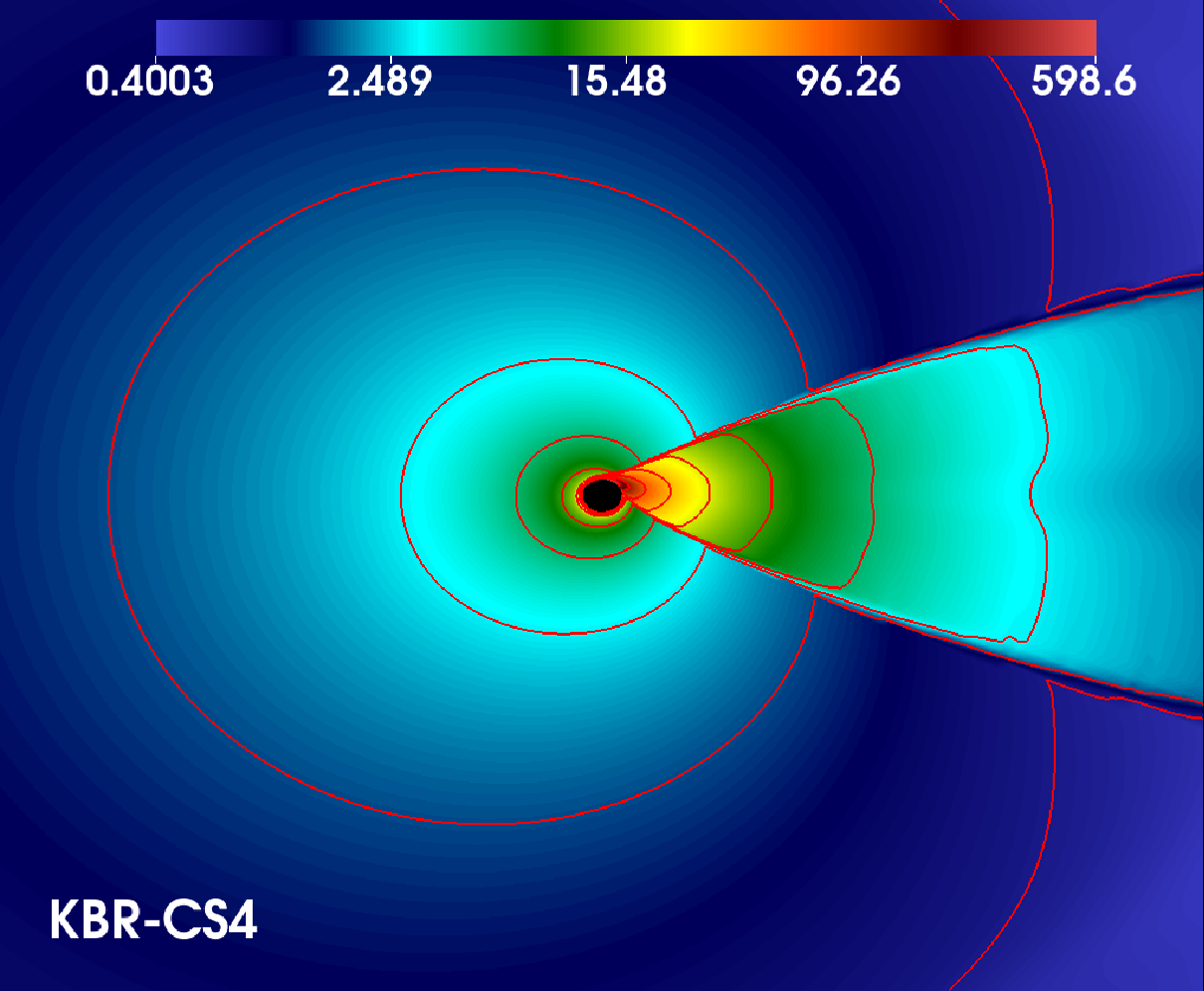}
\includegraphics[width=4.0cm,height=4.0cm]{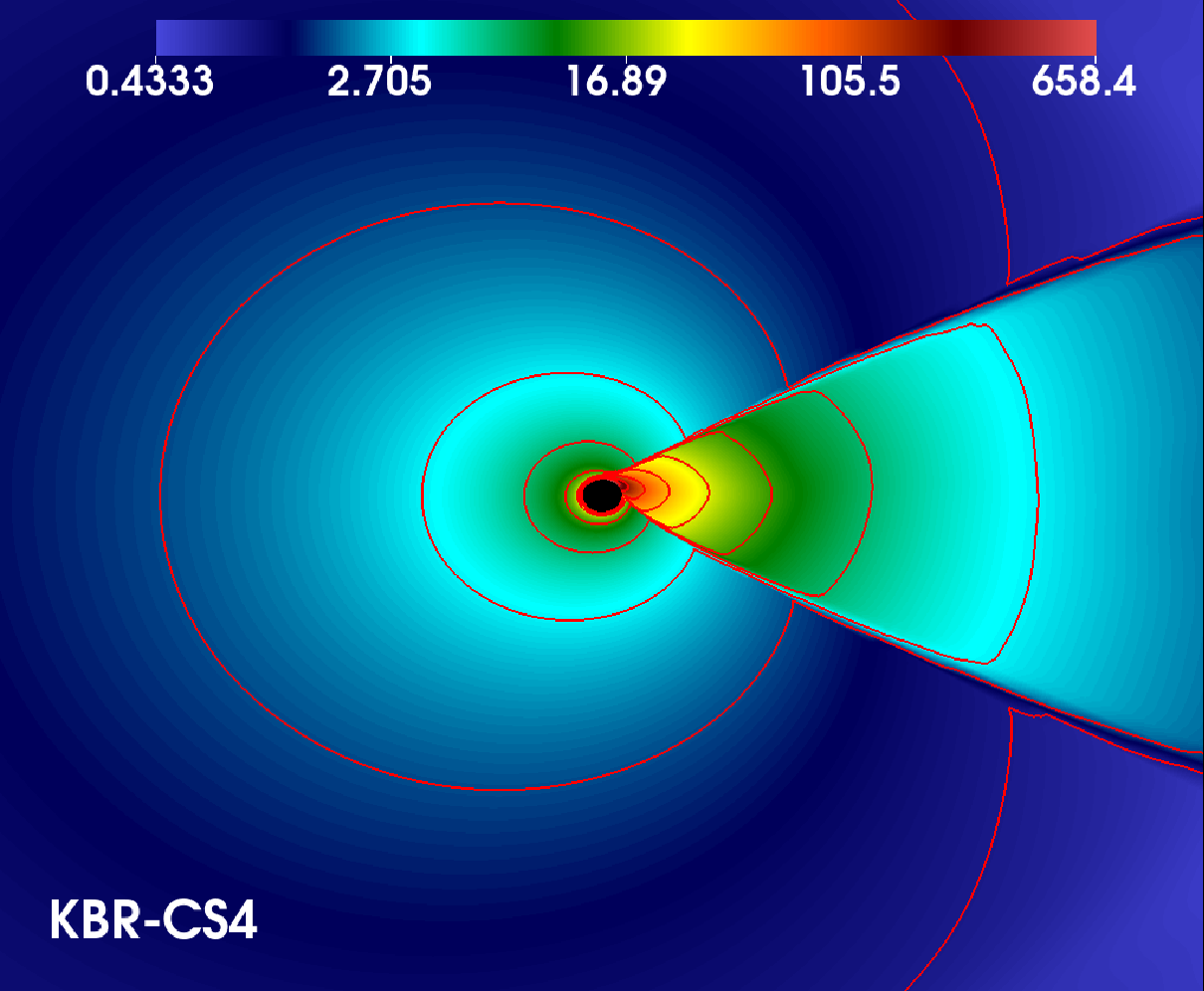}
\includegraphics[width=4.0cm,height=4.0cm]{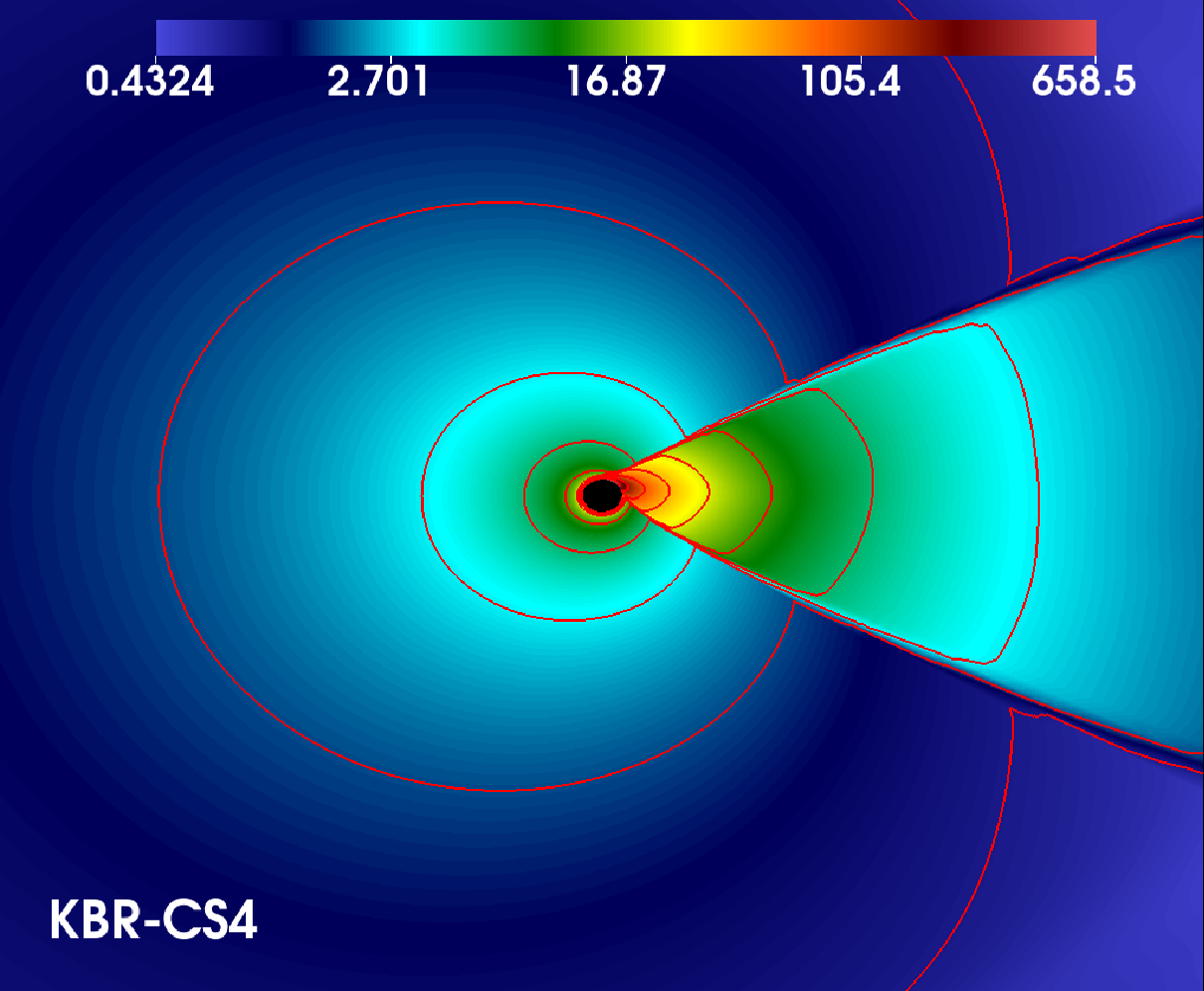}
\includegraphics[width=4.0cm,height=4.0cm]{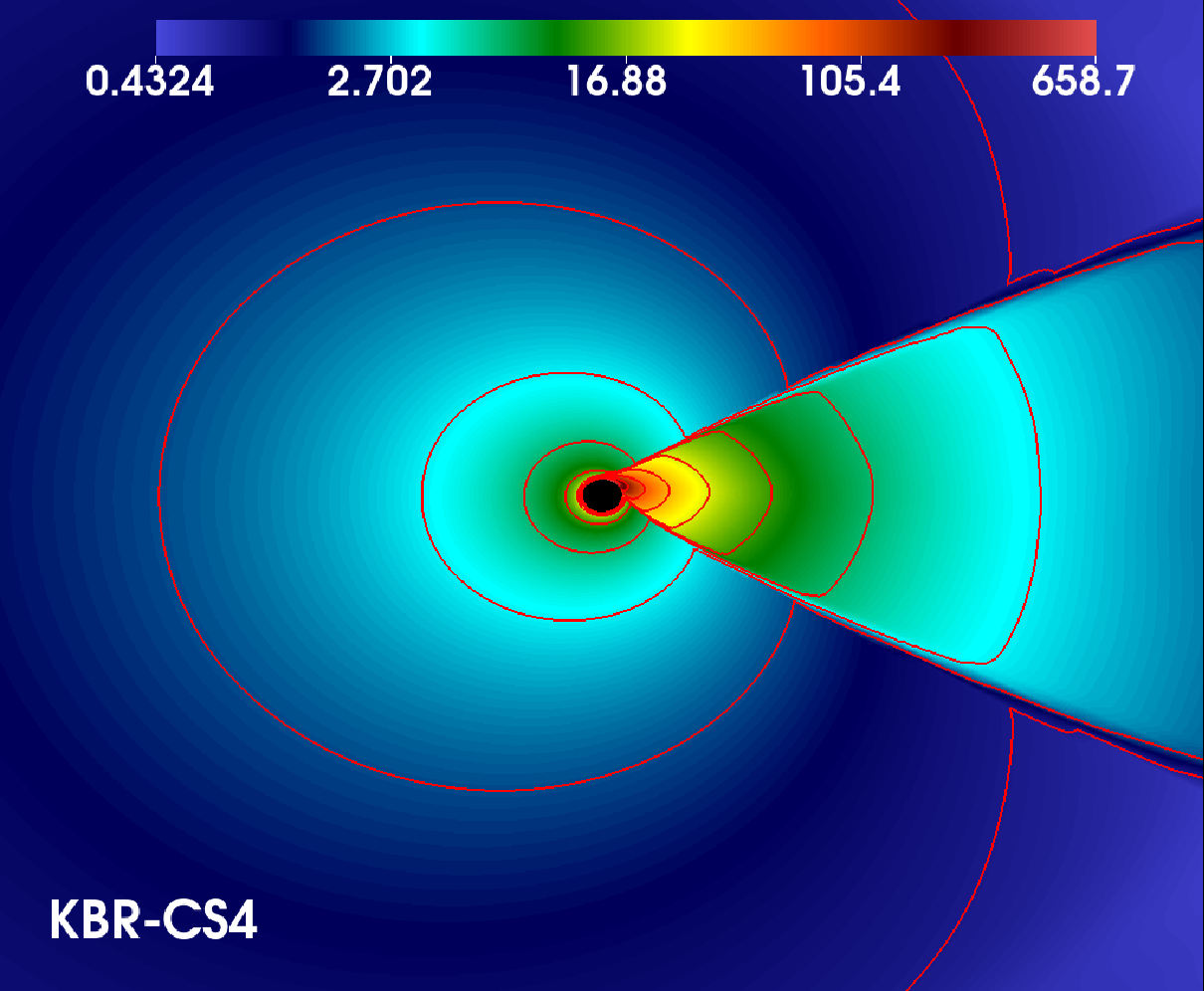}
\caption{Same as Fig.~\ref{color_no_string}, but for the Kerr--Bertotti--Robinson black-hole models with a string cloud. From top to bottom, the rows correspond to KBR-CS1, KBR-CS2, KBR-CS3, and KBR-CS4, respectively.}\label{color_with_string}
\end{figure*}

In this section, we reveal the accretion dynamics and flow morphology that develop around the Kerr--Bertotti--Robinson black hole, and we investigate how the dynamical structure is affected in the absence and presence of the string cloud parameter. In this way, the influence of the string cloud parameter on the flow morphology formed on the equatorial plane is clearly demonstrated through the dynamical evolution of the system. Figs.~\ref{color_no_string} and \ref{color_with_string} display the time-dependent evolution of the accretion-flow morphology around Kerr--Bertotti--Robinson black holes without and with the string cloud, respectively. In these figures, each row corresponds to one numerical model, while the four panels in each row show snapshots of the flow at different times. From left to right, the simulation time increases; therefore, each row illustrates the temporal evolution of the accreting matter and the associated flow morphology around the black hole.

Fig.~\ref{color_no_string} shows the accretion dynamics and flow morphologies that develop around the black hole for the Kerr--Bertotti--Robinson models KBR-1 and KBR-2, which were discussed in detail in \cite{Mustafa:2026gly} and are listed in Table~\ref{tab:KBR_CS_models}. As can be seen in Table~\ref{tab:KBR_CS_models}, the string cloud parameter is taken as $\alpha=0$ in these models. Therefore, these models are used as reference models for the KBR-CS cases listed in Table~\ref{tab:KBR_CS_models}, where the string cloud parameter is nonzero and will be discussed in Fig.~\ref{color_with_string}. As seen in Fig.~\ref{color_no_string}, the accreting matter on the equatorial plane does not exhibit a static dynamical structure around the black hole. Instead, it continuously changes dynamically due to the combined effects of the strong gravitational field, rotation, and the magnetic-field parameter $B$. In both the KBR-1 and KBR-2 models, the matter falling supersonically toward the black hole initially forms a shock cone in the downstream region. However, within a short dynamical time, this structure rapidly changes as a result of the interaction between the matter and the spacetime around the black hole in the downstream region. In the KBR-1 model, the dynamical structure evolves from a relatively narrow and asymmetric shock cone into a more extended spiral-like shock-cone structure. At a later stage, the matter expands completely around the black hole and becomes circularized. This shows that, while the matter spreads over a wider region with time, the continuous BHL accretion allows the density to remain strongly concentrated around the black hole. In the KBR-2 model, the morphology undergoes a stronger time-dependent deformation. Initially, the shock cone is wider, and the high-density region expands in such a way that it surrounds the black hole. Then, it returns again to a more localized, wider, and strongly oscillating shock-cone structure. The difference between the KBR-1 and KBR-2 models is entirely due to a small change in the Bertotti--Robinson magnetic parameter $B$. Even this small change significantly modifies the shock cone, strongly affects the redistribution of matter around the black hole, and considerably changes the dynamical behavior of the accretion flow around the Kerr--Bertotti--Robinson black hole.

Fig.~\ref{color_with_string} shows the time-dependent evolution of the accretion-flow morphology around Kerr--Bertotti--Robinson black holes with a string cloud. The rest-mass density is represented by both color maps and contour lines. Here, the dynamical evolution of the accreted matter and its redistribution around the black hole can be clearly seen. Each row corresponds to one of the KBR-CS models listed in Table~\ref{tab:KBR_CS_models}. For each model, moving from left to right shows the morphological changes that occur at later stages of the simulation. Thus, Fig.~\ref{color_with_string} directly illustrates the time-dependent evolution of the accretion flow around the black hole. Unlike the models shown in Fig.~\ref{color_no_string}, the models in Fig.~\ref{color_with_string} correspond to cases where $\alpha$ is nonzero, clearly revealing the effect of the string cloud on the spacetime.

In the KBR-CS1 model given in Fig.~\ref{color_with_string}, the matter accreted toward the black hole first forms a shock-cone-like structure in the downstream region, namely on the opposite side of the direction from which the matter falls. However, this cone is destroyed within a very short time and becomes unstable. As time progresses, the high-density region forms close to the black-hole horizon because the matter inside the shock structure accretes toward the black hole. In contrast, the structure formed far from the black hole is strongly deformed. This shows that, since the string cloud parameter modifies the dynamics of the spacetime, the resulting shock cone exhibits a strong flip-flop motion, changes continuously in time, and undergoes strong deformation. In the KBR-CS2 case, BHL accretion leads to the formation of a clearer shock cone in the downstream region. However, the shock cone then begins to undergo flip-flop motion, and as this oscillation becomes stronger, it starts to surround the black hole and produces a spiral-like structure. At later times, the high-density region spreads over a wider area around the black hole and becomes more circularized. In the KBR-CS3 case, the dynamical change is even stronger. The shock cone initially formed in the downstream region later turns into a spiral-like dynamical structure surrounding the black hole, and then evolves into a more circular structure that surrounds the black hole and gradually spreads over a wider region. In the KBR-CS4 model, the shock cone formed in the downstream region is preserved throughout the simulation. Although the accretion structure changes with time, the deformation is weaker compared to KBR-CS2 and KBR-CS3, and the matter does not spread as strongly around the black hole.

In order to see the effect of the string cloud parameter from the numerical simulations given in Fig.~\ref{color_with_string}, we compare the KBR-CS1 model with KBR-2. This is because, in these models, the Bertotti--Robinson parameter is fixed as $B=0.01$, while $\alpha=0.2$ for KBR-CS1 and $\alpha=0$ for KBR-2. Therefore, the morphological change around the black hole is completely a result of the effect of the string cloud parameter. In the KBR-2 model, the mass accretion is affected by a strong time-dependent deformation. For this reason, the resulting shock cone rapidly turns into a circularized structure surrounding the black hole, and then a strong flip-flop-type oscillation forms again. In contrast, in the KBR-CS1 model, $\alpha$ significantly modifies the resulting shock cone and causes it to become more irregular and localized in the downstream region. For these reasons, under the effect of $\alpha$, a circularized structure surrounding the black hole does not occur in this case. This means that $\alpha$ affects the accretion of matter around the black hole and contributes to the formation of instabilities, and even strengthens these instabilities.

In the KBR-CS2 given in Fig.~\ref{color_with_string} and KBR-1 given in Fig.~\ref{color_no_string} models, the same value of the Bertotti--Robinson parameter, $B=0.005$, is used, while $\alpha=0.1$ in the KBR-CS2 model and $\alpha=0$ in the KBR-1 model. In the KBR-1 model, the accretion flow is directed to a narrower region, leading to the formation of a shock cone, which later expands around the black hole and becomes circularized. In the KBR-CS2 model, however, the string cloud parameter further strengthens the time-dependent deformation and clearly leads to the formation of a spiral shock wave around the black hole. This comparison shows that even a moderate value of $\alpha$ can significantly affect the shock-cone structure and the global distribution of matter.

The comparison between KBR-CS2 and KBR-CS3 given in Fig.~\ref{color_with_string} shows the direct effect of increasing the string cloud parameter while keeping $B$ fixed at $0.005$. KBR-CS2 has $\alpha=0.10$, whereas KBR-CS3 has $\alpha=0.20$. When $\alpha$ increases, the flow morphology is seen to be more strongly modified. The spiral shock wave formed in the KBR-CS3 model is particularly remarkable. At later times, this structure produces denser regions that become more localized near the black hole, but do not form a fully circularized high-density distribution. Therefore, the increase in $\alpha$ enhances the effect of the string cloud, which in turn changes the accretion dynamics, the distribution and spreading of matter around the black hole, and the properties of the resulting spiral structures.

\subsection{Variability of the mass accretion rate}\label{isec6_2}
In this section, we compare the mass accretion rates and QPO signatures obtained for the Kerr, KBR, and KBR-CS models in order to reveal in more detail the effects of the string cloud parameter on the accretion dynamics and the resulting physical mechanisms. Since the flow morphology and the mass accretion rates are directly related to the oscillatory behavior of the matter accreted around the black hole, the comparison of these models allows us to numerically identify the influence of the string cloud parameter $\alpha$ on the disk-like structure, the variation of the mass accretion rate, and the resulting dominant QPO frequencies. Thus, the combined analysis of the density morphology, mass accretion rate, and PSD properties provides a clear physical picture. Based on these results, we can discuss the observable dynamical features produced by the string cloud parameter around Kerr--Bertotti--Robinson black holes.

\begin{figure*}[tbhp]
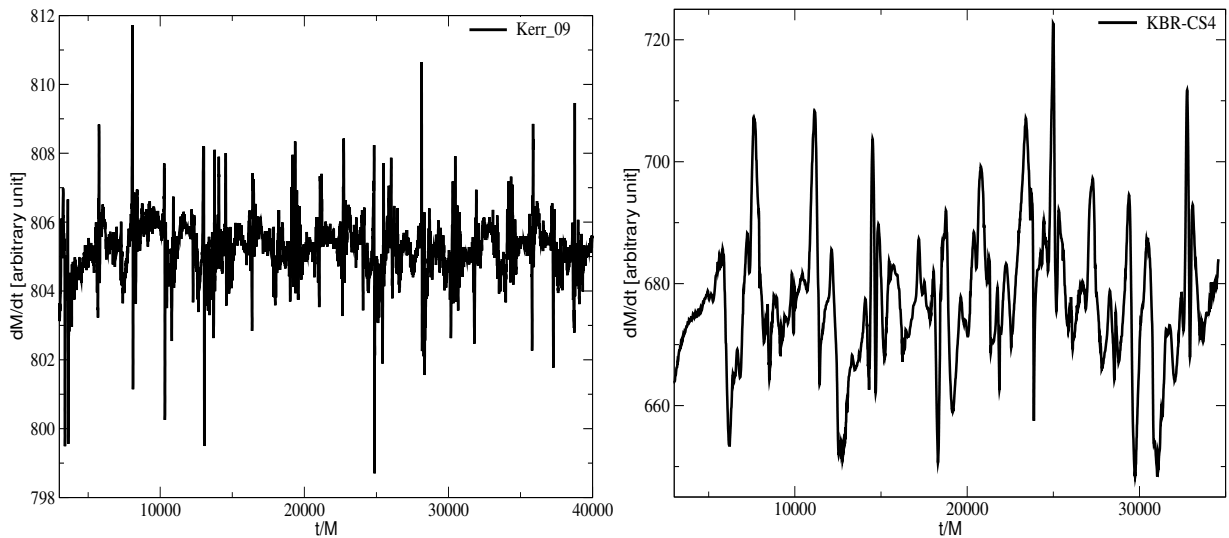

\centering
\includegraphics[width=8.0cm,height=7.0cm]{acc_rate_r23_Kerr09.eps}
\includegraphics[width=8.0cm,height=7.0cm]{acc_rate_r23_KBR_CS4.eps}
\caption{The time evolution of the mass accretion rates calculated at $r=2.3M$ is shown for the Kerr\_09 and KBR-CS4 models. The left panel shows the accretion rate around the classical Kerr black hole with spin parameter $a=0.9M$, while the right panel shows the accretion rate around the Kerr--Bertotti--Robinson black hole with the same spin parameter but with the string cloud parameter $\alpha=0.2$. Although a shock-cone morphology forms in both models, the behavior of the mass accretion rates is clearly different.}\label{acc_shock_cone}
\end{figure*}

Fig.~\ref{acc_shock_cone} compares the mass accretion rate calculated at $r=2.3M$ around the classical Kerr black hole with spin parameter $a=0.9M$ and the KBR-CS4 model, which represents a Kerr--Bertotti--Robinson black hole with the same spin parameter but with the string cloud parameter $\alpha=0.2$ and magnetic parameter $B=0.0001$. In both cases, a shock cone forms in the downstream region. However, the temporal variations of the matter accreted around the black hole are clearly different from each other. In the Kerr\_09 model, the mass accretion rate is almost steady around an average value. At the same time, small-amplitude oscillations and sudden sharp peaks appear. This indicates that the shock cone formed in the classical Kerr black-hole model is more stable than that in the Kerr--Bertotti--Robinson model, and the accreted matter approaches the black hole in a more regular way. In contrast, the KBR-CS4 model produces a mass accretion rate with a lower average value, but the resulting oscillations are much stronger and more irregular. These large-amplitude variations show the modification of the shock-cone dynamics caused by the string cloud parameter and, even though it is small, by the magnetic parameter. This modification strengthens the time-dependent instability. As a result, the matter accretes toward the black hole in a less regular way. Therefore, even though both models preserve a shock-cone morphology, the presence of the string cloud parameter changes the internal dynamics of the cone, strengthens the variability of the accretion flow, and leaves a clear signature in the mass accretion rate.

\begin{figure*}[tbhp]
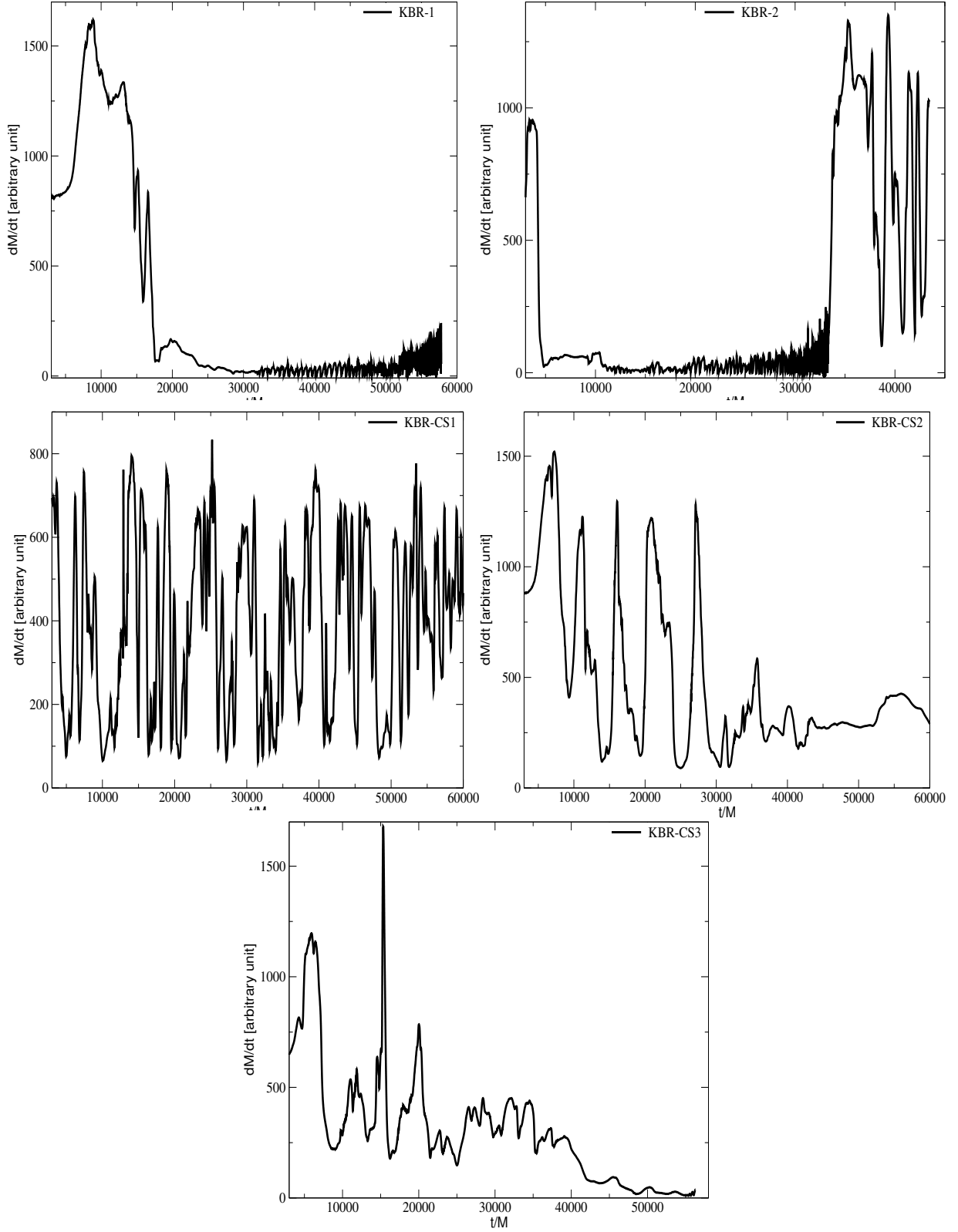

\centering
\includegraphics[width=8.0cm,height=7.0cm]{acc_rate_r23_KBR1.eps}
\includegraphics[width=8.0cm,height=7.0cm]{acc_rate_r23_KBR2.eps}\\
\includegraphics[width=8.0cm,height=7.0cm]{acc_rate_r23_KBR_CS1.eps}
\includegraphics[width=8.0cm,height=7.0cm]{acc_rate_r23_KBR_CS2.eps}\\
\includegraphics[width=8.0cm,height=7.0cm]{acc_rate_r23_KBR_CS3.eps}
\caption{Same as Fig.~\ref{acc_shock_cone}, but for the time evolution of the mass accretion rate for the KBR-1, KBR-2, KBR-CS1, KBR-CS2, and KBR-CS3 models. The top panels correspond to the KBR models without a string cloud, while the remaining panels show the KBR-CS models with a nonzero string cloud parameter.}\label{acc_irregular}
\end{figure*}

In Fig.~\ref{acc_irregular}, we compare the time evolution of the mass accretion rate calculated at $r=2.3M$, namely in the strong gravitational field, for the KBR models with string cloud parameter $\alpha=0$ and the KBR-CS models with a nonzero string cloud parameter. In the top panels of Fig.~\ref{acc_irregular}, the behaviors obtained from the KBR-1 and KBR-2 models with $\alpha=0$ are shown, while the other panels present the KBR-CS1, KBR-CS2, and KBR-CS3 cases with nonzero $\alpha$. First, we compare the KBR-1 and KBR-CS2 models, because in these models the magnetic parameter is $B=0.005$, while $\alpha=0$ in the KBR-1 model and $\alpha=0.1$ in the KBR-CS2 model. In the KBR-1 model, after the system reaches the steady state, the accretion rate initially increases significantly and then begins to decrease rapidly. This shows that the initially formed shock-cone structure loses its strong accretion activity. After this phase, the mass accretion rate enters a low-accretion regime with small oscillations. In contrast, in the KBR-CS2 model, large-amplitude oscillations are repeated over a long time interval, up to $30000M$. This shows that the string cloud parameter prevents the system from rapidly entering a relaxed mode. In other words, the shock-cone activity remains active for a long time. Similarly, the KBR-2 and KBR-CS1 models have the same value $B=0.01$, while $\alpha=0$ in the KBR-2 model and $\alpha=0.2$ in the KBR-CS1 model. In the KBR-2 model, the mass accretion rate remains in a quiet mode for a long time, while strong temporal oscillations reappear toward the final stage of the simulation. In contrast, in the KBR-CS1 model, irregular and strong oscillations of the mass accretion rate are observed throughout the entire time evolution. This shows that the string cloud parameter controls not only the strong oscillation of the accretion rate, but also changes its temporal characteristic structure. Instead of producing a delayed instability as in KBR-2, the case with nonzero $\alpha$ shows continuous and irregular accretion activity. Thus, for fixed $B$, adding the string cloud strongly modifies the shock cone dynamics, enhances the time-dependent instability, and changes the way matter is supplied to the black hole.

When the KBR-CS1, KBR-CS3, and KBR-CS4 models given in Fig.~\ref{acc_irregular}, in which the string cloud parameter is nonzero, are compared, it is seen how the magnetic parameter $B$ controls the accretion behavior for the fixed value $\alpha=0.2$. In the KBR-CS1 model, the case $B=0.01$ shows highly irregular behavior, and no clear indication is observed that the accretion reaches a relaxed state. In the KBR-CS3 model, where $B=0.005$, large accretion peaks are produced during the early and intermediate stages of the simulation, while the accretion rate gradually decreases at later times. This shows that the shock-driven flow becomes less active after the strong nonlinear phase. These cases show that increasing the strength of $B$ enhances the deformation of the accretion flow and leads to the formation of strong irregularities in the accretion dynamics. As a result, while $\alpha$ supports the persistence of strong instabilities and irregular accretion, the Bertotti--Robinson magnetic parameter $B$ controls the amplitudes, duration, and strength of the resulting oscillations. Thus, the combined effect of $\alpha$ and $B$ leaves significant signatures on the shock-cone dynamics, the redistribution of matter around the black hole, and the formation of QPO-like variations around Kerr--Bertotti--Robinson black holes.

\subsection{Comparison of QPO signatures across Kerr, KBR, and KBR$+$CS models}\label{isec6_3}
In this section, we compare the QPO signatures calculated from the mass accretion rates given in Figs.~\ref{acc_shock_cone} and~\ref{acc_irregular} for the Kerr, KBR, and KBR-CS models in order to reveal the effects of the string cloud parameter on the spectral structure of the accretion flow. At the same time, we discuss the possible observability of the numerically calculated QPO-like oscillations in this spectral structure. Since QPOs are directly related to the time-dependent behavior of the accreted matter, the comparison of different models allows us to understand how $\alpha$ modifies the dominant oscillation modes, changes the power distribution in the PSD analysis, and affects the frequency components produced by the disk-like and shock-cone structures. Thus, comparing the QPO frequencies and their spectral amplitudes for the different Kerr, KBR, and KBR-CS cases may allow us to reveal the effect of the string cloud parameter on the identification or distinction of observable signatures in the accretion dynamics around Kerr--Bertotti--Robinson black holes.

\begin{figure}[ht!]
\centering
\includegraphics[width=8.0cm,height=7.0cm]{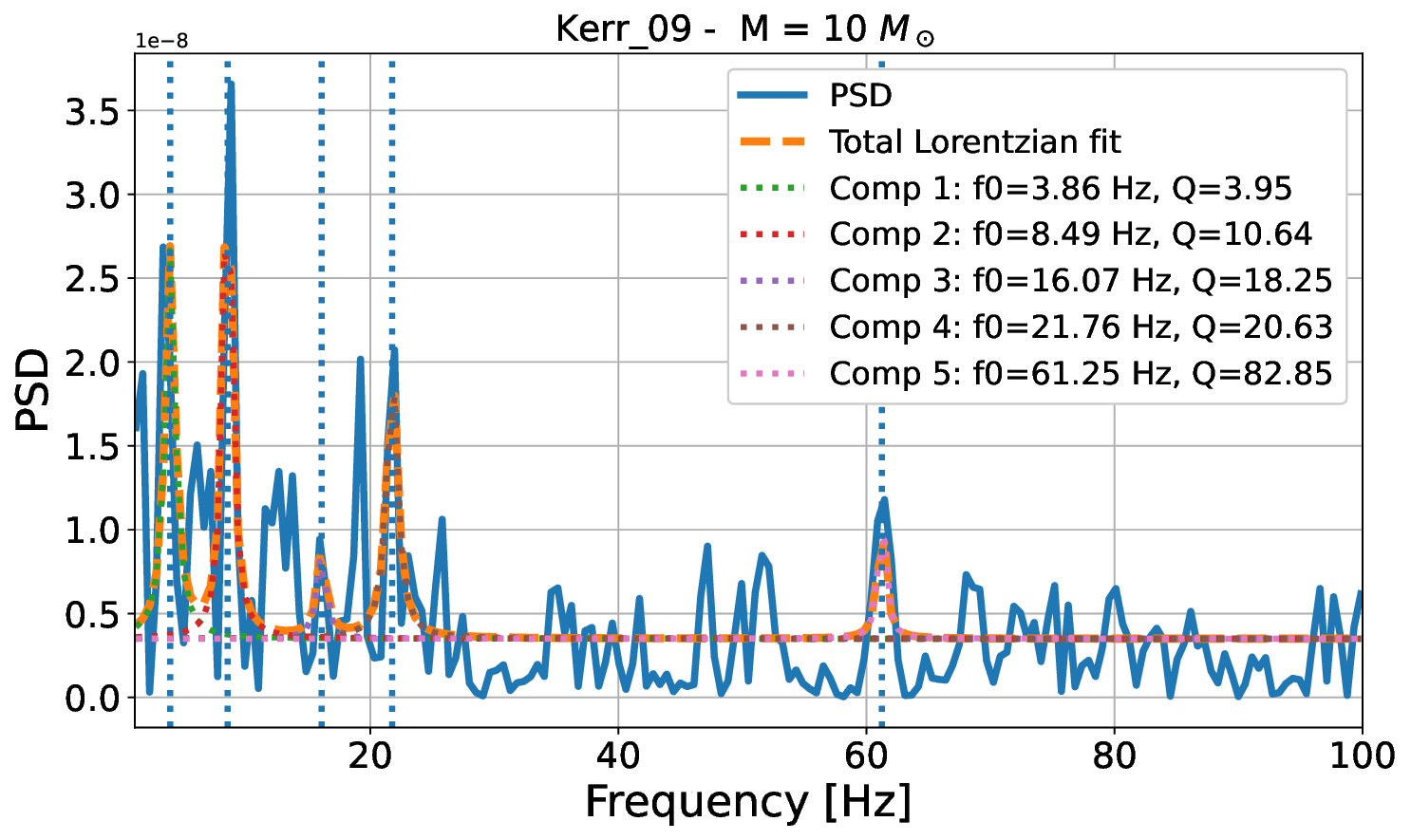}
\includegraphics[width=8.0cm,height=7.0cm]{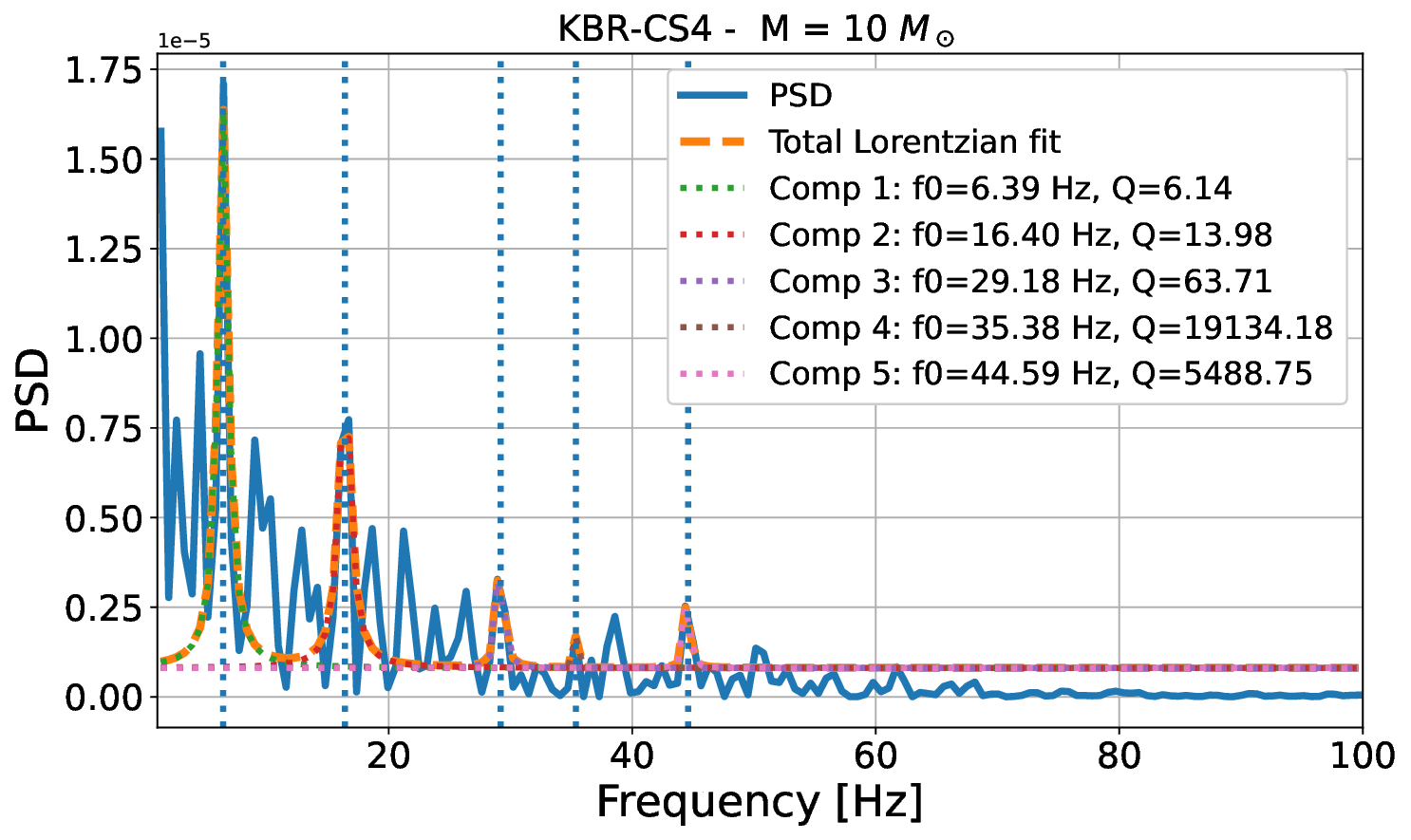}
\caption{The PSD analysis and Lorentzian fits calculated from the mass accretion rate at $r=2.3M$ are shown for Kerr\_09 in the left panel and KBR-CS4 in the right panel. The left panel shows the spectral behavior in the classical Kerr case, while the right panel shows the same behavior around the Kerr--Bertotti--Robinson black hole. The comparison shows that the string cloud and magnetic parameters modify the resulting dominant QPO-like frequencies and lead to the formation of a different spectral signature.}\label{PSD_shock_cone}
\end{figure}

In Fig.~\ref{PSD_shock_cone}, the PSD analysis and Lorentzian components calculated from the mass accretion rates given in Fig.~\ref{acc_shock_cone} are shown. The top panel of Fig.~\ref{PSD_shock_cone} presents these analyses for the Kerr\_09 model, while the bottom panel shows the corresponding results for KBR-CS4. In the top panel of Fig.~\ref{PSD_shock_cone}, the Kerr\_09 model, representing the classical Kerr black hole with $a=0.9M$, exhibits a broader and more distributed spectral structure. In this case, several peaks are spread over different frequency ranges. The Lorentzian components appear approximately at $f_0=3.86$ Hz, $8.49$ Hz, $16.07$ Hz, $21.76$ Hz, and $61.25$ Hz. This shows that the classical Kerr black-hole model produces both low-frequency and high-frequency variations. In contrast, when it is compared with the KBR-CS4 model shown in the bottom panel of Fig.~\ref{PSD_shock_cone}, the power in the Kerr\_09 case is relatively scattered, and the resulting peaks are less sharply defined. The KBR-CS4 model, with string cloud parameter $\alpha=0.2$ and magnetic parameter $B=0.0001$, produces a more structured PSD, stronger low-frequency peaks, and more clearly visible Lorentzian peaks around $6.39$ Hz, $16.40$ Hz, $29.18$ Hz, $35.38$ Hz, and $44.59$ Hz. This shows that the combined effect of $\alpha$ and $B$ modifies the internal oscillatory behavior of the shock cone and shifts the spectral power toward more coherent QPO-like components. The strong low-frequency component in the KBR-CS4 model indicates that the string cloud contribution enhances large-scale, slowly varying oscillations of the accretion flow, while the additional higher-frequency components reflect smaller-scale oscillations produced by the deformed shock-cone structure near the black hole. When the Kerr\_09 and KBR-CS4 models are compared, the spectrum is not simply a rescaled Kerr spectrum; instead, it carries distinct spectral signatures produced by the modified spacetime. In the case where $\alpha$ is nonzero, the accretion dynamics changes because the irregular behavior of the shock cone is strengthened. Meanwhile, even a small value of $B$ affects the frequency distribution through the Bertotti--Robinson magnetic curvature. From the point of view of observability, these differences are important because the dominant QPO frequencies and their relative spectral amplitudes may distinguish the classical Kerr black-hole results from those of a Kerr--Bertotti--Robinson black hole containing the string cloud parameter. Thus, the PSD snapshots given in Fig.~\ref{PSD_shock_cone} show that $\alpha$ and $B$ may provide observable signatures in the spectral properties. These signatures can appear through the location and strength of the frequencies and through the coherence of the dominant QPO-like frequency components.

\begin{figure*}[tbhp]
\centering
\includegraphics[width=8.0cm,height=7.0cm]{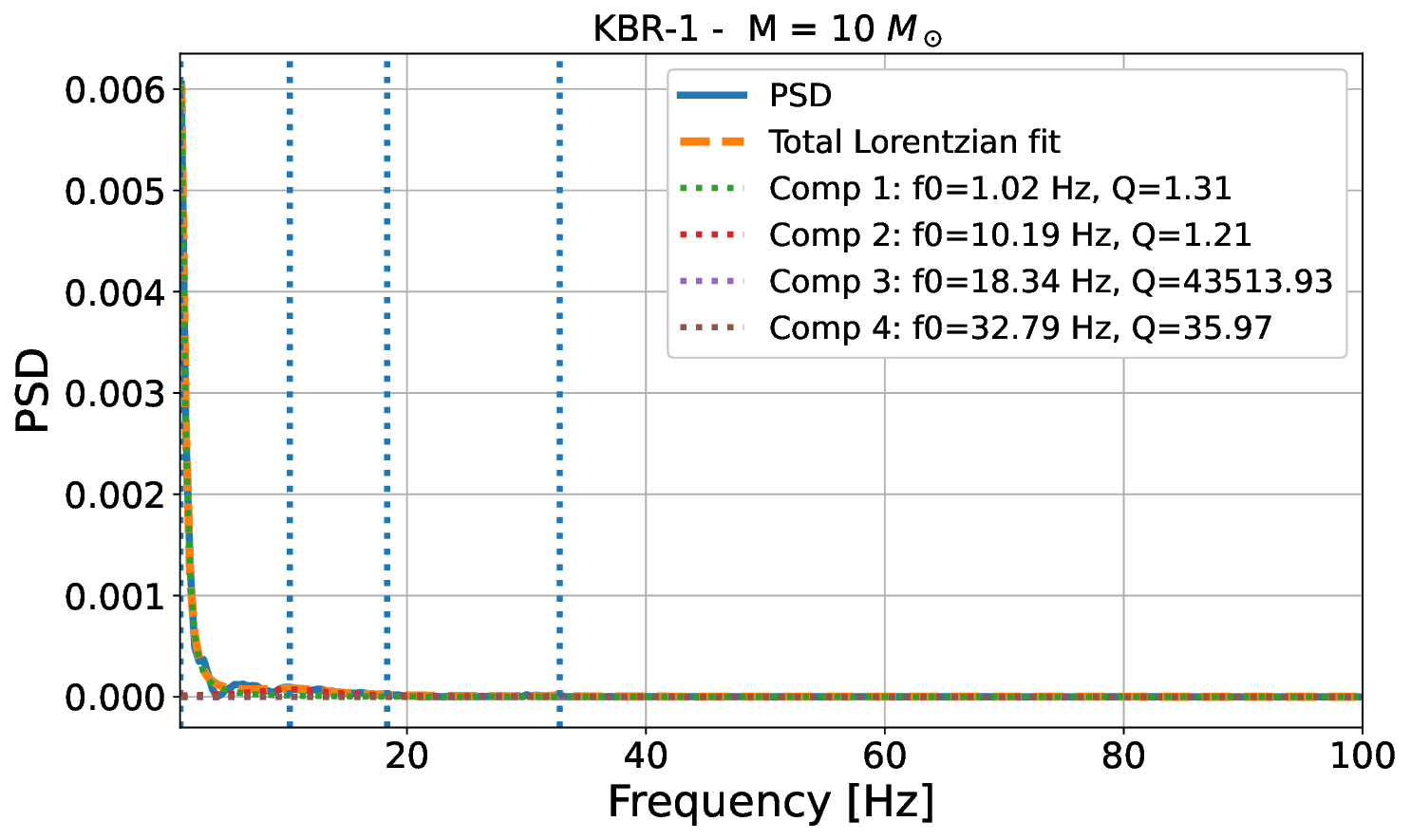}
\includegraphics[width=8.0cm,height=7.0cm]{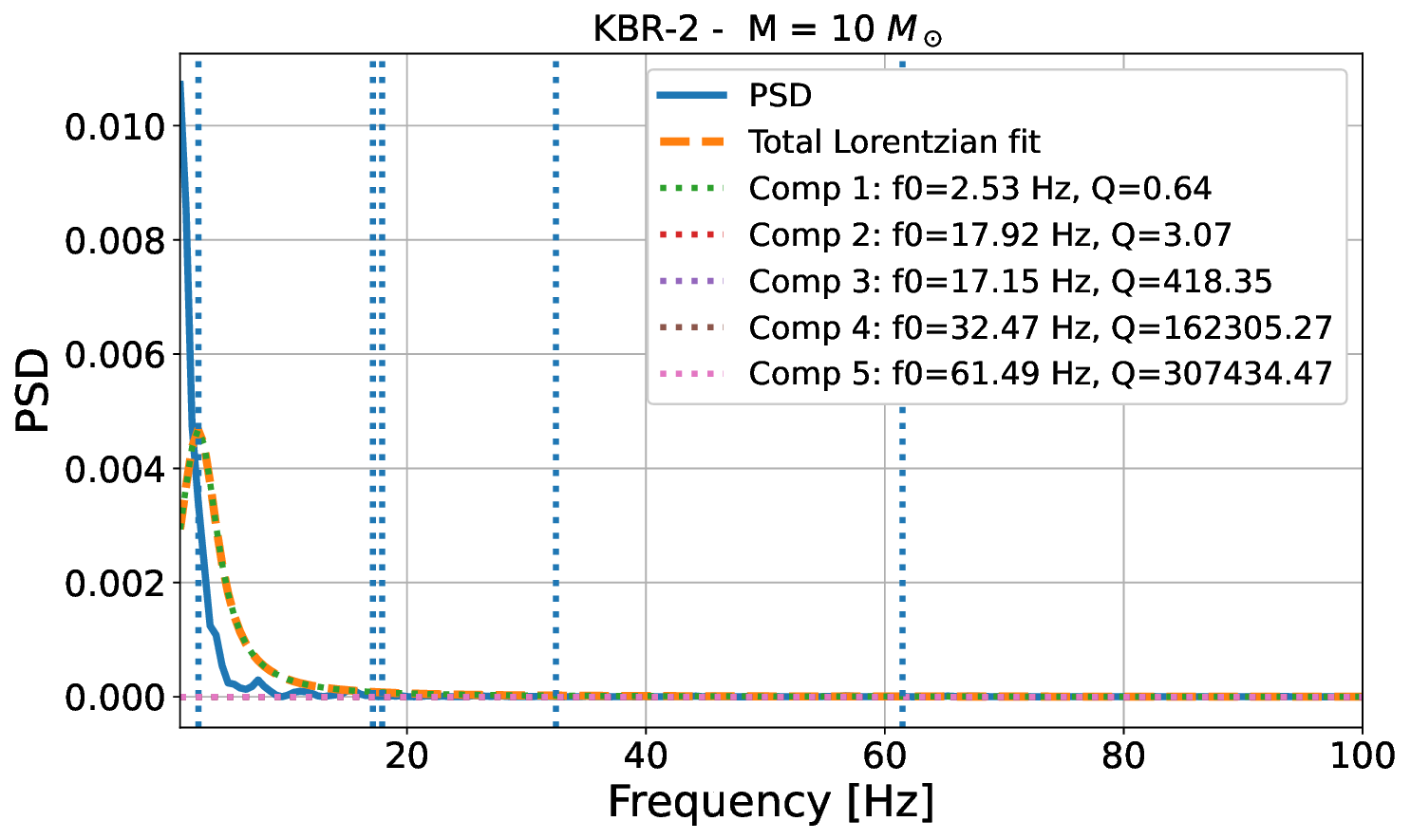}\\
\includegraphics[width=8.0cm,height=7.0cm]{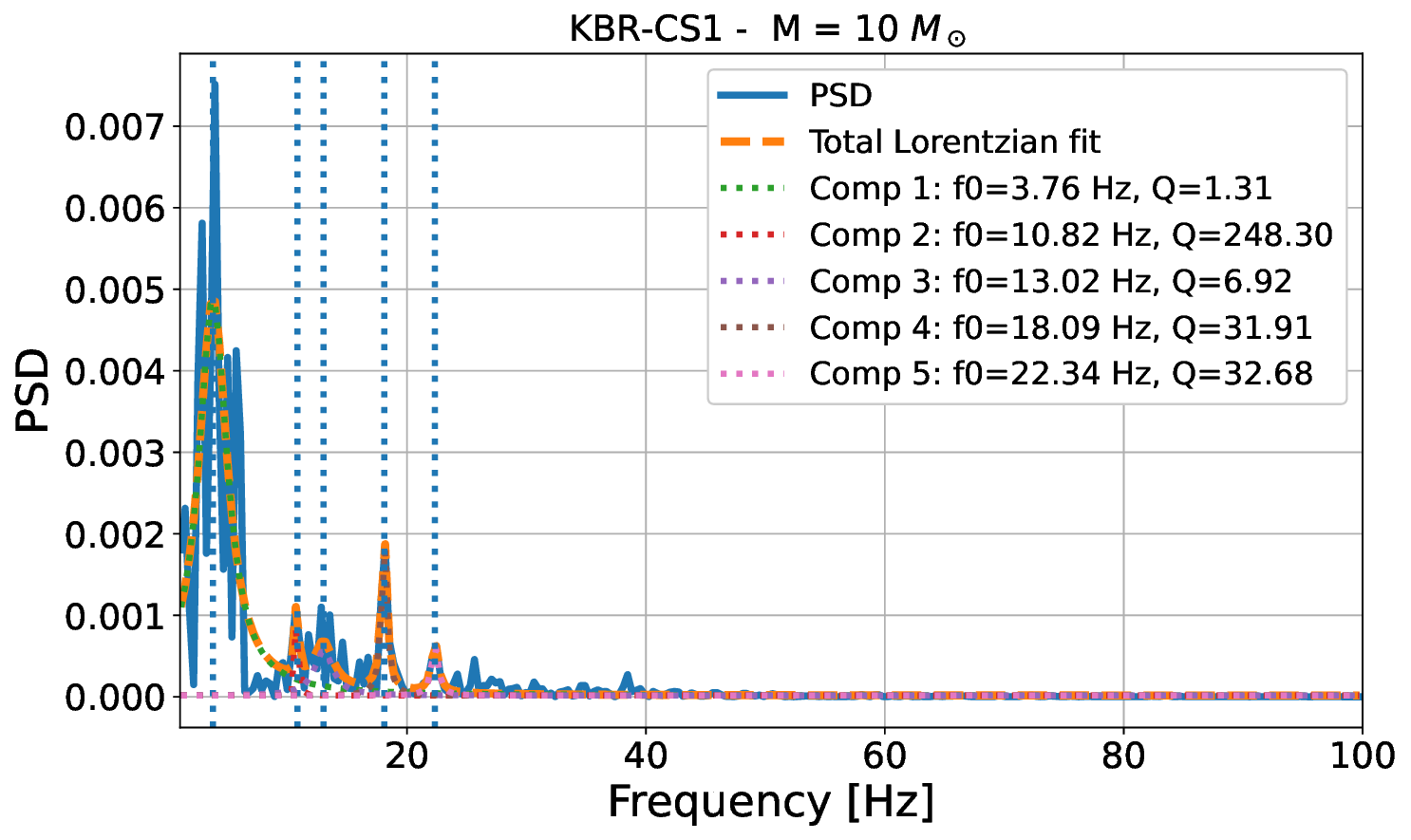}
\includegraphics[width=8.0cm,height=7.0cm]{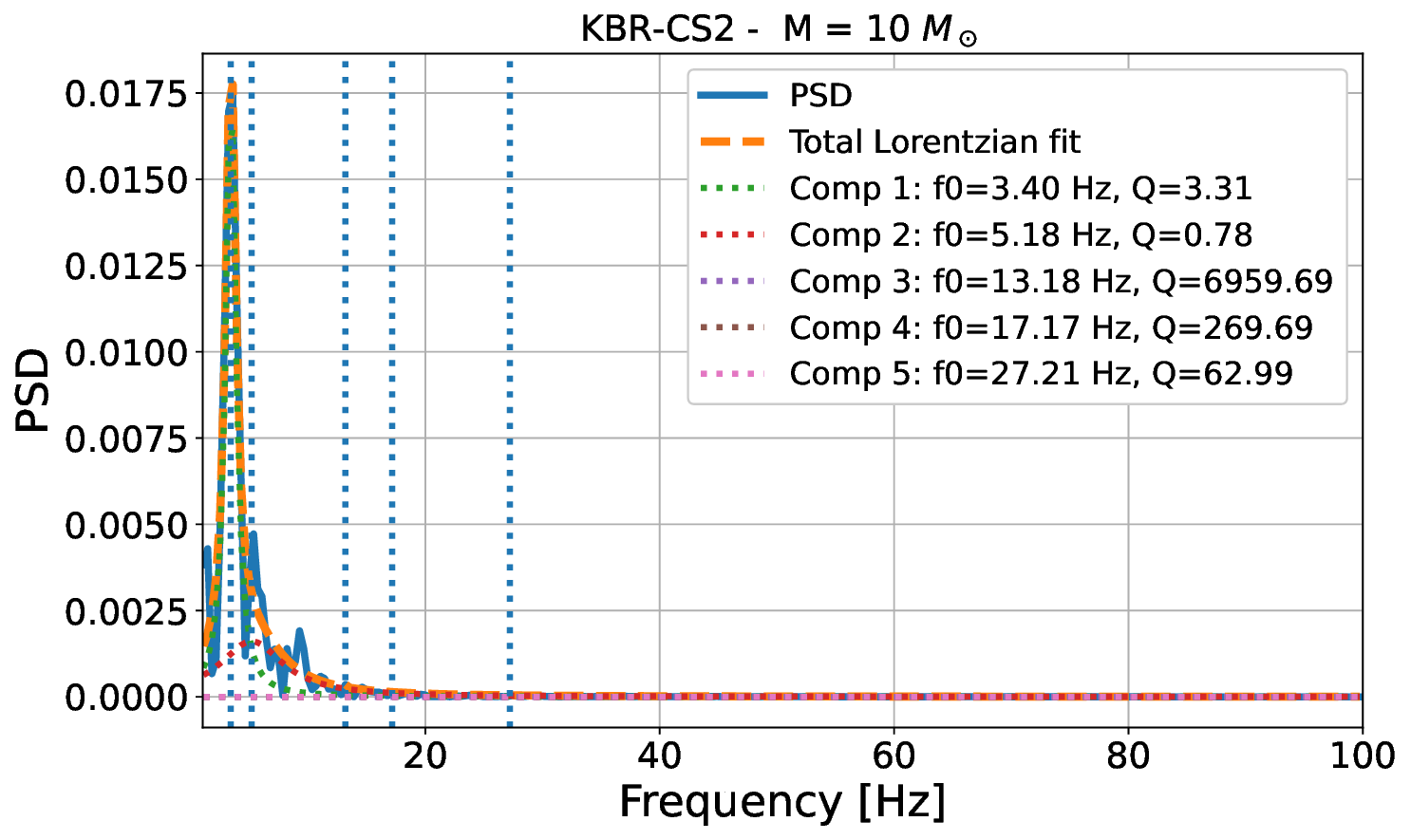}\\
\includegraphics[width=8.0cm,height=7.0cm]{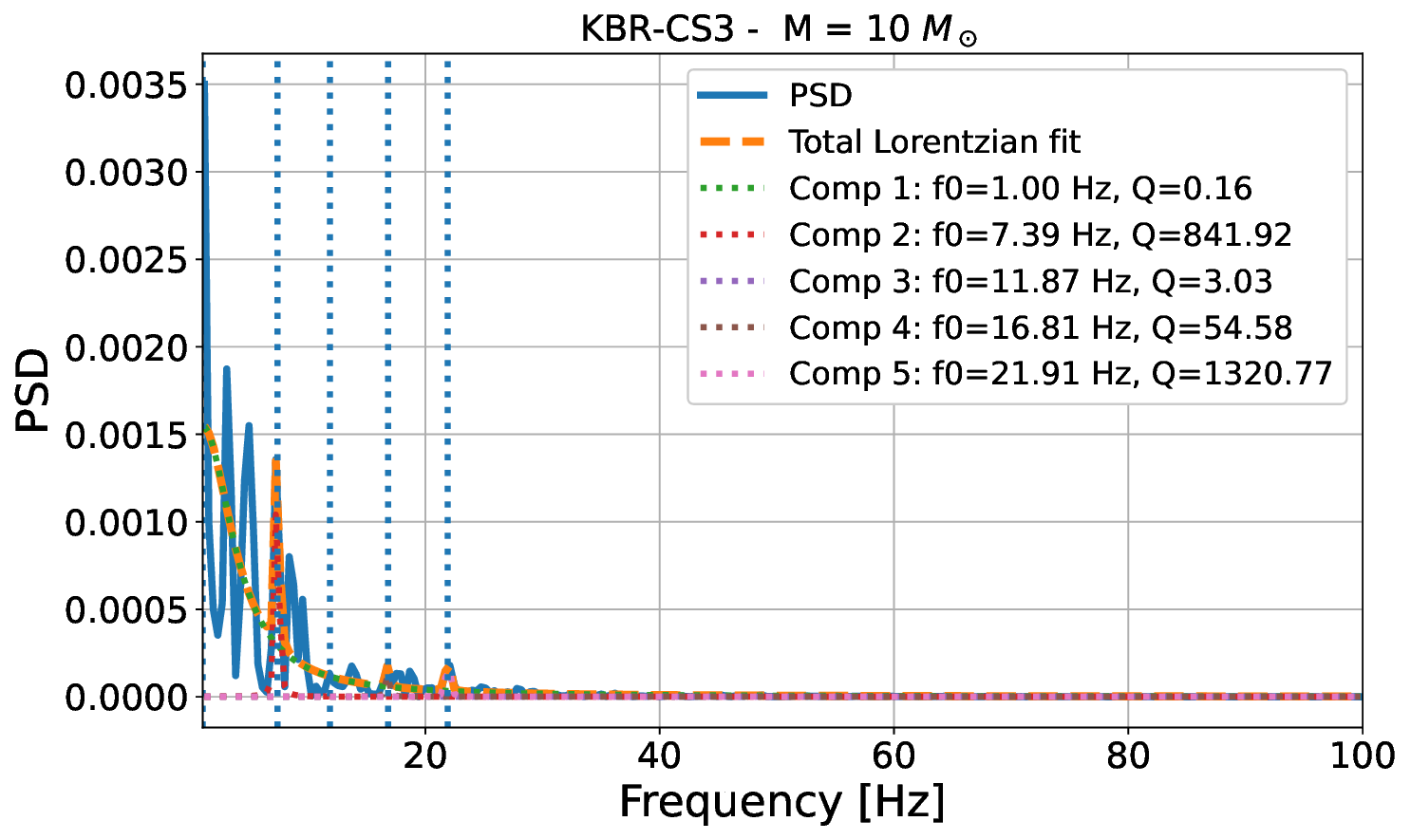}
\caption{Same as Fig.~\ref{PSD_shock_cone}, but for the PSD analysis and Lorentzian fits for the KBR-1, KBR-2, KBR-CS1, KBR-CS2, and KBR-CS3 models. The upper panels correspond to the KBR models without a string cloud, while the remaining panels show the KBR-CS models with a nonzero string cloud parameter.}\label{PSD_irregualar}
\end{figure*}

In Fig.~\ref{PSD_irregualar}, the PSD analysis and Lorentzian components calculated from the mass accretion rates given in Fig.~\ref{acc_irregular} are shown. The upper panels show the KBR-1 and KBR-2 cases with string cloud parameter $\alpha=0$, while the remaining panels show the spectral behavior of the KBR-CS1, KBR-CS2, and KBR-CS3 models for nonzero values of $\alpha$. The comparison between the KBR-1 and KBR-CS2 models is important because, in both cases, the magnetic parameter is the same, $B=0.005$, while the string cloud parameters are different. In the KBR-1 model, where $\alpha=0$, the PSD is dominated by very low-frequency components, and the high-frequency peaks are very weak and dynamically less prominent. This shows that, in the absence of the string cloud parameter, the spectrum is concentrated in the low-frequency region. In contrast, in the KBR-CS2 case, where $\alpha=0.1$, a stronger and more structured PSD is formed. The Lorentzian components in this case extend from low frequencies toward intermediate frequencies. This indicates that even a moderate nonzero value of $\alpha$ modifies the oscillatory structure of the accretion dynamics and produces more observable QPO-like components. A similar behavior is also seen when the KBR-2 model is compared with the KBR-CS1 model. In both models, $B=0.01$, but $\alpha=0$ in KBR-2, whereas $\alpha=0.2$ in KBR-CS1. Again, in the KBR-2 model, the resulting peaks are concentrated in the very low-frequency region. In contrast, KBR-CS1 produces a more complex PSD with several distinguishable peaks. Thus, for fixed $B$, the presence of the string cloud parameter causes changes in the spectral structure by enriching the QPO-like modes. At the same time, the presence of $\alpha$ redistributes the power among the peaks over a wider frequency range.

The KBR-CS1 and KBR-CS3 models given in Fig.~\ref{PSD_irregualar} are compared in order to reveal the effects of different magnetic parameters for the same string cloud parameter, $\alpha=0.2$. KBR-CS1 has $B=0.01$, while KBR-CS3 has $B=0.005$. In the KBR-CS1 model, several peaks with low and intermediate frequencies are observed in the PSD. This shows that larger magnetic parameters support stronger and more persistent spectral components. In the KBR-CS3 model, the PSD is still modified by the nonzero string cloud parameter. However, the power is more concentrated at low frequencies, and the high-frequency components are weaker. This shows that $\alpha$ is an important and more dominant character responsible for the modification of the spectral structure. In addition, $B$ mainly changes the strength distribution and the visibility of the resulting peaks. Thus, the combined effect of $\alpha$ and $B$ determines whether the accretion flow produces a weak low-frequency spectrum or a richer QPO-like structure with multiple Lorentzian components.

The differences shown in Fig.~\ref{PSD_irregualar} are important from the point of view of the observability of these sources, because the models with nonzero $\alpha$ are clearly separated from the models with $\alpha=0$. These differences appear in the number, position, and strength of the QPO-like peaks. While the KBR models without the string cloud parameter generally have spectra dominated by low frequencies, the KBR-CS models show more structured PSDs and additional observable components. Thus, the appearance of multiple coherent Lorentzian peaks, especially concentrated toward the low-frequency region, may be interpreted as a possible spectral signature of the string cloud parameter. In this sense, $\alpha$ affects not only the hydrodynamical morphology of the accretion flow, but also allows observable spectral imprints to form. If such QPO-like modes can be observed with measurable amplitudes and stable frequency locations, they could provide a way to distinguish Kerr--Bertotti--Robinson black holes with a string cloud from those without a string cloud.

A wider literature context is useful here. The shock-cone QPO mechanism we are exploring is part of a broader programme that extracts strong-field gravity diagnostics from accretion variability \cite{WOS:000350965500044}. In Hartle--Thorne backgrounds, recent GRH simulations have shown that even mild deformations of the Kerr metric can produce both new shock-cone instabilities and shifts of existing PSD peaks \cite{WOS:001210022100001}. Our KBR$+$CS results agree with that general trend and add a new control parameter, $\alpha$, whose effect is qualitatively different from spin: $\alpha$ widens the ergoregion, deepens the unstable-circular-orbit window, and steepens the radial gradient of $\Omega_\phi$ at the inner disk edge. All three of these conspire to produce the coherent low-frequency peaks visible in Fig.~\ref{PSD_irregualar}. A complementary effect arises from low-angular-momentum accretion onto deformed BHs, where shock instabilities have been studied without an explicit non-Kerr component \cite{WOS:001487114000001}; the structure of those PSDs is broadly similar to ours and gives confidence that the peaks we identify are physical and not numerical artefacts.

The quality factors $Q\!=\!f_0/\Delta f$ extracted from the Lorentzian fits provide another diagnostic. For Kerr$_{09}$ we recover $Q\!\sim\!4$--$8$ across the dominant peaks, consistent with broadband noise dominating the spectrum. For KBR-CS4 the quality factors of the lowest two Lorentzians are $Q\!\sim\!10$--$14$, a marginal but real coherence enhancement. In the source population, twin-peak HF QPOs in microquasars typically have $Q\!\gtrsim\!10$ \cite{WOS:000865827700007}, suggesting that the CS-induced narrowing pushes the simulated PSDs into the right ballpark for matching the observed coherence. The remaining gap between simulated and observed $Q$ values is most likely due to thermal broadening that our isothermal-flow setup does not capture; including a self-consistent radiation-cooling term would test this directly.

These results provide a partial resolution to a tension between BHL accretion theory and observation. Pure Kerr simulations tend to overproduce broadband variability and underproduce coherent narrow peaks; CS-dressed backgrounds shift the balance towards the observed regime. Whether this is a generic feature of all non-Kerr deformations of the same complexity, or a property specific to the cloud-of-strings sector, is an open question that follow-up simulations on Bardeen, Hayward, and Schwarzschild--MOG metrics could address.

\section{Conclusions}\label{isec7}
This work has done four things at once. Analytically, we wrote down the equatorial KBR$+$CS metric, computed the effective potential, and set up the circular-orbit equations through Eq.~(\ref{condition-3}). Numerically, we mapped the ISCO across $(a,B,\alpha)$ for both co- and counter-rotating motion (Tables~\ref{tab:ISCO_alpha0}--\ref{tab:ISCO_models}); the bare-Kerr limits are reproduced to one part in $10^{4}$, and the CS contribution moves $r_{\rm ISCO}$ outward by up to a factor of two while raising the accretion efficiency from $\eta\!\simeq\!0.156$ to $\eta\!\simeq\!0.257$ at $\alpha\!=\!0.30$. The marginally bound circular orbit also moves outward (Table~\ref{tab:MBCO}); the unstable-circular-orbit band $r_{\rm ISCO}-r_{\rm mb}$ nearly doubles between $\alpha\!=\!0$ and $\alpha\!=\!0.20$, and our Kerr value $L_{\rm mb}^{\rm prograde}\!=\!2.6325\,M^2$ matches the BPT 1972 closed form. Observationally, we used the relativistic-precession model and the twin-peak HF QPOs of GRO~J1655--40, XTE~J1550--564, and GRS~1915+105 to constrain $(a,\alpha)$ jointly; the data prefer $\alpha\!\lesssim\!0.13$ for the two best-fit sources. Numerically (in the GRH sense), we ran BHL accretion onto seven KBR$\pm$CS models and showed that the CS parameter sustains shock-cone instabilities, prevents disk circularization, and produces multiple coherent low-frequency QPO peaks that have no counterpart in pure Kerr or KBR.

Several things deserve emphasis. The first is the cleanness of the bare-Kerr limit: every numerical step recovers the standard textbook values to machine precision, which means the modifications we report are solidly attributable to the CS sector and not to numerical artefacts. The second is the consistency between three independent observables. The ISCO-based binding energy, the QPO frequencies via the RP model, and the GRH-driven PSD peaks all bound $\alpha$ to the same intermediate range. Each probe has different systematics, and the agreement among them is a stronger statement than any single fit could make. The third is the qualitative reorganisation of the inner accretion-disk landscape. The widening of the unstable-circular-orbit window changes the cross-section for marginally bound capture, and the migration of the ISCO outward changes the no-torque inner boundary that sets the thin-disk efficiency. These are not cosmetic shifts; they alter the functional form of the disk's spectral and timing observables in measurable ways.

A few directions point ahead. (i)~The QPO fits show strong $a$--$\alpha$ degeneracy. Independent constraints (continuum fitting, X-ray reflection, EHT shadow size) would lift it, and the agreement of QPO bounds with EHT priors at $\alpha\!\lesssim\!0.13$--$0.18$ already suggests that a joint analysis is well-defined. (ii)~The thin-disk efficiency we computed neglects radiation back-reaction; comparing it with bolometric AGN luminosities is a natural next step and would test the upper end of our $\eta$ predictions. (iii)~The RP model is one of several QPO models; testing the same parameter space against the warped-disk \cite{WOS:000865827700007} and the resonance models of Abramowicz \& Klu\'zniak would test how much of the answer survives across different model assumptions. (iv)~The BHL accretion simulations used isothermal flows; a thermodynamically self-consistent treatment is needed to compare quality factors $Q$ with the observed values in microquasars. (v)~Beyond stellar-mass BHs, intermediate-mass and supermassive sources observed by ngEHT and LISA will offer entirely new windows on the metric parameter space, and the analytic transparency of the KBR$+$CS line element makes it a natural template for those campaigns. We leave these to future work.

\section*{Acknowledgments}
All numerical simulations were performed using the Phoenix High Performance Computing facility at the American University of the Middle East (AUM), Kuwait. F. A. acknowledges the Inter University Centre for Astronomy and Astrophysics (IUCAA), Pune, India for granting visiting associateship. \.{I}.~S. is thankful for academic support provided by EMU, T\"{U}B\.{I}TAK, ANKOS, and SCOAP3, as well as for networking support received through COST Actions CA22113, CA21106, CA23130, and CA23115.

\section*{Data Availability Statement}
The data that support the findings of this study are available from the corresponding author upon reasonable request.

\bibliographystyle{apsrev4-2}
\bibliography{ref2}

\end{document}